\documentclass[prb,twocolumn]{revtex4}
\usepackage[dvips]{graphicx} 
\usepackage{latexsym}
\begin{document}

\newcommand{\fig}[2]{\includegraphics[width=#1]{#2}}

\title{Broad relaxation spectrum and the field 
theory of glassy dynamics for pinned elastic systems}

\author{Leon Balents}
\affiliation{Department of  Physics, University of California,
Santa Barbara, CA 93106--4030}
\author{Pierre Le Doussal}
\affiliation{CNRS-Laboratoire de Physique Th{\'e}orique de
l'Ecole Normale Sup{\'e}rieure, 24 rue Lhomond, 75231 Cedex 05, Paris
France.}

\date{\today}

\begin{abstract}
  We study thermally activated, low temperature equilibrium dynamics
  of elastic systems pinned by disorder using one loop functional
  renormalization group (FRG). Through a series of increasingly
  complete approximations, we investigate how the field theory reveals
  the glassy nature of the dynamics, in particular divergent barriers
  and barrier distributions controling the spectrum of relaxation
  times. First, we na\"ively assume a single relaxation time $\tau_k$
  for each wavevector $k$, leading to analytical expressions for
  equilibrium dynamical response and correlations.  These exhibit two
  distinct scaling regimes (scaling variables $T k^\theta \ln t$ and
  $t/\tau_k$, respectively, with $T$ the temperature, $\theta$ the
  energy fluctuation exponent and $\tau_k \sim e^{c k^{-\theta}/T}$)
  and are easily extended to quasi-equilibrium and aging regimes. A
  careful study of the dynamical operators encoding for fluctuations
  of the relaxation times shows that this first approach is
  unsatisfactory.  A second stage of approximation including these
  fluctuations, based on a truncation of the dynamical effective
  action to a random friction model, yields a size ($L$) dependent
  log-normal distribution of relaxation times (effective barriers
  centered around $L^{\theta}$ and of fluctuations $\sim
  L^{\theta/2}$) and some procedure to estimate dynamical scaling
  functions.  Finally, we study the full structure of the running
  dynamical effective action within the field theory. We find that
  relaxation time distributions are non-trivial (broad but not
  log-normal) and encoded in a closed hierarchy of FRG equations
  divided into levels $p=0,1,\ldots$ corresponding to vertices
  proportional to the $p$-th power of frequency $\omega^p$.  We show
  how each level $p$ can be solved independently of higher ones, the
  lowest one ($p=0$) comprising the statics.  A thermal boundary layer
  ansatz (TBLA) appears as a consistent solution.  It extends the one
  discovered in the statics which was shown to embody droplet thermal
  fluctuations.  Although perturbative control remains a challenge,
  the structure of the dynamical TBLA which encodes barrier
  distributions opens the way for deeper understanding of the field
  theory approach to glasses.
\end{abstract}

\maketitle

\def\asinh{{\rm asinh \ }}
\def\atan{{\rm atan \ }}

\section{Introduction}
\label{introduction}

Extremely slow dynamics is a ubiquitous property of complex and
disordered materials. Despite many decades of research, current
understanding of such {\sl glassy} motion is limited to
phenomenological models,\cite{fisher_huse} mean-field theory
,\cite{Cuku,Cuku2,Cukureview} and abstract caricatures in terms of the
dynamics of small numbers of degrees of freedom in a complex energy
landscape.\cite{aging_traps,sibani} Besides exact solutions in one
dimension (e.g. for the random field Ising model \cite{rfim1d}) little
is known about the non-equilibrium behaviour of realistic models.
True disordered materials, from spin glasses to supercooled liquids to
the pinned elastic medium, involve extensive numbers of local degrees
of freedom such as atoms and spins moving {\sl collectively } in a
random environment,\cite{Vihaoc,anderson,blatter_vortex_review} with
either external or self-induced randomness. The pinned elastic medium
being the simplest model involving such physics we study it as a
prototype. It is of interest by itself for numerous experimental
systems such as vortex lattices in
superconductors,\cite{blatter_vortex_review,Legi} interfaces in
magnets,\cite{nattermann_book_young,creepexp} charge density
waves,\cite{gruner_revue_cdw} Wigner crystals.\cite{andrei_wigner_2d}
The equation of motion is
\begin{eqnarray}
        \eta \partial_t u_{rt} = c \nabla^2_r u_{rt} + f(u_{rt},r) +
        \zeta(r,t), \label{eom}
\end{eqnarray}
where $u(r)$ is a height (or displacement) field, $\eta$ a bare
frictional damping coefficient, $c$ is the elastic modulus, $f(u,r)$
is the quenched random pinning force, and $\zeta(r,t)$ is a thermal
noise.  Here $r$ is he $d$-dimensional internal coordinate of the
elastic object.  Both $f$ and $\zeta$ are Gaussian random variables
with zero mean and second moment
\begin{eqnarray}
&& \overline{f(u,r) f(u',r')} = \Delta(u - u') \delta^d(r-r')  \\
&& \langle \zeta(r,t) \zeta(r',t') \rangle  = 2 \eta T \delta^d(r-r')
\delta(t-t'), \label{rvars}
\end{eqnarray}
where $T$ is the temperature and we set Boltzmann's constant $k_B=1$.
The value of $\eta$ generally sets the relaxation time scale, e.g.
here $\eta = t_0 c \Lambda^2$, $t_0$ being the microscopic time scale
and $\Lambda$ the short scale momentum cutoff.  In general, one may be
interested in a variety of thermal and sample-to-sample fluctuations
of the system, as well as various responses of the system to external
probes.  We will focus on the simplest of the latter, described by
(linear) response functions,
\begin{eqnarray}
&&  {\cal R}_{r r' t t'} = \left\langle \frac{\partial
    u_r(t)}{\partial \zeta_{r'}(t')} \right\rangle  
\end{eqnarray}
in a given disorder realisation, and its disorder average,
\begin{eqnarray}
&&  \overline{ {\cal R}_{r r' t t'} } = R_{r-r', t t'} .
\end{eqnarray}
At equilibrium, both the single sample and the averaged response
functions become time translationally invariant, ${\cal R}_{r-r', t
  t'}={\cal R}_{r-r', t-t'}$ and $R_{r-r', t t'}=R_{r-r', t-t'}$.

The equatiuon (\ref{eom}) has
the usual Langevin form, and guarantees the existence of a stable
equilibrium probability distribution, provided, as assumed here, that
$f(u,r) = - \partial V(u,r)/\partial u$ is of gradient form.  The
equilibrium distribution (strictly speaking defined in a finite size
sample) has the Boltzmann form $p(u) \propto \exp (-H[u]/T)$, with
\begin{equation}
  H[u] = \int \! d^d{\bf r} \left[ {c \over 2}|\nabla u|^2 + V(u({\bf
      r}),{\bf r}) \right]. \label{eq:ham}
\end{equation}
Three universality classes of special interest are usually considered:
(i) short range disorder $\Delta(u)$ which describes e.g.  random bond
(RB) disorder for magnetic domain walls (ii) long range disorder which
describes e.g. random field (RF) disorder, and (ii) random periodic
(RP) $\Delta(u)$ which describes pinned density waves or lattices.
Although these systems differ in their details, e.g. in their
roughness exponent $u \sim r^\zeta$, they do not yield qualitatively
different behaviour in their dynamical response studied here.

The aim of the present paper is to develop an approach based on the
renormalization group (RG) to study the low temperature dynamics of
pinned elastic systems described by (\ref{eom}). Although we focus on
equilibrium dynamics, some of our considerations are also relevant for
non-equilibrium relaxation. It was shown that to describe the statics
at equilibrium one needs to follow the full correlator of the random
potential (or the random force) using a functional RG (FRG) method in
a $d=4-\epsilon$
expansion.\cite{fisher_functional_rg,balents_frg_largen} Several
extensions describe correlated disorder,\cite{balents_loc} the driven
dynamics near depinning
\cite{narayan_fisher,nattermann_depinning,frg2loop} and the small
applied force, thermally activated, creep regime.\cite{chauve_creep}
However until now the FRG has not been used to study the dynamical
response and correlations in equilibrium or aging regimes, nor to
probe the crucial question of the distribution of the relaxation
times.  These are important quantities directly probed in experiments
where the system is often dominated by fluctuations or not able to
reach equilibrium within the experimental time scales. We investigate
this problem in three stages, of increasing accuracy (and,
unfortunately, complexity), only in the last stage attempt is made to
be exhaustive.  A companion paper (Ref.~\onlinecite{staticslong}) is
devoted to the statics. A short account of both works can be found in
Ref.~\onlinecite{us_short}.  Some of the present considerations concerning
approximate schemes have also been discussed independently
in Ref.~\cite{denis} .

The first question investigated in the present paper is the validity
of the single time scale approximation within the RG. Specifically, in
the first part of our study (Section II) we use, as was done in
previous works,\cite{narayan_fisher,nattermann_depinning,chauve_creep}
the simplifying hypothesis that the relaxation of each internal mode,
of wavevector $k$, is controlled by a single relaxation time scale
$\tau_k$. This allows to obtain closed equations, within the one loop
FRG, for the general two time response and correlations, as measured
in aging experiments. It yields, in the equilibrium regime on which
focus from then on, interesting analytical expressions for the
equilibrium response and correlation which exhibit two distinct
scaling regimes with scaling variables $k^\theta \ln t$ and
$t/\tau_k$, respectively ($\theta=d-2+2 \zeta$ is the energy
fluctuation exponent, and $\tau_k \sim e^{c k^{-\theta}/T}$).
However, several features of these results are found to be
unsatisfactory, such as the non-monotonicity of the response as a
function of wavevector. A more complete description including time
scale fluctuations thus appears necessary.

That sample to sample fluctuations should play an important role both in
the statics and dynamics of disordered glasses is indeed expected
from phenomenological arguments, e.g. the droplet scenario,\cite{fisher_huse,creep_pheno} 
which appears to describe simpler models such as
(\ref{eq:ham}) relatively well, at least in low
dimensions.\cite{dropevid} Let us recall its
main conclusions. In its simplest form, it
supposes the existence, at each length scale $L$, of a small number of
excitations of size $\delta\Phi\sim L^\zeta$ above a ground state,
drawn from an energy distribution of width $\delta E \sim L^\theta$
with constant weight near $\delta E=0$. While typically the 
elastic manifold is localized near a ground state, disorder averages of
static thermal fluctuations at a given scale are dominated by rare
samples/regions with two nearly degenerate minima. For example,
as a simple but remarkable consequence, the
$(2n)^{\rm th}$ moment of $u$ fluctuations is expected to behave as
\begin{equation}
  \overline{(\langle u^2\rangle - \langle u\rangle^2)^n} \sim
  c_n (T/L^\theta) L^{2n\zeta}. \label{eq:twowell}
\end{equation}
The droplet picture supposes that the long-time equilibrium dynamics
is dominated by thermal activation between these quasi-degenerate
minima controlled by barriers of typical scale $U_L\sim L^\psi$.
Little is known about the distribution of these barriers, but there is
some evidence \cite{drossel_barrier,mikheev_barrier_2d,drossel_barrier_3d}
 that $\psi\approx \theta$.  Even
a modest distribution of barriers, however, due to the Arrhenius law
$\tau_L \sim e^{U_L/T}$, yields relaxation time scales with an
extremely broad distribution as $T\rightarrow 0$. Some probes of this
broad distribution are the relaxation time moments which may be
defined in a variety of ways.  One begins by defining the relaxation
time moments in a single sample,
\begin{eqnarray}
  \label{eq:samplemoment}
  && \langle t^n \rangle_L = L^{-(d+2)} \int_0^\infty \! dt \int_{r r'}^L t^n
  {\cal R}_{ r r' t} ,  
\end{eqnarray}
which, for a particular disorder realization, describe the response of
the center of mass coordinate to a (spatially) uniform force.  For
$n=1$ ($\langle t\rangle_L$) this gives one definition of the
relaxation time of a single sample.  A set of disorder-averaged
moments may be obtained then by directly averaging the above objects,
$\overline{\langle t^n \rangle_L}$, giving the averaged response of
the system:
\begin{eqnarray}
  && \overline{\langle t^n \rangle_L} = \left.  
    q^2 \int_0^\infty \! dt \, t^n R_{q,t} \right|_{q = 1/L} .
\end{eqnarray}
Alternatively, the {\sl distribution} (from sample to sample) of the
unaveraged relaxation time $\langle t\rangle_L$ is described by a
second set of moments, $\overline{\langle t \rangle_L^n}$.  In
general, one may construct many such objects scaling dimensionally as
$t^n$ but with different physical content.  Mathematically, this is
accomplished by averaging arbitrary products of the single sample
moments, i.e.  $\overline{\langle t^{p_1} \rangle_L \cdots \langle
  t^{p_N}\rangle_L}$, with $\sum_{j=1}^N p_j = n$.  Any of these
``$n$-th'' moments may behave as $\sim e^{\alpha(n) U_0 L^\theta/T}$
with $\alpha(n)\geq n$, or grow even faster with $\ln\overline{\tau^n}
\gg L^\theta$, nor is it clear that the different definitions for a
given $n$ exhibit the same growth.  Indeed, we will ultimately find
different operators in a dynamical field theoretic formulation
corresponding to each different type of moment, and some indications
that indeed different growth rates obtain for each of these.  A theory
of these timescales is crucial to understanding both equilibrium
response and correlations and to near-equilibrium phenomena such as
creep.\cite{blatter_vortex_review,creep_pheno,Legi,creepexp}
Extensive calculations are possible in certain zero dimensional toy
models.\cite{toy} Although some analytical results have been obtained
for $d\geq 1$ within mean-field limits,\cite{mezpar,frgN,infinited} 
these do not include thermal activation over divergent barriers ($U_L
\sim L^\theta$).  The FRG
\cite{fisher_functional_rg,narayan_fisher,nattermann_depinning,frg2loop}
on the other hand, extended to non-zero temperature
,\cite{balents_loc,chauve_creep}  seems to capture, already at the
level of the single time scale approximation the existence of these
growing barriers. However until now neither the rare events nor
fluctuating barriers have been obtained in this approach.

In the main part of our study (Sections III and IV) we thus
investigate how relaxation time distributions appear within the FRG.
We first show that the equation of motion (\ref{eom}) generates under
coarse graining a {\it random friction} term $\eta(r)$ (equivalently a
random relaxation time $\tau \sim 1/\eta$).  It is then natural to
define, as a toy model, a random friction model which, in the absence
of pinning disorder possesses a manifold of fixed points indexed by
the full coarse grained probability distribution of the friction. A
highly non-trivial question is how this distribution will flow under
RG due to feedback from non-linear terms when pinning disorder is
reintroduced. We consider this question at the one loop level, in two
stages.

In Section III we present a highly simplified analysis, but with the
merit of explicitly exhibiting the broadening of the barrier
distribution and allowing for some analytical expressions. It yields
asymptotically a log-normal distribution of relaxation times, i.e.  a
nearly gaussian distribution of effective barriers centered around
$L^{\theta}$ and of typical fluctuations $\sim L^{\theta/2}$. Such a
log-normal tail corresponds to the moment exponents
$\alpha(n)=2n^2-n$.  This is compared with numerical results
\cite{drossel_barrier,mikheev_barrier_2d,drossel_barrier_3d} in the
case of a directed polymer, where the width was fitted to $\sim
L^{\theta}$. The question of the width is important since a width
$\sim L^{\theta/2}$ is not expected to be large enough to modify the
creep exponent, as can be seen from reexamining the calculations of
Ref. ,\cite{chauve_creep} while a $\sim L^{\theta}$ width
\cite{mikheev_barrier_2d} would pose a problem to this order. We
derive, within the same approximate scheme (Appendix~\ref{general}),
closed equations for correlations and response functions (the fact
that the width grows very fast can be exploited in a resummation of
the fastest growing terms in the dynamical part of the Martin Siggia
Rose functional). We find that the broadening is sufficiently fast to
invalidate some of the previous analysis, e.g. the existence of a
$t/\tau_k$ scaling regime).

Section IV contains the full systematic analysis of the running
dynamical effective action. It is found that relaxation times
distributions are determined by a closed hierarchy of FRG equations,
each level $p$ corresponding to an increasing power of frequency
$\omega^p$ can be solved independently of higher ones, the lowest one
being the statics $p=0$.  This hierarchy involves functions
parameterizing the local cross correlations between pinning disorder
and random relaxation times. The previous approximation corresponds to
projecting these FRG functions to their values at zero, while in fact
the full set of non-linear differential equations obeyed by these
functions need to be solved, a formidably complex task.  A thermal
boundary layer ansatz (TBLA) appears as a consistent solution.  It
extends the one discovered in the statics which was shown to reproduce
droplet theory type behaviour in thermal fluctuations. Here it yields
a natural growth for moments of relaxation times measured by
non-trivial exponents $\alpha(n)\neq 2n^2-n$ determined by eigenvalue
problems. Although perturbative control remains a challenge, the
structure of the dynamical TBLA which encodes for barrier
distributions opens the way for deeper understanding of the field
theory approach to glasses.

The detailed outline of the paper is as follows. In Section
\ref{sec:onetime} we recall the standard results of the FRG for the
equilibrium dynamics using a single relaxation time approach. We then
give a qualitative derivation of the two scaling regimes for the
equilibrium response and correlation functions.  The detailed
equations obeyed by these functions are derived using a Wilson scheme
in Appendix \ref{sec:onetimedetails} and their analytical form is
analyzed in the equilibrium regime in Appendix
\ref{sec:onetimedetails} and in the aging regime in
Appendix~\ref{sec:onetimedetails2}. In Section \ref{sec:fterm} we go
beyond the single relaxation time approach. The random friction model
is introduced in Section \ref{sec:rf}.  We then incorporate pinning
disorder in an approximate way in Section \ref{pinningdisorder},
analyze the resulting distribution of relaxation times and show that
it become broad.  The breakdown of the $\omega \tau_k$ scaling is
analyzed in Section \ref{breakdown}. In Section
\ref{sec:fullstructure} we discuss the systematics of the structure of
the dynamical effective action. It does contain the statics which its
recalled, together with its thermal boundary layer ansatz solution, in
its equilibrium dynamics formulation in Section~\ref{sec:staticstbla}.
Then in Section~\ref{sec:dynamicstbla}-\ref{sec:terms-assoc-with} we
display the hierarchical structure of the FRG equations and how a
generalized thermal boundary later structure appears as a consistent
solution determining the growth of the moments of the relaxation times
through non-trivial eigenvalue problems.  We conclude in
Section~\ref{sec:conclusion} with some general remarks.  Finally, a
set of appendices elaborate on various details and calculated related
to the main text.

\section{Single time-scale approximation}
\label{sec:onetime}\

At conventional pure critical points, scaling emerges directly from
the existence of a RG fixed point.  Moreover, in an epsilon expansion,
the {\sl scaling functions} are obtained to leading order by a simple
matching procedure (REF).  The {\sl static} equilibrium FRG for the
random
elastic problem is, aside from the complication of a functional fixed
point, very similar to such an ordinary RG calculation.  The important
distinction is the non-analyticity of the $T=0$ fixed point function
$\Delta^*(u)$
which at finite temperature results in a narrow boundary layer for
small $u \lesssim \tilde{T}_l \epsilon$, whose width continuously
decreases
under the FRG as the running effective temperature $\tilde{T}_l$ (see
below) flows
to zero. The corresponding growth of the mean-squared curvature
of the effective potential felt by the manifold is a hint of
unconventional behavior not present in ordinary critical theories.

The influence of this divergence is very dramatic in the dynamics.  In
this section, we attempt to naively extend the conventional RG
approach to calculating response functions to the random manifold
problem.  This conventional approach implicitly assumes the existence
of a single time-scale (at a given wavevector), as we shall soon see.
This assumption leads to somewhat unsatisfactory results for the
response
function, forcing us to reconsider the distribution of
time-scales in Section ~\ref{sec:fterm}.  Although we will ultimately
conclude that the single time-scale calculation is fundamentally
incorrect, it is useful to review the methodology of
this approach.

We begin by reviewing the basics of the FRG.  We employ the
Martin-Siggia-Rose (MSR) formalism,\cite{msr}  in order to use
field-theoretic
techniques.  The MSR approach is based on the generating functional
$Z[h,\hat{h}] $ for the disorder-averaged correlation and response
functions,
\begin{equation}
Z[h,\hat{h}]  =  \int Du D\hat{u} e^{- S[u,\hat{u}] + \int_{r t}
\hat{h}_{rt} u_{rt} + h_{rt}
i \hat{u}_{rt}}, \label{zdef}
\end{equation}
where the dynamics in (\ref{eom}) is encoded in the action
$S[u,\hat{u}]  =  S_0[u,\hat{u}] + S_{int}[u,\hat{u}]$, with
\begin{eqnarray}
S_0[u,\hat{u}] & = & \int_{rt} ~ i \hat{u}_{rt}
\left( \overline\eta \partial_t - \nabla^2\right) u_{rt}
 \nonumber \\
        && - \overline\eta T \int_{rt} (i \hat{u}_{rt}) (i
\hat{u}_{rt})
\label{msr1} \\
S_{int}[u,\hat{u}]  & = & - \frac{1}{2} \int_{rtt'} (i \hat{u}_{rt})
(i \hat{u}_{r t'})
\Delta(u_{rt} - u_{rt'})
\label{msr2}
\end{eqnarray}
where $\hat{h},h$ are source fields, and we have put a overbar on the
friction coefficient $\eta\rightarrow\overline\eta$ to indicate the
mean, i.e that it is for now a constant uniform in space. As justified
below we have set $c=1$.  We use the Ito convention to regularize
equal-time response
functions, i.e. $R_q(t,t)=0$ and $R_q(t^+,t)=1/\overline{\eta}$.  The
disorder averaged
response and correlations are given by
\begin{eqnarray}
&& R_q(t,t') = \overline{  \frac{\delta \langle u_q(t) \rangle }{\delta
h_{q}(t')} }= \langle u_q(t) i \hat{u}_{-q}(t')\rangle_S,
\label{respdef}\\
&& C_q(t,t') = \overline{\langle u_q(t) u_{-q}(t')\rangle} =  \langle
u_q(t) u_{-q}(t')\rangle_S. \label{corrdef}
\end{eqnarray}

The FRG in its Wilsonian formulation begins by introducing an
ultraviolet (short distance) cut-off $\Lambda$ on the spatial Fourier
wavevectors.  In the FRG, this cut-off is progressively reduced to
$\Lambda_l = \Lambda e^{-l}$ ($0<l<\infty$).  At each stage of the RG,
the spatial Fourier components of $u,\hat{u}$ are divided into
``slow'' and ``fast'' modes, with momenta in the range $0<k<\Lambda_l
e^{-dl}$ and $\Lambda_l e^{-dl} <k<\Lambda_l$, respectively.  The fast
modes are then integrated out, working perturbatively in $S_{int}$ to
one loop order, and $l$ is increased by $dl$.  This leads, in the
limit of infinitesimal rescaling $dl \rightarrow 0$, to a smooth
renormalization of the effective action for the remaining slow modes,
and hence to continuous FRG equations for the running
disorder correlator $\Delta_l(u)$.

Naive power counting (see e.g.
Refs.~\cite{fisher_functional_rg,balents_frg_largen}) indicates that
all terms beyond those in (\ref{msr1})-(\ref{msr2}) are
irrelevant, so we for the moment neglect their generation under the
FRG.  The flow of the random pinning correlator $\Delta_l(u)$ has been
derived many times previously
.\cite{narayan_fisher,nattermann_depinning,balents_loc,chauve_creep}
It is better expressed in terms of the dimensionless rescaled pinning
force correlator $\tilde{\Delta}_l(u)$ defined such that:
\begin{equation}
\Delta_l(u)= \frac{\Lambda^{\epsilon}}{ A_d } e^{-\epsilon l}  e^{2
\zeta l} \tilde{\Delta}_l(u e^{-\zeta l})
\label{DeltaEq}
\end{equation}
with $A_d = S_d/(2 \pi)^d= 1/(2^{d-1} \pi^{d/2} \Gamma(d/2))$, and
reads:
\begin{eqnarray}
&& \partial_l \tilde{\Delta}(u) = (\epsilon-2 \zeta) \tilde{\Delta}(u)+
\zeta u \tilde{\Delta}'(u)
+ \tilde{T}_l \tilde{\Delta}''(u) \\
&& + \tilde{\Delta}''(u) (\tilde{\Delta}(0) - \tilde{\Delta}(u) ) -
\tilde{\Delta}'(u)^2
\label{frgeq}
\end{eqnarray}
Here the fluctuation dissipation theorem ensures that the temperature
$T_l=T$ is uncorrected but the effective dimensionless temperature
$\tilde{T}_l = A_d T \Lambda^{d-2} e^{-\theta l}$ itself flows to
zero, controlled by the energy fluctuation exponent $\theta = d-2 + 2
\zeta$, the temperature being formally irrelevant. Here and in the
following we do not not make any spatial or temporal rescalings of
coordinates or momenta.

Study of the one loop FRG equation shows that, with the proper value
for the roughness exponent $\zeta \sim O(\epsilon)$ depending on the
universality class (RB,RF or RP), the dimensionless disorder
correlator converges {\sl non-uniformly} to a nonanalytic
``fixed-point'' function $\Delta^*(u)$ formally of order $\sim
O(\epsilon)$ as $l \rightarrow \infty$, whose functional form is not
important for this discussion (see however Section IV for much
more details). The non-uniformity of this convergence
is due to a boundary-layer centered on $u=0$, whose width decreases
continuously with scale.\cite{balents_loc,chauve_creep} In particular
one can show that, to this order \cite{balents_loc,chauve_creep}:
\begin{eqnarray}
&& \lim_{l \to +\infty} \tilde{T}_l  \tilde{\Delta}_l''(0) \to - \chi^2
\end{eqnarray}
Thus asymptotically the curvature of the correlator diverges with the
scale
(here $\chi=|\Delta'^*(0^+)|$ is a constant depending of the
universality class,
e.g. for periodic systems $\chi=\tilde{\chi} \epsilon$, $\epsilon =
4-d$.

To one loop, all that remains is the renormalization of the mean
friction coefficient, since the elastic modulus is fixed by Galilean
invariance\cite{balents_frg_largen,balents_loc,chauve_creep}\ and the temperature by the
Fluctuation-Dissipation-Theorem (FDT).  This was determined in
Refs.~.\cite{narayan_fisher,nattermann_depinning,balents_loc,chauve_creep}:
\begin{equation}
  \partial_l \overline\eta = \Gamma_l \overline\eta, \label{meanrg}
\end{equation}
where $\Gamma_l = - \tilde{\Delta}_l''(0) \sim_{l \to + \infty}
\chi^2/\tilde{T}_l$
thus grows with the scale as:
\begin{eqnarray}
&& \Gamma_l \sim \tilde{\beta} e^{\theta l} \\
&& \tilde{\beta} = T^*/T
\label{gammal}
\end{eqnarray}
where $\tilde{\beta}$ is the reduced bare inverse temperature
and $T^* = \chi^2 \Lambda^{2-d}/A_d$ a non-universal
temperature scale. (\ref{meanrg}) implies {\sl activated scaling}
\cite{balents_loc,chauve_creep}
since the friction coefficient, which plays the
role of a time scale $\tau = \eta/\Lambda^2$,  grows exponentially with
the length $e^l$:
\begin{equation}
  \overline\eta_l = \overline\eta_0 \exp\left[ {\tilde{\beta}\over
      \theta}(e^{\theta l}-1 ) \right].
\end{equation}
(in $d=4$ one has $\Gamma_l=\tilde{\beta} e^{2 l}/l^2$
and $\tilde{\beta} = \tilde{\chi}^2 \Lambda^{-2}/(T A_4)$).

Activated dynamics leads to ambiguities in a single time-scale
approach, as can be seen from a simple matching argument.  We consider
for simplicity the equilibrium dynamics, in which the response and
correlation functions are time-translationally invariant (TTI).  It is
then convenient to work in terms of both frequency $\omega $   and
wavevector.  The usual RG considerations lead one naively to expect
that the response function obeys a relation,
\begin{equation}
  R_k(\omega) = e^{2l} R_{ke^l}(\omega \tau_l),
\end{equation}
where $\tau_l = e^{2l} \overline\eta_l/\overline\eta_0$.  We now
obtain two {\sl inequivalent} scaling forms by matching.  In
particular, if we choose $ke^l = 1$ (we set $\Lambda=1$ for now), we
find
\begin{equation}
  R_k^{(1)}(\omega) = {1 \over k^2} {\cal R}^{(1)}(\omega \tau_k),
  \label{sf1}
\end{equation}
with
\begin{equation}
  \tau_k = {1 \over k^2} e^{{\tilde{\beta} \over
\theta}(k^{-\theta}-1)}.
\label{tauk}
\end{equation}
If, however, we choose $\omega \tau_l=1$, we find asymptotically
\begin{equation}
  R_k^{(2)} = (\frac{1}{\tilde{\beta}} \ln(\frac{1}{\omega})
)^{2/\theta}
{\cal R}^{(2)}(\frac{1}{\tilde{\beta}} k^\theta |\ln\omega|).
\label{sf2}
\end{equation}
Note that these two forms {\sl cannot} be related by redefining the
two scaling functions ${\cal R}^{(1/2)}$, as can be seen from the fact
that $\ln (\omega \tau_k) \sim \ln\omega + (\chi/\theta)k^{-\theta} =
(\chi/\theta + k^\theta|\ln\omega|)/k^\theta$, which is {\sl not} a
function of $k^\theta |\ln\omega|$ only.

The inequivalence of (\ref{sf1},\ref{sf2}) appears to point to the
existence of two scaling regimes, which we will call the $X$ and $Y$
scaling limits.  The first is formally defined by defining the scaling
variable $Y=\omega \tau_k$.  With $Y$ fixed and $\omega, k \rightarrow
0$, one obtains the scaling regime in (\ref{sf1}).  The second
scaling regime obtains with $X = k^\theta |\ln\omega|$ fixed and
$\omega, k \rightarrow 0$.  To reconcile the two regimes, note that
$\ln Y \sim (\chi/\theta+X)/k^\theta$, so that for fixed $X$, as
$\omega,k \rightarrow 0$ (the $X$ scaling limit) $Y\rightarrow
\infty$!  The second scaling regime ((\ref{sf2})) thus appears to
occur at the ``boundary'' ($Y=\infty$) of the first.

We have indeed verified this behaviour, and obtained the analytical
scaling forms similar to those in (\ref{sf1},\ref{sf2}) using FRG
techniques
{\sl under the assumption of a single characteristic time scale
  parameterized by $\overline\eta$}. These calculations are performed
in Appendix A for equilibrium, and in Appendix B for the more
general nonequilibrium situation.
Interestingly, the general equations for the response
function in this approach share some similarities to those
arising in the (infinitely connected and/or large $N$) mean field
limit of a number of model glasses.  As discussed in Appendices A and B
their
solutions exhibit several time regimes with various aging scaling forms
and also show differences compared to mean field.

Many features of these results, however, point to problems with the
single time scale assumption, as also discussed in Appendix A.
The real-time response function is
found to be an {\sl increasing} function of wavevector at fixed time
in the logarithmic ($X$) scaling regime.  This somewhat unexpected
(and possibly unphysical) behavior is apparently a very general
consequence of the mere {\sl existence} of two distinct scaling limits,
and hence is inevitable
given the single time scale approach.  More significantly, the
appearance of a sharply-defined $\tau_k$ in the {\sl mean} response
function (in the $Y$ regime) is difficult to understand on physical
grounds.  Even in (random) models involving only a small number of
degrees of freedom, while a given sample may be characterized by a
longest relaxation time, the sample to sample variations of this would
generally lead, as espoused in the introduction, to the disappearance
of such a time in the mean response.  In the collective elastic model
considered here, interactions between the enormous number of modes
with differing wavevector (and hence differing relaxation rates) would
only worsen the situation.

\section{Broad distributions of time scales: simplified approach}

\label{sec:fterm}

\subsection{distribution of relaxation times and the f-term }

Up to this point, we have assumed that the dynamics at each scale can
be described by a single friction coefficient $\tau_l =
\overline{\eta}_l$, which corresponds to a sharply defined time scale
for relaxation.  On general grounds, however, we should expect
extremely broad distributions of relaxation times.  This follows
simply from the Arrhenius law,
\begin{equation}
\tau_k = \tau_{l=\ln (k/\Lambda)} \sim \tau_0 \exp( U_k /T ),
\end{equation}
which estimates the time required to overcome an energy barrier of
height $U_k$ at scale $k$.  At low temperature, even a modestly wide
distribution of $U_k$ gives rise to extremely broadly distributed
$\tau_k$.  If this distribution is sufficiently broad, it is no longer
adequately characterized by its average, and indeed many physical
quantities may depend upon the precise form of the
distribution.

We now investigate how this distribution can be incorporated into the
FRG treatment within the MSR formalism.  We will consider a spatially
varying friction coefficient $\eta(r)$, with the equation of motion
(\ref{eom}) modified to
\begin{eqnarray}
\eta(r) \partial_t u_{rt} = c \nabla^2_r u_{rt} + f(u_{rt},r) +
\zeta(r,t),
\label{eom2}
\end{eqnarray}
where, in order to maintain the stationary equilibrium Boltzmann
probability
distribution function, the noise correlations are modified to
\begin{equation}
\langle \zeta(r,t) \zeta(r',t') \rangle  =  2 \eta(r) T
\delta^d(r-r')\delta(t-t'). \label{noise2}
\end{equation}

For simplicity, we will initially take $\eta(r)$ to be identically and
independently distributed at each $r$, according to the distribution
$P(\eta)$.  The distribution naturally enters the MSR theory via its
characteristic function, which we parameterize by $F(z)$:
\begin{eqnarray}
\int_0^{+\infty} d \eta P(\eta) e^{- z \eta} = e^{- F[z]} \label{Fdef}.
\end{eqnarray}
The Taylor series expansion of $F(z)$ thereby gives the connected
cumulants of $\eta$,
\begin{equation}
F[z] = \sum_{m=1}^\infty \eta^{(m)} {{(-1)^{m+1}} \over
m!} z^m,
\label{moments}
\end{equation}
where $\eta^{(m)} = [\overline{\eta^m}]_C$ is the $m^{\rm th}$
cumulant (connected) moment of $\eta$.
For the continuum field $\eta(r)$, the analogous expression is
\begin{eqnarray}
\overline{exp( - \int_r \eta(r) z(r))} = exp( - \int_r F[z(r)]).
\label{etaavg}
\end{eqnarray}
We initially assume no cross-correlations between $\eta(r)$ and
$f(u,r)$, though these can to some extent be generated in perturbation
theory.  The single time scale model studied in the previous Section
(and Appendices A and B)
with $\eta(r) = \overline\eta$ corresponds
to
$F(z) = \overline\eta z$.  Although this is not essential, we shall
assume here an initially narrow (but not $\delta$-function)
distribution $P(\eta)$. As shown in Appendix C, even if initially
$F=\overline\eta z$, a
higher-order analysis shows that a non-trivial distribution is
generated under coarse graining.

In the MSR formalism, the modified equation of motion,
(\ref{eom2}-\ref{noise2}), is described by the action:
\begin{eqnarray}
S_0[u,\hat{u}] & = & \int_{r t} \overline{\eta} ~ i \hat{u}_{rt}
\partial_{t} u_{rt}
\!-\! i \hat{u}_{rt} c \nabla^2_{r} u_{rt}
\!-\! \overline{\eta} T i \hat{u}_{rt} i \hat{u}_{rt} , \nonumber \\
&&
\label{smod1} \\
S_{int}[u,\hat{u}]  & = &          \int_{r}
\tilde{F}[ \int_t
( i \hat{u}_{rt} \partial_{t} u_{rt} - T i \hat{u}_{rt} i \hat{u}_{rt})]
\nonumber \\ & & - \frac{1}{2} \int_{r,t,t'} (i \hat{u}_{rt})
(i \hat{u}_{r t'})
\Delta(u_{rt} - u_{rt'}).
\label{smod2}
\end{eqnarray}
We have defined $F[z] = \overline{\eta} z + \tilde{F}[z]$ where
$\tilde{F}[z]$ starts with higher powers of $z$, to do perturbation
theory using the average friction coefficient.  One immediate remark
is that the statistical tilt/translational symmetry (STS)
holds,\cite{dropevid} thus $c$ will not be corrected and so we set it
to $c=1$. Note that the $\overline{\eta}$ kinetic term could equally
well be considered as an interaction term, in the spirit of a
``perturbation theory in $i \omega$'' with bare propagator simply
$R_{q,\omega}=1/q^2$ (in real time $R_q(t,t')=1/q^2 \delta(t^{-} -
t')$). Note also that for each realization, the instantaneous response
function satisfies
\begin{equation}
R({\bf r},{\bf r'},t-t'=0+) = {1 \over \eta(r)} \delta({\bf r-r'}).
\end{equation}
Averaging over disorder gives $R_k(t-t'=0^+) = \overline{1/\eta}$.

Although it may seem obvious, it is important to stress at this stage
that the renormalized relaxation time moments are measurable
quantities.  One can define the renormalized moments in the usual way
from the effective action $\Gamma$.  Taking the same form as
(\ref{smod2}), one has
\begin{eqnarray}
  \label{eq:gammaef}
&&  \Gamma_{int} = -\frac{\eta^{(2)}}{2} \int_{r tt'} \hat{u}_{rt}
\dot{u}_{rt} \hat{u}_{rt'}\dot{u}_{rt} +\cdots
\end{eqnarray}
with of course many higher order terms describing the higher friction
coefficient moments, momentum dependence of vertices, etc.  As usual
in field theory, correlation functions are exactly evaluated at tree
level using this effective action -- thus the $\eta^{(2)}$ term here
has the meaning of a fully-renormalized second moment on the scale of
the system size (or infrared momentum cutoff).  One may then consider
the physically defined relaxation time from (\ref{eq:samplemoment})
and construct its second moment:
\begin{eqnarray}
  \label{eq:t2}
&&   \overline{\langle t\rangle^2_L} = \frac{1}{L^{2(d+2)}} \int\! dt
dt' \, t t' \int_{r_1 r'_1 r'_2 r_2} \hspace{-0.2in}\overline{\langle u_{r_1 t} \hat{u}_{r'_1
    0} \rangle \langle u_{r_2 t'} \hat{u}_{r'_2 0} \rangle}.\nonumber
\\ &&
\end{eqnarray}
On physical grounds, in equilibrium, we expect that the latter product
of response functions in two ``replicas'' is the same as considering
the product of two subsequent responses in a single replica, provided
the two response measurements are taken far apart.
\begin{eqnarray}
  \label{eq:factor}
&&   \overline{\langle u_{r_1 t} \hat{u}_{r'_1
    0} \rangle \langle u_{r_2 t'} \hat{u}_{r'_2 0} \rangle} \nonumber
\\
&& = \lim_{\tau\rightarrow \infty} \overline{\langle u_{r_1 t+\tau}
  \hat{u}_{r'_1 \tau} u_{r_2 t'} \hat{u}_{r'_2 0} \rangle}. 
\end{eqnarray}
This latter four-point function can be calculated using the effective
action above.  One finds
\begin{eqnarray}
  \label{eq:int4pt}
&&   \lim_{\tau\rightarrow \infty} \int_{r_1 r'_1 r'_2 r_2} \overline{\langle u_{r_1 t+\tau}
  \hat{u}_{r'_1 \tau} u_{r_2 t'} \hat{u}_{r'_2 0} \rangle} 
\\
&& = L^{2d} R_{q_0 t} R_{q_0 t'} + \eta^{(2)} L^d (R_{q_0}
*\dot{R}_{q_0})_t  (R_{q_0} *\dot{R}_{q_0})_{t'}, \nonumber
\end{eqnarray}
where $q_0\sim 1/L$ is the infrared momentum cutoff, and $*$
denotes the convolution in the time domain.  Integrating over the time
coordinates and using the result of statistical translational symmetry
$R(q_0,0) = 1/q_0^2$, one then obtains
\begin{eqnarray}
  \label{eq:t2res}
  &&   \overline{\langle t\rangle^2_L} \sim (\overline{\eta}L^2)^2 +
  \eta^{(2)} L^{4-d}.
\end{eqnarray}
For a not too broad distribution of friction coefficients with
scale-independent $\eta^{(2)}$, the correction due to $\eta^{(2)}$ is
vanishing for $d>4$, and small compared to the first ``disconnected''
term (scaling as $\overline{\langle t\rangle_L}^2$) for any $d$.
However, for the glassy dynamics studied here, we will find
$\eta^{(2)}$ is {\sl exponentially} larger than $\overline{\eta}^2$ as
a function of $L$, so that in fact the second term is dominant.  Thus
the second moment of the physical relaxation time $\overline{\langle
  t\rangle^2_L}$ indeed measures the coupling constant $\eta^{(2)}$ as
promised.

\subsection{no pinning disorder: the random friction model}
\label{sec:rf}

We now turn to the FRG analysis of the modified action in
(\ref{smod1}-\ref{smod2}).  We first consider {\sl only} the effects
of
randomness in $\eta$, {\sl neglecting} the pinning disorder $\Delta$.
This defines a {\sl random friction} model described by the MSR action
with $\Delta=0$.  Remarkably, the random friction model represents an
infinite
manifold of fixed points parameterized by $F[z]$.  Indeed, a
diagrammatic treatment explicitly shows the absence of renormalization
of $F[z]$ order by order in $\tilde{F}$.  Despite this absence of
renormalization, the random friction model represents a non-trivial
interacting field theory.

\begin{figure}
\centerline{\fig{6cm}{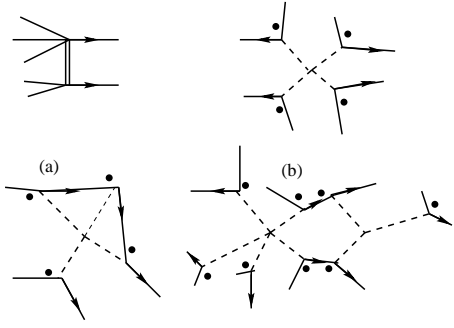}}
\caption{Top: graphical representation of the disorder vertex (double
lines) and
of the $F$ term vertex (the dots represent the time derivatives) where
the
arrows represent a $\hat{u}$ response field and solid line a $u$ field
(arrows are along inreasing time). Bottom: (a) corrections of the $F$
term to itself at $T=0$
which vanish (b) corrections to order $F^2$ which also vanish (as all
orders do - see text).
\label{graph1}}
\end{figure}

\begin{figure}
\centerline{\fig{6cm}{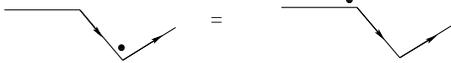}}
\caption{ Shifting of a time derivative from an internal line (in the
middle)
to an external one (right) along a line of response functions at $T=0$
(works for disorder as well as $F$ vertices) \label{graph11}}
\end{figure}

For simplicity, we sketch this here for $T=0$.  In this limit, the
vertex is
\begin{eqnarray}
\int_r \tilde{F}[ \int_t  i \hat{u}_{rt} \partial_{t} u_{rt} ].
\end{eqnarray}
Diagrams occuring in the expansion of $F$ are indicated in
Fig. \ref{graph1} (a) and (b).  The fields connected by dotted lines
occur at the same spatial point, solid lines with and without arrows
indicate
$u$ and $\hat{u}$ fields, respectively.
Considering a product of the form
\begin{eqnarray}
\int_{r_1} (\int_{t_1} i \hat{u}_{r_1 t_1} \partial_{t_1} u_{r_1
t_1})^{n_1}
\int_{r_2} (\int_{t_2} i \hat{u}_{r_2 t_2} \partial_{t_2} u_{r_2 t_2}
)^{n_2} ,
\end{eqnarray}
the only possible contractions contain products of the type
$< \partial_{t_j} u_{r_j t_j} i \hat{u}_{r_k t_k} >$ with no time
loop allowed. Thus all relative time integrals factor and
one is left with products of integrals of the type
$\int_t \partial_t R(r_j - r_k,t)$ which vanish since
the response function vanishes for $t<0$ and $t \to +\infty$
.\cite{footnoteq0}
Thus $F$ does not correct $F$.   However $F$ itself produces new terms
such as:
\begin{eqnarray}
&& \int_{r,t} i \hat{u}_{r t} \partial^2_{t} u_{r t} \\
&& \int_{r,t_1,t_2} i \hat{u}_{r t_1} \partial^2_{t_1} u_{r t_1}
i \hat{u}_{r t_2} \partial^2_{t_2} u_{r t_2},
\end{eqnarray}
obtained by time gradient expansions, with nonvanishing coefficients
(of the form $\sim \int_t t \partial_t R(r_j - r_k,t)$), as well as
similar terms with higher order time derivatives (note that
similar terms containing also higher order spatial gradients
are also generated, but we will not consider them as important here
\cite{footnotek}). One can embed these new terms into a
new function:
\begin{eqnarray}
\int_{r} F_2[\int_t i \hat{u}_{r t} \partial^2_{t} u_{r t}]
\end{eqnarray}
and so on -- the full systematics of these new terms will be examined
later in Appendix \ref{general}
and Section (\ref{sec:fullstructure}). It is important to note for consistency that
there is also no feedback from higher-derivative terms such as $F_2$
back into $F$.  Graphically, as in Fig. \ref{graph1} (a) and (b), one
can perform time integration by parts along each line joining several
vertices which leads to terms $\int_t \partial_t^p \hat{u}_{t}
\partial_t^q u_t$ with $p+q>1$. Indeed a very useful rule is represented
graphically on Fig \ref{graph11}: one can simply shift the time
derivative
along any internal response line to the external one
(at $T=0$ any diagram is a tree of such lines) since, schematically:
\begin{eqnarray}
&& u_t \langle i \hat{u}_t \partial_{t_1} u_{t_1} \rangle i
\hat{u}_{t_1} =
u_t \partial_{t_1} R_{t_1,t} i \hat{u}_{t_1} \\
&& = - u_t \partial_{t} R_{t_1,t} i \hat{u}_{t_1} \to
\partial_{t}  u_t R_{t_1,t} i \hat{u}_{t_1}
\end{eqnarray}
after integration by parts.  In the Fourier domain, this rule is just
conservation of frequency along all solid lines, since the
interactions are all fully non-local in time and therefore do not
carry frequency.

To conclude, the apparent non
renormalization of $F[z]$ makes it tempting to define {\it a manifold of
fixed point theories} indexed by $F[z]$.  These fixed points are quite
interesting and non-trivial.  For instance, the computation of the
averaged
response function at $T=0$ can be mapped exactly onto the problem of
calculating the partition function for a self-avoiding walk. This is
developed
further in Appendix D.

\subsection{pinning disorder: distribution of barriers}

\label{pinningdisorder}

We now consider the combined effects of the pinning disorder and
distribution of time scales.  Because the pinning disorder can be
defined in a purely static theory (using the equilibrium Bolzmann
partition function), its renormalization is unaffected by the $F$
term.  However, the converse is not correct.  Due to the
non-renormalization of $F$ in the random friction model, we must
consider only terms of order $F^p \Delta^q$, with $q \geq 1$.  The
leading non-vanishing terms correcting $F$ at $O(\Delta)$ are linear in
$F$, and are indicated diagrammatically in
Figs.~\ref{graph2},\ref{graph3}.   They are computed in detail in
Appendix \ref{app:fterms}\ but one easily sees the structure of the
result,
thanks to the property of shifting internal time derivatives (dots in
the figures) to the external ones, e.g. that the three graphs in
Fig.~\ref{graph2}~
have identical values.

There is a subtle distinction between the contributions in
Fig.~\ref{graph2}~ and those in Fig.~\ref{graph3}.  In particular, in
the diagram of Fig.~\ref{graph3}, the pinning vertex suffers
contractions between both of its independent time variables (i.e.
graphically both ends of the double-line) and {\sl the same} time
variable (dotted leg) of the f-term vertex.  The locality of the
response function therefore implies that the two internal times of the
pinning force correlator are constrained to be nearby, justifying a
temporal gradient expansion and hence giving a leading contribution
proportional to $\Delta''(0)$.  In the diagrams of Fig.~\ref{graph2},
by contrast, the two times of the pinning vertex are contracted with
{\sl different} legs of the F-term.  These diagrams therefore generate
in fact more general terms involving $\Delta''(u_{t}-u_{t'})$ with
free integration over $t$ and $t'$.  If $|u_t - u_{t'}|$ is not
extremely small (within the boundary layer), this is a small
correction lacking the singular temperature dependence.  It is thus
not clear at this stage whether or not the graphs in
Fig.~\ref{graph2}\ should in fact be interpreted as renormalizations
of the F-term.

\begin{figure}
\centerline{\fig{3cm}{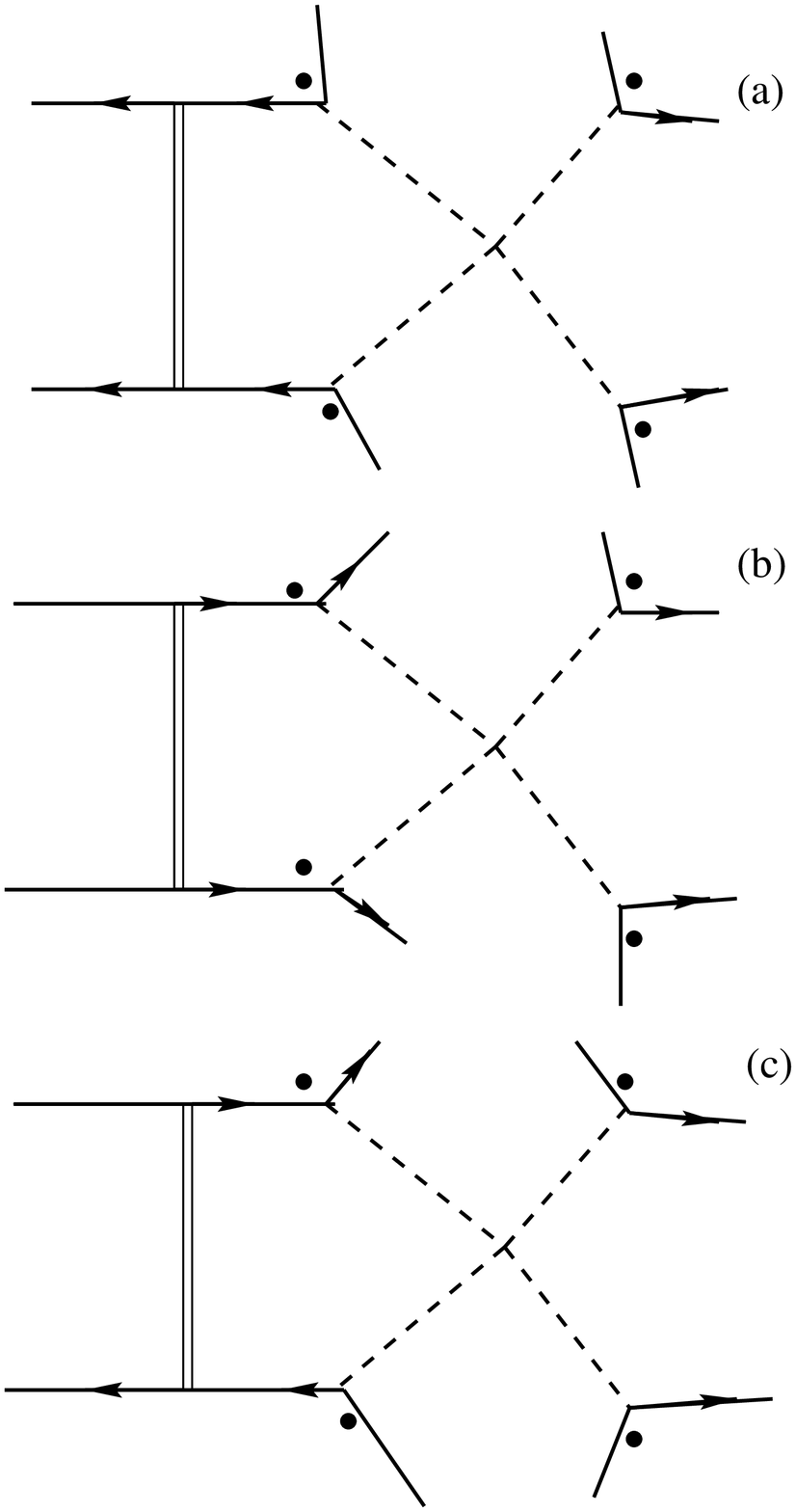}}
\caption{ Graphs involving pinning disorder which correct the $F$ term
proportionally to $F''$ \label{graph2}}
\end{figure}

\begin{figure}
\centerline{\fig{3cm}{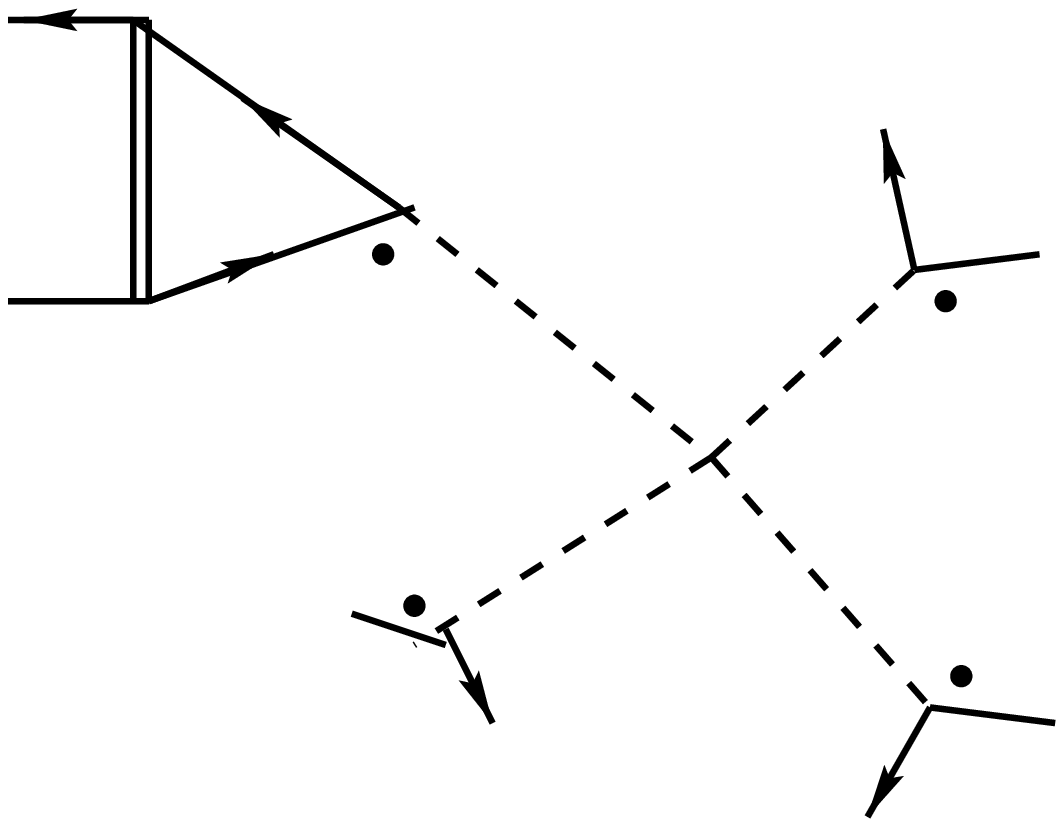}}
\caption{ Graphs involving pinning disorder which correct the $F$ term
proportionally to $F'$ \label{graph3}}
\end{figure}

In fact, the true situation is more delicate, and will be returned to
in Sec.(\ref{sec:fullstructure}). For the moment, however, we will shut our eyes to this
complication, and gain some physical insight by 
%
taking into account both sets of
diagrams as renormalizations of the f-term.
Their sum, integrated in the
momentum shell, gives the following correction to $F$:
\begin{eqnarray}
\delta F[z] =  - \Delta''(0) S_d \Lambda_l^{d-4} dl ( z F'[z] + 2 z^2
F''[z]) . \label{eq:Fflow}
\end{eqnarray}
Here the $F'$ and $F''$ terms comes from the diagrams in
Fig.~\ref{graph3},\ref{graph2}, respectively.  In agreement with
(\ref{meanrg}) it can be rewritten as
\begin{eqnarray}
  \partial_l F_l[z] = (\partial_l \ln \overline{\eta}_l) ( z F_l'[z] + 2
  z^2 F_l''[z]) . \label{fteq1}
\end{eqnarray}
Note that the disputed diagrams in Fig.~\ref{graph2}~ do not
contribute to the mean relaxation time due to the second derivative of
$z$, so that $\overline\eta$ is unambiguous.  The RG equations for the
connected moments, $\eta^{(n)}$, of $\eta$ are thereby obtained using
(\ref{moments}) as
\begin{eqnarray}
  \eta^{(n)}_l \sim \eta^{(n)}_0
  (\frac{\overline{\eta}_l}{\overline{\eta}_0})^{2 n^2 - n} .
\end{eqnarray}

It is more convenient and physical to introduce the random barrier
$U = \ln\eta$,
and the barrier corresponding to the average relaxation time
$U_l = \ln \overline\eta_l$. Changing to the energy variable
$v= \ln z$, and letting $G_l(v) = F_l(e^v)$ gives
\begin{equation}
\partial_{U_l } G_l[v] = 2 G''_l[v] - G'_l[v] , \label{diffdrift}\
\end{equation}
i.e a diffusion with drift equation.  Some physical understanding of
$G(v)$ can be obtained from the two extreme limits,
\begin{equation}
  G(v) \sim \cases{\overline{\eta} e^v & $v \rightarrow -\infty$ \cr v
    - \ln P(0) & $v \rightarrow \infty$ \cr},
\end{equation}
as can easily be found from (\ref{Fdef}), assuming a constant
probability density for small barriers, $0<P(0)<\infty$.
More generally, using (\ref{Fdef}) the diffusing and drifting
``density'' $G_l$ is related to the barrier probability distribution via
\begin{equation}
G_l[v] = - \ln \int dU P_l(U) e^{- e^{U + v}}. \label{peqn}
\end{equation}
Formally, the solution of (\ref{diffdrift}) is given by
\begin{equation}
  G_l(v) = \int \! dw \, {1 \over \sqrt{8\pi U_l}} \exp\left[ -
    {{(v-U_l-w)^2} \over {8U_l}}\right] G_0(w), \label{heatkernel}
\end{equation}
and inverting the results via (\ref{peqn}) to obtain $P_l(U)$.
A simple approximation may be applied in the regime of large $U \gg
U_l$ and $U_l \gg 1$, in which $G_l(v) \ll 1$.  In this case,
it is valid (and justified a posteriori since the distribution of
barriers become broad) to replace
$e^{- e^{U + v}}$ by $\theta(U < - v)$ and thus
one gets that $P_l(U) \approx G'_l(-U)$.  This yields (via
(\ref{heatkernel}) or directly differentiating (\ref{diffdrift}))
\begin{eqnarray}
P_l (U) \approx \frac{1}{\sqrt{8 \pi U_l}}
\exp( - \frac{(U + U_l)^2}{8 U_l} ), \qquad U \gtrsim
U_l. \label{gaussiantail}
\end{eqnarray}
Note that this asymptotic form reproduces all cumulants, $\eta^{(n)}_l
= (\overline{\exp(n U)})_C \sim \exp((2 n^2 -n) U_l)$ as expected.


(\ref{gaussiantail}) is clearly not exact.  Indeed, a breakdown of
(\ref{gaussiantail}) is inevitable on physical grounds, since the
mean/typical barrier cannot be negative!  It suggests a distribution
of barriers with a width proportional to $\sqrt{U_l}$, and hence a
peaked distribution (since the mean barrier $\propto U_l$).
Nevertheless, it does represent a very broad (in fact log-normal)
renormalized distribution of characteristic times $\eta$.  While there
is no reason to believe that such a log-normal tail is exact, the true
distribution of relaxation times will certainly be very broad, with
significant consequences for the average response functions.

\subsection{Breakdown of $\omega \tau_k$ scaling}

\label{breakdown}

\begin{figure}
\centerline{\fig{4cm}{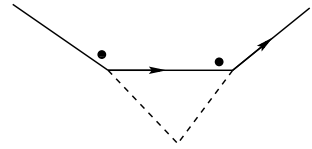}}
\caption{Correction to the $(i \omega)^2$ term in the response function
coming from the second moment $\eta^{(2)}$ of the relaxation time
distribution \label{graph12}}
\end{figure}

The first consequences of this broad distribution occur in the
variance $\eta^{(2)}$ of the relaxation time, and hence at $O(\omega^2)$ 
in the response function.  We therefore examine
more carefully the $O(\omega^2)$ terms in the dynamical action, but
for the moment still neglecting the full functional dependence of these
terms (i.e. on $u_{t}-u_{t'}$).  In the kinetic part of the
action (representing relaxation times and their fluctuations), we
include the
following terms:
\begin{eqnarray}
&& S_{\rm kin} = \int_{rt} \left[ \overline{\eta} ~ i \hat{u}_{rt}
\partial_t u_{rt}  + D ~ i \hat{u}_{rt}
\partial_t^2 u_{rt}  \right] \nonumber \\
&& - {\eta^{(2)} \over 2} \int_{ r t_1 t_2} i \hat{u}_{r t_1}
\partial_{t_1} u_{r t_1}
i \hat{u}_{r t_2} \partial_{t_2} u_{r t_2}.
\end{eqnarray}
One has in general, defining $\langle t^n \rangle_R
= \int_t t^n R_k(t)/\int_t R_k(t)$:
\begin{eqnarray}
&& \langle t \rangle_R = \overline{\eta} \\
&& \langle t^2 \rangle_R - \langle t \rangle_R^2 = \overline{\eta}^2
- 2 D \label{conn1} 
\end{eqnarray}
There is no generic constraint on the sign of $D$. If the
inverse response function contained only the
two above terms ($\overline{\eta}$ and $D$), then 
causality requires $D$ to be positive (
similar to an inertial term) .\cite{footnote10}
Since in general these
are only trucation of an infinite series of terms in power of
$i \omega$,
the only constraint is causality, i.e. that all poles in
$\omega$ lie on the same side of the real axis.
These three couplings satisfy the following closed RG flow equations to
first order in $\Delta$:
\begin{eqnarray}
\partial_l D_l & = & \Gamma_l D_l -
 \Gamma_l  \Lambda_l^{-2} \overline{\eta}_l^2 - A_d \Lambda_l^{d-2}
\eta^{(2)}_l \\
\partial_l \overline{\eta}_l & = & \Gamma_l  \overline{\eta}_l, \\
\partial_l \eta^{(2)}_l  & = & 6 \Gamma_l \eta^{(2)}_l  ,
\end{eqnarray}
where $\Gamma_l = - \tilde{\Delta}_l''(0)  \sim \tilde{\beta} e^{\theta
l}$
and the correction to $D$ from $\eta^{(2)}$ is the graph represented in
Fig. (\ref{graph12}).

For $\eta^{(2)}=0$ the equations for $D_l$ and $\eta_l$, which can be
obtained e.g.  from (\ref{Rflow1}, \ref{sigtti}) by expansion to
second order in $i \omega$, are consistently solved with $D_l \sim
\Lambda_l^{-2} \overline{\eta}_l^2$, in the limit of large $l$, where
$\overline{\eta}_l = \overline{\eta}_0 \exp[ \tilde{\beta}(e^{\theta
  l} -1)/\theta]$, consistent with the Taylor expansion of the
putative scaling function $g(y=i \omega \tau_k)$ given in (\ref{sf3a})
based on the single time scale analysis (\ref{sf3a}).

For $\eta^{(2)} > 0$, however, the broad distribution of relaxation
times completely alters the situation.  From the above equations,
$\eta^{(2)}_l \sim \eta^{(2)}_0
(\overline{\eta}_l/\overline{\eta}_0)^6$, and hence $\eta^{(2)}_l \gg
\overline{\eta}_l^2$ (the mere exponential prefactors are negligible)
at large $l$.  Thus the feedback of $\eta^{(2)}$ in $D$ dominates the
renormalization of $D$, and at large $l$ one finds:
\begin{eqnarray}
D_l \sim \frac{A_d \Lambda_l^{d-2} }{6 \Gamma_l} \eta^{(2)}_0
(\frac{\overline{\eta}_l}{\overline{\eta}_0})^6 
\end{eqnarray}

Thus, allowing for fluctuations in relaxation times invalidates the
$\omega\tau_k$ scaling form already at order $\omega^2$! 

It is still possible, within the approximation scheme of the
present Section, to obtain an equation for the disorder averaged
response function. This is explored further in Appendix 
\ref{general}. 

\section{Distribution of time scales: full structure of the dynamical 
field theory}

\label{sec:fullstructure}

We have established the mechanism for breakdown of the unphysical
$\omega\tau_k$ scaling regime and described the indications of a broad
distribution of timescales within the FRG.  However, to properly
determine this distribution and its consequences, e.g. on the mean
response function, requires a much more complete analysis.  While we
have unfortunately so far been unable to carry this program to
completion, in this section we will detail the formal structure within
which this analysis must be carried out.  In particular, we shall see
that the distribution of relaxation times and its consequences is
encoded within the {\sl boundary layer} (BL) regime already present in
the
statics.  An understanding of the equilibrium dynamics is therefore
contingent first upon an understanding of the static BL,
and we first describe the rather complex structure therein.  Following 
this discussion, we show how the BL regime recurs in dynamical theory, 
and show how it can be formulated to describe broad distributions and
non-trivial scaling of the moments of the relaxation time.

\subsection{statics thermal boundary layer}

\label{sec:staticstbla}

In the appropriate limit the dynamical theory should reproduce the
results for the corresponding statics quantity. We can therefore 
benefit from the knowledge of the thermal boundary layer in the
statics. To do so let us review how the static disorder correlations
are encoded in the dynamical formalism. 

At equilibrium, the part of the dynamical action containing 
the static disorder correlations comprises those terms with no explicit
time derivatives, and reads:
\begin{widetext}
\begin{eqnarray}
&& S_{int} = - \frac{1}{2} \int_{r t_1 t_2} i \hat{u}_{rt_1} i \hat{u}_{rt_2} \Delta(u_{rt_1} -
u_{rt_2}) - \frac{1}{6} \int_{r t_1 t_2 t_3} i \hat{u}_{rt_1} i \hat{u}_{rt_2}
i \hat{u}_{rt_3} S^{(3)}_d(u_{rt_1},u_{rt_2},u_{rt_3}) - ..
\end{eqnarray}
This form is easily understood as arising from the cumulants of the
pinning force. The relation was given for the second cumulant in
(\ref{rvars}) and for higher ones it reads:
\begin{eqnarray}
&& \overline{f(u_1,r_1) .. f(u_k,r_k)}^c 
 = (-)^k S^{(k)}_d(u_1,\cdots,u_k) \delta^d(r_1,\cdots,r_k)
\label{conncumdis}
\end{eqnarray}
with $S^{(2)}_d(u,u') \equiv \Delta(u-u')$ as (\ref{rvars}).
Dues to statistical translational invariance $S^{(k)}(u_1,\cdots,u_k) = S^{(k)}(u_1+v,\cdots,u_k+v)$
and satisfy reflection symmetry 
$S^{(k)}(-u_1,\cdots,-u_k) = (-)^k S^{(k)}(u_1,\cdots,u_k)$.
The cumulants higher than second are generated by
coarse graining, and are thus included here from the start.

The static problem being defined from the equilibrium Boltzmann measure (cf 
\ref{eq:ham}), deals not with the distribution of the random force but with that
of the random potential
\begin{eqnarray}
&& \overline{V(u_1,r_1) .. V(u_k,r_k)}^c 
 = (-)^k S^{(k)}(u_1,\cdots,u_k) \delta^d(r_1,\cdots,r_k) .
\label{connstat}
\end{eqnarray}
Since $f(u,r) = - \partial_u V(u,r)$ one has:
\begin{eqnarray}
&& \Delta(u) = - R''(u) \\
&& S^{(3)}_d(u_1,u_2,u_3) = \partial_1 \partial_2 \partial_3 S^{(3)}(u_1,u_2,u_3) \label{deriv} 
\end{eqnarray} 
and so on. 

It is straighforward to derive the one loop FRG equation in the Wilson
scheme for these cumulants using the dynamical formulation. They are
conveniently expressed using rescaled cumulants:
\begin{eqnarray} \label{resc} 
&& S^{(k)}_d[u_{a_1},\cdots,u_{a_k}] = A_d^{1-k} \Lambda_l^{d + k (\zeta - \theta)} 
\tilde S^{(k)}_d[u_{a_1} \Lambda_l^\zeta,\cdots,u_{a_k} \Lambda_l^\zeta]
\nonumber
\end{eqnarray}
and read,
for the second and third cumulant:
\begin{eqnarray}
\label{DelEqn} && \partial_l \tilde \Delta(u) = (\epsilon - 2 \zeta + \zeta u \partial_u) \tilde \Delta(u) 
+ \tilde T_l \tilde \Delta''(u)
+ 2 \tilde S_{100}(0,u,0)
- \tilde \Delta'(u)^2 - \tilde \Delta''(u) ( \tilde \Delta(u) -  \tilde \Delta(0))
\\
&& \partial_l \tilde S(u_1,u_2,u_3) = (-2 + 2 \epsilon - 3 \zeta + \zeta u_i \partial_{u_i})  \tilde S(u_1,u_2,u_3)
+
\frac{1}{2} \tilde T_l (\tilde S_{200}(u_1,u_2,u_3) + \tilde
S_{020}(u_1,u_2,u_3) \nonumber \\ 
&& +
\tilde S_{002}(u_1,u_2,u_3) ) - \frac{1}{24} \tilde \Delta(0) ( \tilde S_{200}(u_1,u_2,u_3) + \tilde S_{020}(u_1,u_2,u_3) +
\tilde S_{002}(u_1,u_2,u_3) ) \nonumber \\
&&
- \frac{1}{4} \tilde \Delta(u_1 - u_2)
\tilde S_{110}(u_1,u_2,u_3)  - \frac{1}{4} \tilde \Delta''(u_1 - u_2)
( \tilde S(u_1,u_1,u_3) - \tilde S(u_1,u_2,u_3))
\nonumber \\
&& - \frac{1}{4} \tilde \Delta'(u_1 - u_2)
(  \tilde S_{010}(u_1,u_2,u_3) - \tilde S_{100}(u_1,u_2,u_3) +
\tilde S_{010}(u_1,u_1,u_3) + \tilde S_{100}(u_1,u_1,u_3) )
\end{eqnarray}
\end{widetext}
where we have denoted $S^{(3)}_d=S$ and we have suppressed explicitly the feedback 
of the fourth cumulant $S^{(4)}_d$ into the third one. One can check that this
gives exactly the derivatives (\ref{deriv} ) of the one loop FRG 
equations for the static correlators $R$ and $S^{(3)}$ displayed in (6,7)
in Ref.~\cite{us_short}. These relations (\ref{deriv} ) should indeed
be preserved by RG at equilibrium.

As discussed in Ref.~\onlinecite{staticslong} when all arguments of
these functions are distinct and order one conventional scaling holds.
That is, at large scales for which $T_l \to 0$, the functions $\tilde
S^{(k)}$ approach well defined nonanalytic fixed point forms $\tilde
S^{(k)*}$. Moreover, naively these can be organized in an
$\epsilon=4-d$ expansion in which $\tilde S^{(k)*} \sim \epsilon^k$,
$k \geq 3$. Naively this would allow the truncation of the hierarchy
of FRG equations for the $\tilde S^{(k)}$, neglecting feedback of the
$k > p$ cumulants with an accuracy of $O(\epsilon^p)$. However, the
convergence to these values is highly non-uniform as mentionned in
Section II since at non-zero temperature these functions remain
analytic at $u=0$. A detailed analysis of the static hierarchy of FRG
equations relating these cumulants revealed the existence of a thermal
boundary layer (TBL) of the form:
\begin{eqnarray}
&& \tilde{\Delta}(u) = \tilde{\Delta}^*(0) - \tilde T_l f(\tilde u) \label{tblf} \\
&& \tilde u = \epsilon \tilde{\chi} u/\tilde{T} _l
\end{eqnarray}
for $\tilde u = O(1)$, $\tilde{T}_l \ll \epsilon^2$ and $f$ an
analytic function with $f(x) \sim |x|$ at large $x$ to match the
cusp of the zero temperature solution. For higher cumulants 
the very unconventional TBL scaling implies that it is no longer
legitimate to neglect the feedback of higher cumulants (the $n$-th
cumulants gets a feedback from the $n$ and $n+1$ ones). Therefore,
we are unable to truncate and solve the hierarchy of FRG equations.
Instead, in Ref.~\onlinecite{staticslong} we argued for the consistency
of a thermal boundary layer {\it ansatz} (TBLA), which for the force
cumulants reads:  
\begin{widetext}
\begin{eqnarray}
&&  \tilde S^{(k)}_d(u_1, \cdots u_k) = \cases{ f_{k}  
+ (\tilde \chi \epsilon)^{k-2}
T_l s_d^{(k)}(\tilde u_1, \cdots \tilde u_{k})   & $k$ even \cr
\hspace{.295in} (\tilde \chi \epsilon)^{k-2}
T_l s_d^{(k)}(\tilde u_1, \cdots \tilde u_{k})   & $k$ odd },
\label{bl}
\end{eqnarray}
\end{widetext}
where $s_d$ are well defined functions of order one
in the TBL $\tilde u \sim 1$. 
The set of ($l$-dependent) constants,
\begin{eqnarray}
  && f_{2p} = \tilde{S}_d^{(2 p)}(0, \cdots, 0)/(\tilde\chi\epsilon)^{2p}
\label{f2p}
\end{eqnarray}
with $f_{2} = \tilde \Delta(0)/(\tilde\chi\epsilon)^2$,
have the meaning of the linearized random force cumulants within the
zero temperature Larkin description. As discussed in
Ref.~\onlinecite{staticslong}  
the crucial difference with the naive dimensional reduction
result, where the $f_{2p}$ are unrenormalized, is that they
get feedback from the TBL functions and acquire non-trivial
asymptotic values. 

The TBLA encodes a huge amount of physics -- in particular, all the
distributions of minima degeneracy responsible for large 
thermal fluctuations in the droplet picture, as detailed
in Ref.~\onlinecite{staticslong}. For instance, 
averages such as (\ref{eq:twowell}) can be estimated 
using the TBLA, the coefficients $c_n$ being in principle
determined by the functions $s^{(k)}$. 

This already non-trivial structure must now be generalized to 
intrinsically dynamical quantities.

\subsection{dynamical hierarchy of kinetic coefficients}

\label{sec:dynamicstbla}

In a conventional dynamical renormalization group in the MSR formalism
a succession of individual terms are added to the action corresponding
to increasingly high frequency kinetic coefficients, e.g. for a
particle the Stokes drag, inertial mass, ... For the disordered
elastic manifold however we recognize that these kinetic coefficients
have a broad distribution characterized by an infinite set of
cumulants and cross-correlations, which moreover can be non-trivial
functions of displacement field differences. The latter dependence was
neglected in the approximate treatment of Section III. The need for
treating it was already indicated in the ambiguities in the diagrams
of Fig. 3.  Each of these cumulants and cross-correlations appears as
a distinct interaction {\sl function} in the MSR action.

By symmetry (time translation and STS,
statistical reflection, causality) alone, the set of all such
interactions contributing to the effective action at zero temperature
can be written as
\begin{eqnarray}
&& S = \sum_{n=1}^\infty  \sum_{ P= \{ p_i^k \} }
\int_{r t_1 \cdots t_n} \! i \hat{u}_{r t_1} \cdots i \hat{u}_{r t_n}
\nonumber \\
&&  S^{(n)}_{P}(u_{r t_1},..u_{r t_n})
\prod_{k=1}^{+\infty} \prod_{i=1}^{n}
(\partial_{t_i}^{k} u_{r t_i})^{p_i^k} \label{eq:fullST0}
\end{eqnarray}
where $p_i^k \geq 0$ and from STS symmetry, the $S(u_1+ u,..u_n+ u) =
S(u_1,..u_n)$ are translationnally invariant, and statistical
reflection implies the full action is also
invariant under $(\hat{u}, u) \to (- \hat{u}, - u)$.  The random force
correlators correspond to 
\begin{equation}
  S^{(n)}_{P=0}(u_1,\cdots,u_n) = S_d^{(n)}(u_1,\cdots,u_n),
\end{equation}
where $P=0$ above indicates the function with $p_i^k=0$ for
all $i,k$.  Other terms correspond to intrinsically dynamical
cumulants.  

It is instructive to begin the characterization of such terms at $T=0$
by listing all possible forms in order of increasing number $m$ of
time derivatives and order of cumulant (i.e. the number of independent
times which equals the number of $\hat u$ fields at
$T=0$). Each term in (\ref{eq:fullST0}) can be assigned
$m=\sum_{i=1}^n \sum_{k=1}^{+\infty} k p^k_i$.  For 
organisational purposes it is convenient to rewrite the action
of (\ref{eq:fullST0})  in a schematic (but transparent)
notation, first expanding in number of cumulants:
\begin{widetext}
  \begin{eqnarray}
    && S = i\hat{u}_1 (k^2 + \Sigma( \partial_1 )) u_1 + S_{int} \\
    && \Sigma(s) = \overline{\eta} s + D s^2 + .. \label{eq:sigmas}\\
    && S_{int} = - \frac{1}{2}i\hat{u}_1 i\hat{u}_2 {\cal S}_{12} 
    - \frac{1}{6} i\hat{u}_1 i\hat{u}_2 i\hat{u}_3 {\cal S}_{123} - .. 
  \end{eqnarray}
  with $s=i \omega$.  Here the subscripts $1,2,...$ refer to different
  times being independently integrated over in the action at same space
  point $r$ (further integrated on).  The ${\cal S}_{12...}$ are then
  functions of the $u_1,u_2,...$ and their time derivatives. We then
  expand each of these in increasing number $m$ of time derivatives:
  \begin{eqnarray}
    &&
    {\cal S}_{12}  = 
    \Delta(u_{12}) + (\dot{u}_1 - \dot{u}_2) G(u_{12}) + \dot{u}_1 \dot{u}_2
    A(u_{12}) + (\dot{u}_1^2 +  \dot{u}_2^2) B(u_{12}) + (\ddot{u}_1 -
    \ddot{u}_2) C(u_{12}) + \ldots  \label{eq:cals12}\\
    && {\cal S}_{123}  = S(u_1,u_2,u_3) + \frac{1}{3} ( \dot{u}_1
    H(u_{1};u_{2},u_{3}) + \dot{u}_2  H(u_{2};u_{3},u_{1}) + \dot{u}_3 H(u_{3};u_{1},u_{2}) ) 
 + \dot{u}_1 \dot{u}_2 W(u_{1},u_{2};u_{3}) + .. \\
&& \ldots \label{eq:cals123}
  \end{eqnarray}
As discussed above, each new term in
(\ref{eq:cals12},\ref{eq:cals123}) corresponds to statistical
properties of the random kinetic coefficients and forces in a
renormalized equation of motion, in particular,
\begin{eqnarray}
  \label{eq:reneom}
  &&  \cdots+ D(u,r)\ddot{u} + \eta(u,r)\dot{u} = \nabla^2 u + f(u,r) +
  g(u,r)\dot{u}^2 + \cdots + \zeta(r,t),
\end{eqnarray}
with
\begin{eqnarray}
  \label{eq:crosscorrels}
  && \overline{D(u,r)}=D,  \qquad \overline{\eta(u,r)f(u',r')}^c =
  -G(u-u')\delta(r-r'), \qquad \overline{\eta(u,r)\eta(u',r')}^c =
  A(u-u')\delta(r-r'), \\
  && \overline{g(u,r)f(u',r')}^c =  B(u-u')\delta(r-r'), \qquad
  \overline{f(u,r)D(u',r')}^c =  C(u-u')\delta(r-r'), \\
  && \overline{\eta(u_1,r_1)f(u_2,r_2)f(u_3,r_3)}^c =
  \frac{1}{3} H(u_1;u_2,u_3)\delta(r_1-r_2)\delta(r_2-r_3) \\
&& \overline{\eta(u_1,r_1) \eta(u_2,r_2) f(u_3,r_3)}^c = \frac{1}{3} W(u_1,u_2;u_3)
\end{eqnarray}
\end{widetext}
In the approximate treatment of Sec.~\ref{sec:fterm}, $\eta^{(2)}$
hence corresponds to $A(u)$ approximated as $A(0)$.  Note that it is
the {\sl small argument} behavior of $A(u)$ (and its higher cumulant
analogues) that is related to the physically interesting second
(higher) relaxation time moment $\eta^{(2)} \sim \overline{\langle
  t\rangle^2}^c$ ($\eta^{(n)} \sim \overline{\langle t\rangle^n}^c$).
Hence these relaxation times are encoded within the BL regime of these
functions.  This was also apparent from the na\"ive renormalization of
$\eta$ by $\Delta''(0)$, also a BL quantity.  We will return to the
problem of the dynamic BLs in $G(u),A(u),\ldots$ momentarily.

Although it is convenient as above for the purpose of {\sl
  enumerating} terms in the dynamical action to first separate by
cumulant index $n$ and then by number of time derivatives $m$,
conceptually we analyze them in the opposite scheme, i.e. collecting
all terms of a given $m$, and organising these afterward in order of
$n$.  This scheme is clearly convenient insofar as the first term
($m=0$) of each of the ${\cal S}_{12\cdots n}$ corresponds to the
$n$-th term of the static cumulant hierarchy, so that the set of terms
with $m=0$ satisfies a closed hierarchy of FRG equations independent
of those with $m>0$.  We now demonstrate diagrammatically that, at
zero temperature, a similar property holds for $m>0$.  In particular,
all terms of any given $m$ will satisfy a closed hierarchy of FRG
equations containing only terms with $m'\leq m$.  Thus, one may
imagine (dream of?) solving the FRG equations up to level $m$, then
using this solution to complete a closed set of FRG equations for
level $m+1$, and iteratively solving for higher and higher $m$.

This closure relies on the rule of conservation of powers of
frequency, established in Sec.~\ref{sec:fterm}.  Recall that this
occurs because at $T=0$, the correlation function vanishes, and all
contractions take the form of {\sl causal} response functions.  Thus
no closed time loops can appear in any diagram.  This implies that
internal time derivatives which appear in any diagram appear as
factors of frequency of some external leg to which they are connected.
In any case, this rule implies that, since all terms in the action
have $m\geq 0$, terms with $m'>m$ can never reduce their number of
time derivatives by contraction with another vertex at $T=0$, and
hence cannot renormalize $m$-vertices.  This is true to all orders for
diagrams with any number of loops.  The frequency conservation rule
implies in fact a more detailed result.  If the quadratic terms in
$\Sigma(s)$ ($\eta,D,\ldots$) are regarded themselves as coupling
constants \cite{footnotepert} (with $m=1,2,\ldots$), 
then each term in the FRG equation for any
quantity at level $m$ is a product of factors for which the total
frequency level (i.e. $\sum m_i$ for all terms $i$ in the product) is
exactly $m$.  Thus a static quantity (e.g. $\Delta$) can renormalize a
dynamic one (e.g. $G$,$\eta$) only in combination (i.e. multiplied by)
another dynamic quantity, and so on.  

\begin{figure}
\centerline{\fig{9cm}{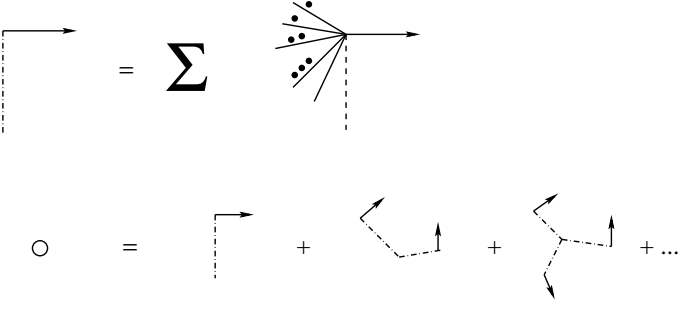}}
\caption{ Compact notation for a generic vertex at $T=0$.  \label{graph20}}
\end{figure}

\begin{figure}
\centerline{\fig{6cm}{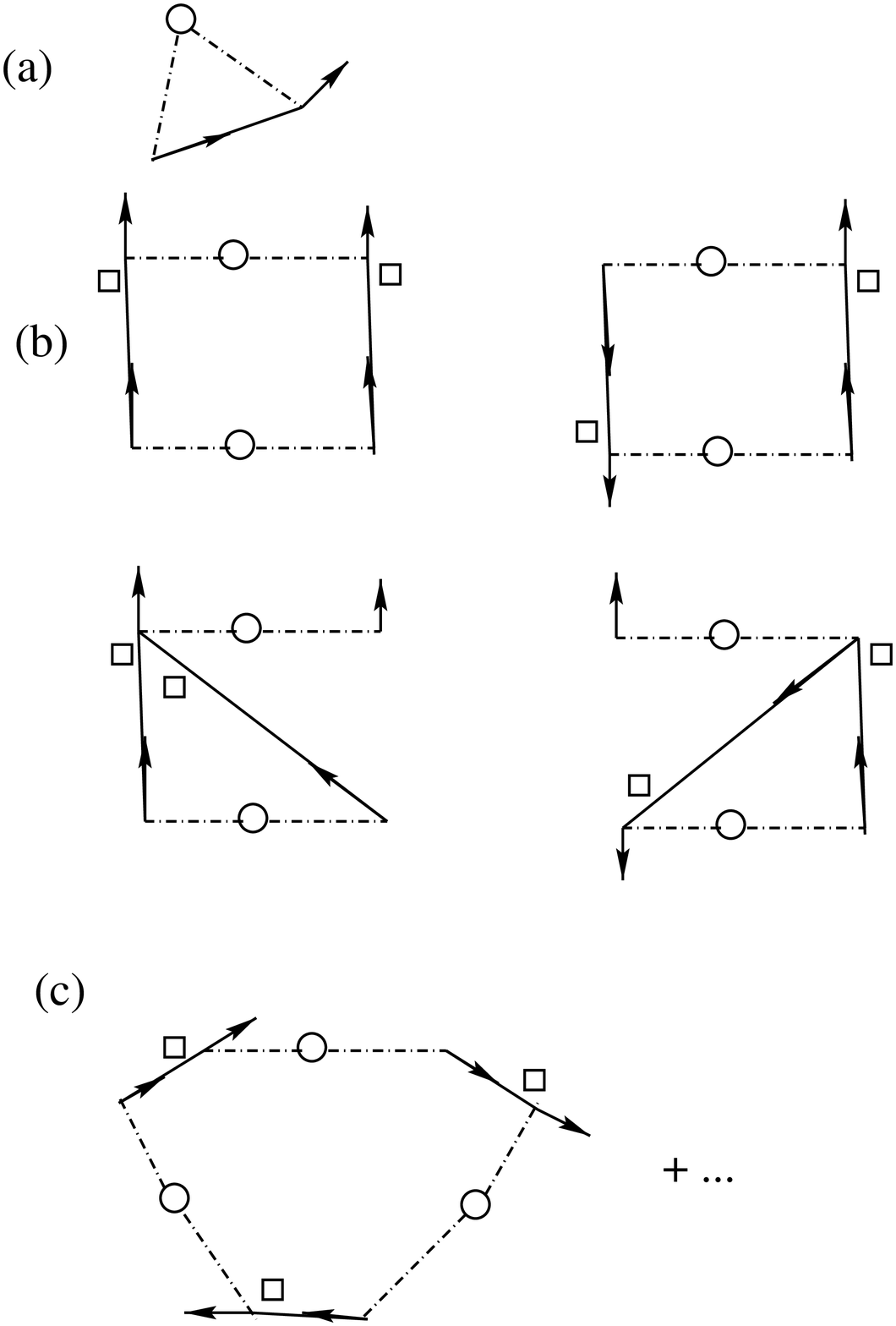}}
\caption{One loop diagrams which correct the effective action at $T=0$:
  the internal lines contain the full response function and the graphs
  are 1P irreducible. Graph (a) is a ``tadpole''.  Graphs (b) and (c)
  (and higher orders) correct terms with $n \geq 1,2,3$ respectively.
  \label{graph21}}
\end{figure}

One can also establish a set of rules to understand how cumulants with
different $n$ are connected in the FRG equations.  At $T=0$, this
process is highly constrained, since each contraction involves one
response function, which removes one $\hat{u}$, it is straightforward
to count the possible connections.  We will restrict our attention to
one loop diagrams, anticipating future nonperturbative exploration
using the exact RG,\cite{ERG,staticslong} in which only these appear
(and in any case only these are consistently treated in the Wilsonian
scheme of this paper). The counting is illustrated for such one loop
diagrams in Figs.~\ref{graph20},\ref{graph21}.  One readily sees that
when $N$ vertices are combined in this manner, the resulting vertex
which is renormalized in the effective action contains a total number
of independent times (or $\hat{u}$ factors) $n=\sum_{i=1}^N n_i -N$,
due to the $N$ response functions appearing in the loop.

With these rules in mind, we can describe the structure of the FRG
hierarchy as far as the feeding of terms of a given $m,n$ into other
$m',n'$.  We note symbolically by $S^n_m$ the terms with $n$ response
fields and $m$ time derivatives.  
The term $\overline{\eta}$ is $S^1_1$, the response
function is the quadratic part of $S^1_m$ (we note
$R^{-1}=\text{quad}(S^1_m)$) and the cumulants $\eta^{(n)}$ are
included in $S^n_n$. From the above discussion,
neglecting rescaling terms, the structure of the FRG equations reads
\begin{widetext}
\begin{eqnarray}
  &&
  \delta S_m^{(n)} = S_m^{(n+1)}  +
  \sum_{m'=0}^m \sum_{n'=1}^{n+1} S_{m'}^{(n')} S_{m-m'}^{(n+2-n')}
  + \sum_{m'+m''\leq m}\sum_{n'+n''\leq n+2}
  S_{m'}^{(n')} S_{m''}^{(n'')}  S_{m-m'-m''}^{(n+3-n'-n'')} + \cdots
  . \label{eq:schematicFRG}
\end{eqnarray}
\end{widetext}

It is straightforward to see that this series contains a finite number
of terms for any given $m,n$.  Let us suppose that an $N$-loop term
exists, such that each vertex making it up has $n_i$ time
integrations.  Suppose of these $N$ vertices, $n_i>1$ for $N'$ of
them and $n_i=1$ for the remaining $N-N'$.  Then $n=\sum_{i=1}^N n_i
-N = N-N' + \sum_{i=1}^{N'} n_i -N = \sum_{i=1}^{N'} (n_i-1)$.  Hence
at most $N' \leq n$.  Now the remaining $N-N'\geq N-n$ vertices have
only one time integration.  Since there are no allowed local terms
without time derivatives, these must each have $m_i \geq 1$, i.e. $m
\geq N-n$.  Turning this around, $N\leq m+n$, so that the series of
one loop diagrams terminates at at most $(m+n)^{\rm th}$ order.
Clearly from (\ref{eq:schematicFRG}), each order contains a finite
number of terms, so that the one loop FRG equations are finite.

\subsubsection{terms proportional to frequency $m=1$}

\begin{figure}
\centerline{\fig{5cm}{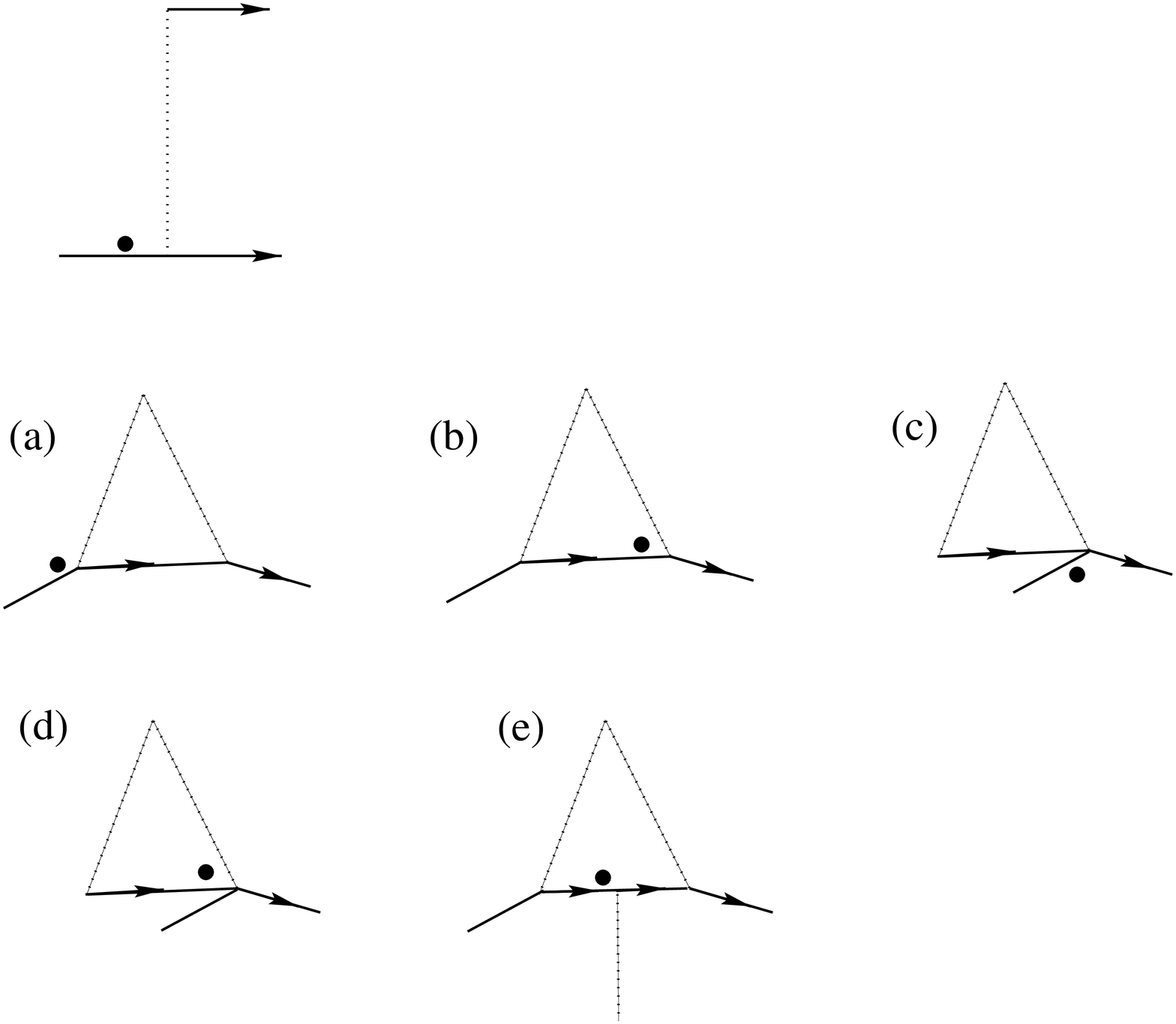}}
\caption{Shown is the $G$ vertex (top image with no alphabetic label),
  and diagrammatic corrections to $\overline{\eta}$.  Graphs (a-d) are
  contributions from tadpoles of the G vertex (note that (a) and (c)
  cancel by the same mechanism as dimensional reduction, and that (d)
  vanishes upon integration by parts on internal line). Graph (e) is
  the contribution from $\overline{\eta} \Delta$.
 \label{graph22}}
\end{figure}

\begin{figure}
\centerline{\fig{4cm}{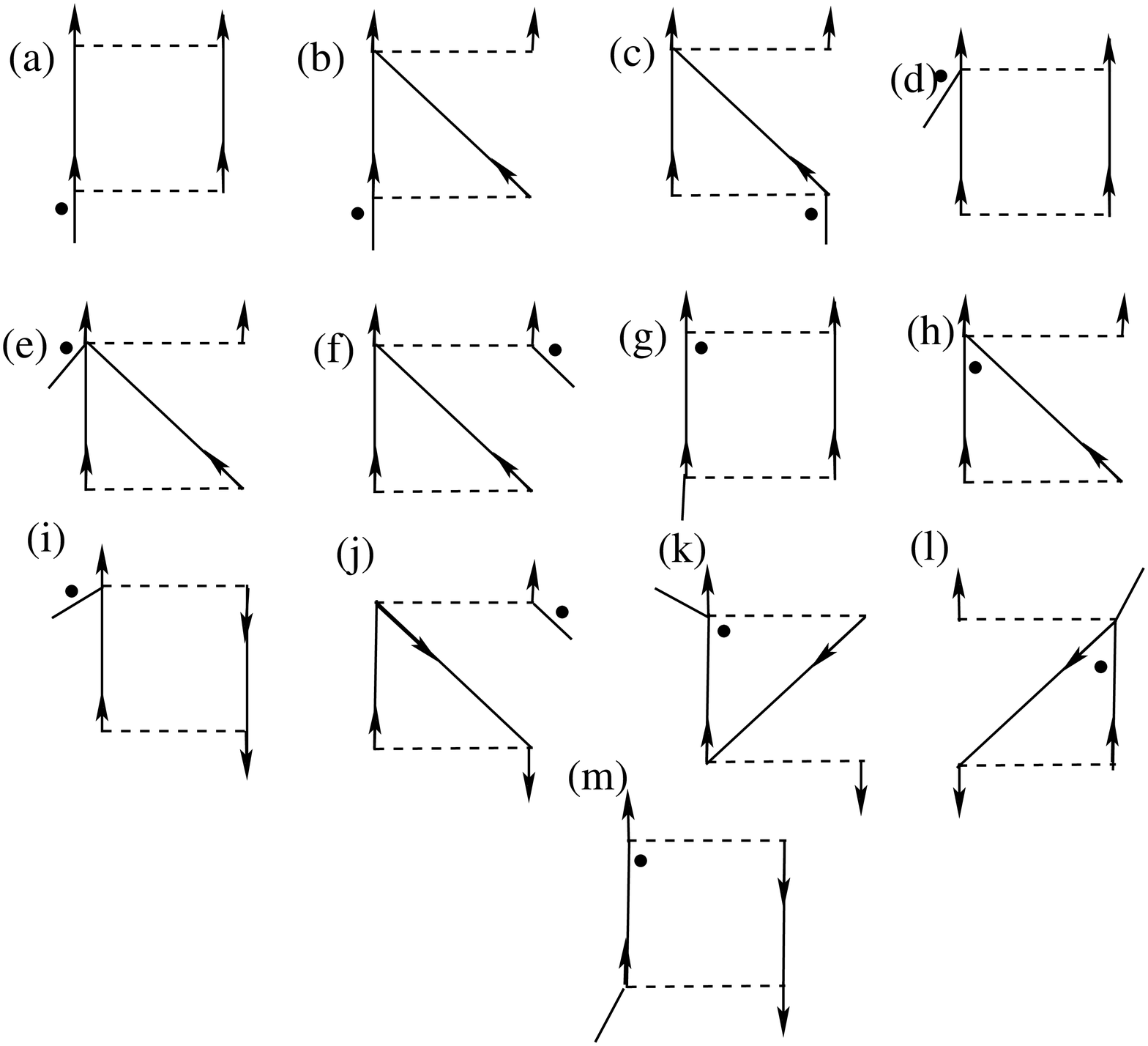}}
\caption{Corrections to G. \label{graph23}}
\end{figure}

We will now examine level $m=1$ and $m=2$ of the hierarchy. For $m=1$
we will restrict to study the FRG equation for terms with $n \leq 2$
for which we need terms up to $n=3$:
\begin{widetext}
\begin{eqnarray}
  \label{eq:dynints}
   S^{(1-3)}_1 & = &  \int_{r t} \overline{\eta} \,i
  \hat{u}_{rt}  
      \dot{u}_{rt} - \frac{1}{2} \int_{r t_1
      t_2} 
   i \hat{u}_{rt_1} i \hat{u}_{rt_2} (\dot{u}_{rt_1} - \dot{u}_{rt_2})
    G(u_{rt_1}-u_{rt_2}) 
-\frac{1}{6} \int_{rt_1t_2t_3}   i \hat{u}_{rt_1}  i \hat{u}_{rt_2}
     i \hat{u}_{rt_3} \dot{u}_{rt_1}
    H(u_{rt_1},u_{rt_2},u_{rt_3}) \nonumber \\
&&
  \end{eqnarray}
where $G(-u)=-G(u)$.

The renormalisation of $\overline{\eta}$ and $G$ is determined by a
standard if cumbersome one loop calculation performed in the
Appendix~\ref{sec:one-loop-hierarchy}. The corresponding graphs are
represented in Fig. \ref{graph22} and \ref{graph23}.  From dimensional
analysis and the structural form of the FRG equation
(\ref{eq:schematicFRG}) we see that and $G$ and $H$ as single
frequency $m=1$ terms will be fed by $O(\overline{\eta} \Delta^2)$ and
$O(\overline{\eta} \Delta^3)$ respectively. Hence, given the rapid
growth of $\overline{\eta}$ with scale, we expect these functions to
be at least growing as $\overline{\eta}$ with scale. We thus defined
rescaled functions:
\begin{eqnarray}
&& G_l(u) = \overline{\eta}_l \, 
\frac{\Lambda_l^{2-d} e^{\zeta l}}{A_d} \,  \tilde{G}(u
e^{-\zeta l}) \label{resceta} \\
&& H_l(u_1,u_2,u_3) = \overline{\eta}_l \, 
\frac{\Lambda_l^{4-2d} e^{2 \zeta l}}{A_d^2}  \, \tilde{H}(u_1
e^{-\zeta l}, u_2 e^{-\zeta l},u_3 e^{-\zeta l}) \nonumber
\end{eqnarray}
in terms of which one finds the flow equation \cite{depinning} for $\overline{\eta}$
and $\tilde{G}(u)$:

\begin{eqnarray}
&& \partial_l \overline{\eta} = 
(\tilde{G}'(0) - \tilde{\Delta}''(0) ) \overline{\eta} \label{deta}\\
&& \partial_l \tilde{G} = (-2 + \epsilon - \zeta) \tilde{G} + \zeta u \partial_u
\tilde{G} - 2 \tilde{\Delta}'' \tilde{G} + (\tilde{\Delta}(0)-\tilde{\Delta})
\tilde{G}'' - 3 \tilde{\Delta}' \tilde{G}'
- \tilde{G}'(0)  \tilde{G} - \tilde{G}'(0) \tilde{\Delta}' \nonumber \\
&& +  \tilde{\Delta}' (2\tilde{\Delta}''(0) +
2\tilde{\Delta}'' ) + \tilde S_{110}(0,0,u) 
+ \frac{1}{3} (\tilde H_{010}(u,0,0) - 2 \tilde H_{001}(0,u,0) - \tilde H_{100}(0,u,0) ) , 
 \label{dG}
\end{eqnarray}
\end{widetext}
Because of the above rescaling (\ref{resceta}) no explicit 
$\overline{\eta}$ appear in (\ref{dG})

Since $\tilde{G}'(0)$ appears on the same footing as 
$\tilde{\Delta}''(0) \sim 1/\tilde T_l$ in (\ref{deta}) 
it is natural to expect it to grow unboundedly with scale in the
same fashion. Indeed inspection of (\ref{dG}) reveals that
$\tilde G$ is fed by a term explicitly proportional to 
$\tilde{\Delta}''(0)$ which can consistently be balanced by
the $\tilde{G}'(0)$ term appearing in the first line of the
same equation. Therefore we are led to expect that
$\tilde{G}$ itself, like $\tilde{\Delta}$ should exhibit a 
thermal boundary layer and the effect of temperature will be
essential in understanding the structure properly. We 
will come back after taking a brief look at the
$m=2$ terms.

\subsubsection{terms proportional to square frequency $m=2$}

The $m=2$ terms, restricting to $n \leq 2$ -th order cumulant
read:
\begin{widetext}
\begin{equation} 
S^{(1-2)}_2=  \int_{r t} i \hat{u}_{rt} D  \ddot{u}_{rt} 
 - \frac{1}{2} \int_{r t_1 t_2} \!\!\!\!\! i \hat{u}_{rt_1}
i \hat{u}_{rt_2}
[ \dot{u}_{rt_1} \dot{u}_{rt_2}
A(u_{rt_1}\!-\!u_{rt_2}) +  ( \dot{u}_{rt_1}^2 \!+\!
\dot{u}_{rt_2}^2) B(u_{rt_1}\!-\!u_{rt_2}) +
(\ddot{u}_{rt_1}\! -\! \ddot{u}_{rt_2})
C(u_{rt_1}\!- \!u_{rt_2}) ]
\end{equation}
\end{widetext}

The renormalisation of $D$ and of the functions $A$,$B$ and $C$
via a one loop calculation is performed in the Appendix. It turns
out to be convenient to define:
\begin{eqnarray}
&& B_1(u) = B(u) - \frac{1}{2} C'(u)
\end{eqnarray}
which simplifies the equations, the physics being explained
below. We define rescaled quantities as follows:
\begin{eqnarray}
&& D = \Lambda_l^{-2} \tilde D \\
&& A(u) = 
\frac{\Lambda_l^{-d} }{A_d} \,  \tilde{A}(u
e^{-\zeta l}) \quad, \quad B_1(u) = 
\frac{\Lambda_l^{-d} }{A_d} \,  \tilde{B_1}(u
e^{-\zeta l}) \nonumber \\
&& C(u) = \frac{\Lambda_l^{-d} e^{\zeta l}}{A_d} \,  \tilde{C}(u
e^{-\zeta l}) \label{resc2} \\
&& W_l(u_1,u_2,u_3) = \overline{\eta}_l^2 \, 
\frac{\Lambda_l^{2-2d} e^{ \zeta l}}{A_d^2}  \, \tilde{W}(u_1
e^{-\zeta l}, u_2 e^{-\zeta l},u_3 e^{-\zeta l}) \nonumber
\end{eqnarray}

and one finds:
\begin{widetext}
\begin{equation}
 \partial_l  \tilde D = (-2 - \tilde{\Delta}''(0)) \tilde D  - \tilde A(0) + \tilde C'(0)
- \overline{\eta}^2 (2 \tilde G'(0) - \tilde{\Delta}''(0)) \label{dD}
\end{equation}
together with the FRG coupled flow equations for $\tilde A(u)$,$\tilde
B(u)$ and $\tilde C(u)$ which read:
\begin{eqnarray} 
 \partial_l \tilde A  & = & 
\big( - d + \zeta u \partial_u - 2 \tilde \Delta''(0)  -
  4 \tilde \Delta''  \big) \tilde A 
-4 \tilde \Delta' \tilde A'
+ ( \tilde \Delta(0) - \tilde \Delta ) \tilde A'' \label{frgA} \\ && 
+ \overline{\eta}^2 \big(
2\,\tilde G'(0)\,\tilde G' +
   5\, \tilde G'^2 - 4 \tilde G'\,\tilde \Delta''(0) - 8 \tilde G'\,\tilde \Delta'' +
  2 \tilde \Delta''^2 + 4\,\tilde G\,\tilde G'' -
  4 \tilde \Delta'\,\tilde G'' \big)  \nonumber
\end{eqnarray}

\begin{eqnarray} 
 \partial_l \tilde  B_1 & = &  \big( - d + \zeta u \partial_u 
- 3 (\tilde \Delta'' + \tilde \Delta''(0)) ) \tilde B_1 + 3 \tilde \Delta'' \tilde B_1(0) 
- 4 \tilde \Delta' \tilde B_1'   + (\tilde \Delta(0)- \tilde \Delta) \tilde B_1'' 
+ \tilde A(0) \tilde \Delta'' \label{frgB} \\
& & + \overline{\eta}^2 \big( 
- \tilde G'(0) \tilde G'
- \tilde G'^2 - \tilde G \tilde G'' 
+ 2 \tilde G' \tilde \Delta''(0)
+ \tilde G'(0) \tilde \Delta''
+ \tilde G' \tilde \Delta'' + \tilde \Delta' \tilde G''
- \tilde \Delta''(0) \tilde \Delta'' )  \nonumber
\end{eqnarray}

\begin{eqnarray} 
   \partial_l \tilde C & = & 
\big (-\zeta - d  + \zeta u \partial_u   - \tilde \Delta''(0)
  - 2 \tilde \Delta''
\big) \tilde C + \tilde \Delta' \tilde A(0)  - \tilde \Delta' \tilde C'(0) -
   3 \tilde \Delta' \tilde C'  +
 (\tilde  \Delta(0) - \tilde \Delta)\,\tilde C''  
 + 2 \tilde D \,\tilde \Delta'\,(\tilde \Delta''(0) + \tilde \Delta'') \nonumber 
\\
&& + \overline{\eta}^2 \big(
- 2\,\tilde G\,\tilde G'(0) + 4 \tilde \Delta'\,\tilde G'(0)
   - 2\,\tilde G\,\tilde G' + 2 \tilde \Delta'\,\tilde G'
+ 2 \tilde G\,\tilde \Delta''(0)
  - 3 \tilde \Delta'\,\tilde \Delta''(0) + 2 \tilde G\,\tilde \Delta''  -
  2 \tilde \Delta'\,\tilde \Delta''  \big) \label{frgC}
\end{eqnarray}

\end{widetext}

From these equations we expect exponential growth of
$\tilde D$, $\tilde{A}$, $\tilde{B}$, $\tilde{C}$ 
at least as fast as $\overline{\eta}^2$ due to the
feeding terms. We expect from the considerations of
Section \ref{sec:fterm} that the growth is actually faster.
Indeed considering the $A$ flow equation at the origin gives
$\partial_l \tilde A(0) = - 6 \tilde \Delta''(0) \tilde A(0)$ keeping
the largest terms of order $1/\tilde T_l$ and neglecting
feeding terms. Note the similarity to the result of Section III.
While we expect this qualitative behaviour, the
precise nature of the growth is more subtle due to the
fact that at non-zero temperature $A(0)$ no longer satisfies a
closed equation. We will discuss this in more details below.

Let us close this Section by noticing that all
non-trivial terms in the r.h.s. of the above set 
of FRG equations ($\beta$-functions) for
$G$, $A$, $B$ and $C$ cancel when one chooses:
\begin{eqnarray}
&& \tilde G(u) = \tilde \Delta'(u) \label{manifold} \\
&& \tilde A(u) = - \overline{\eta}^2 \tilde \Delta''(u) \\
&& \tilde C = \tilde D \tilde \Delta'(u) \\
&& \tilde B_1 = \tilde D \frac{1}{2} \tilde \Delta''(u) \\
&& \tilde H(u_1,u_2,u_3) = 3  \tilde S_{100}(u_1,u_2,u_3) \\
&& \tilde W(u_1,u_2,u_3) = - 3 \tilde S_{110}(u_1,u_2,u_3)
\end{eqnarray}
and furthermore the flow of the kinetic coefficients
simplify into:
\begin{eqnarray}
&& \partial_l \overline{\eta} = 0 \\
&& \partial_l D = 0  
\end{eqnarray}
This remarkable property, which serves as a useful
check on the RG equations, can be understood
in terms of a simple integrable model which is
studied in the Appendix \ref{marvelous} (which has very different
physics from the one studied here).

\subsection{dynamical thermal boundary layer}

\subsubsection{dynamical action at non-zero temperature and FDT}
\label{sec:dynamical-action-at}

\begin{figure}
\centerline{\fig{9cm}{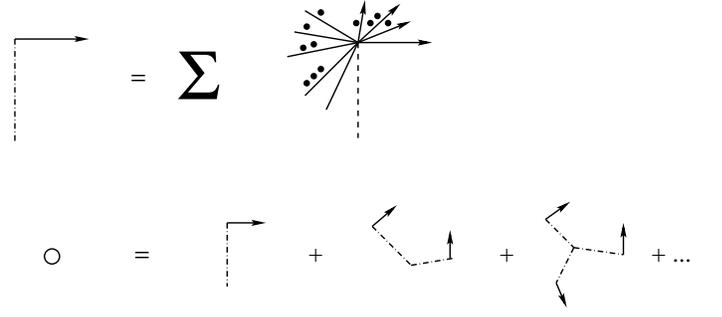}}
\caption{Compact notation for a generic vertex at $T>0$.  \label{graph25}}
\end{figure}

At $T>0$ two new effects must be taken into account not present at
$T=0$.  First, in addition to the kinetic coefficients studied above
one must take into account a variety of thermal noise terms. In the
dynamical action this corresponds to terms with two or more $\hat
u$ field having the same time argument.  Second, new thermal
contractions are possible using the non-zero correlation function
$\langle u u\rangle$ of the Gaussian theory.  We first focus on the
former, and discuss the additional contractions at the end of
this subsubsection.  The action takes the general form:
\begin{eqnarray}
&& S = \sum_{n=1}^\infty  \sum_{ P= \{ p_i^k \} , R= \{ r_i^k \}}
\int_{r t_1 \cdots t_n} \! i \hat{u}_{r t_1} \cdots i \hat{u}_{r t_n}
\nonumber \\
&&  S^{(n)}_{P,R}(u_{r t_1},..u_{r t_n})
\prod_{k=0}^{+\infty} \prod_{i=1}^{n}
(\partial_{t_i}^{k} i \hat{u}_{r t_i})^{r_i^k}
(\partial_{t_i}^{k} u_{r t_i})^{p_i^k} \nonumber
\\
&&
\end{eqnarray}
with $p_i^0=0$ and the $r_i^k \geq 0$ are integers. The
$S^{(n)}_{P,R}$ are translationnally invariant functions. Compared to
the $T=0$ action it has additional powers of $\hat{u}_i$ and possibly
their time derivatives (such vertices are shown in
Fig.~\ref{graph25}).  There is a temperature homogeneity degree
$s=\sum_k \sum_i r_i^k$ such that the term is $\sim T^s$. The standard
thermal noise corresponds to $S^{(1)}_{P=0,R} = - \eta T$, with
$r^k_i=2\delta_{k0} \delta_{i1}$.

In order to maintain the Fluctuation Dissipation Theorem (FDT) there are
relations between these vertices. A useful symmetry which constrains
the allowed form of these terms is:
\begin{eqnarray}
&& i \hat u_{rt} \to i \hat u_{r,-t} - \lambda_r \dot u_{r,-t} \label{lambda} \\
&& u_{rt} \to u_{r,-t}
\end{eqnarray}
(we mean $\dot u_{-t}=u'(-t)$). For actions with no explicit time dependence, such as considered here,
one can then later make a change of variables $t \to - t$ in integrals over
times. We
apply this to the bare action (\ref{msr1} - \ref{msr2}). 
Consider first the infinitesimal variation of the interaction term:
\begin{eqnarray}
&& \delta S_{int} = - \int_r \lambda_r \int_{t_1,t_2} i \hat u_{rt_1} \dot u_{rt_2} 
\Delta(u_{rt_1} - u_{rt_2}) + O(\lambda^2) \nonumber \\
&& =
- \int_r \lambda_r \int_{t_1,t_2} i \hat u_{rt_1} 
\partial_{t_2} R'(u_{rt_1} - u_{rt_2}) + O(\lambda^2) ,
\end{eqnarray}
using $\Delta(u)=-R''(u)$, a consequence of
potentiality. This integrates to a boundary term which is a function
only of the coordinates $u$ and corresponds to the energy difference
between the initial and final times(see below).
Hence the interaction term is invariant for an arbitrary
($r$-dependent) $\lambda_r$. Unfortunately, this large symmetry is
quadratically broken by (\ref{msr1}). First, the variation of the elastic
term vanishes (up to boundary terms) only for spatially constant 
$\lambda_r=\lambda$. Thus the full action for $\eta=0$ has a continuous 
global $\lambda$-symmetry (\ref{lambda}). This can be used
e.g. to put constraints on the terms appearing in the FRG equation
order by order in $\eta$ .\cite{randommass}
More importantly, the remaining terms in the action are only
invariant under a discrete transformation, specifically 
(\ref{lambda}) with:
\begin{eqnarray}
&& \lambda = \frac{1}{T} \label{symfdt}
\end{eqnarray}
Note that they are, however, {\it exactly} invariant (no boundary term).

Having established the invariance of the bare model under the symmetry (\ref{symfdt})
we know that it should be preserved under renormalisation. We must 
thus understand the consequences of this symmetry for a more general
effective action. The relation to FDT is apparent since, performing the transformation
in the path integral defining the response function one finds:
\begin{eqnarray}
&& R_{t_2-t_1} = \langle i \hat u_{t_1} u_{t_2} \rangle 
\\
&& = \langle ( i \hat u_{- t_1} - \dot u_{- t_1}) \dot u_{- t_2} \rangle
= R_{t_1-t_2} + \frac{\dot C_{t_1-t_2}}{T}
\end{eqnarray} 
i.e. the FDT relation for two point functions. The same is obtained
from considering the action of the symmetry (\ref{symfdt}) on a general form (i.e.
non-local in time)
for the quadratic part of the effective action functional.\cite{kpz}

We now discuss more precisely the conditions on the boundary
terms which relate this symmetry to the FDT. 
This is simplest to see first in the context of the
theory before averaging over disorder. Let us define
the Ito path integral:
\begin{eqnarray}
&& Z(u_f,t_f;u_i,t_i) = \int_{u(t_i)=u_i}^{u(t_f)=u_f} D \hat u D u e^{-S_V} \\
&& \int du_f Z(u_f,t_f;u_i,t_i) = 1 \label{norm}
\end{eqnarray}
the (normalized) conditional probability to find the system in state $u_f$ at time $t_f$
given that it is in state $u_i$ at time $t_i$. Here $S_V$ is the
MSR dynamical action in a given disorder realisation. By construction the
Boltzmann measure is the stationary distribution for this
$Z(u_f,t_f;u_i,t_i)$ regarded as an evolution operator:
\begin{eqnarray}
&&  \int du_f Z(u_f,t_f;u_i,t_i) e^{- (H(u_i) -H(u_f)) /T } = 1 \label{boltz}
\end{eqnarray}
Note that if under the transformation above $S_V \to S_V + \delta S_V$
where $\delta S_V=\delta S_V[u_i,u_f]$ 
is a function only of $u_i$ and $u_f$ (boundary term)
one has:
\begin{eqnarray}
&& Z(u_f,t_f;u_i,t_i) = Z(u_i,t_f;u_f,t_i) e^{- \delta S_V} , \label{sym}
\end{eqnarray}
since $t$ is changed to $-t$ and thus boundary times 
$t_i$ and $t_f$ must be interchanged. Interchanging $u_i$
and $u_f$ in (\ref{boltz}) and using ( \ref{sym})
the normalization condition (\ref{norm}) is found to 
hold only if:
\begin{eqnarray}
&& \delta S_V[u_i,u_f] = \frac{1}{T} ( H(u_{t_f}) - H(u_{t_i}) )
\end{eqnarray} 
If on the other hand $\delta S_V$ depends also on the
time derivatives at the boundary, then the FDT may
or may not be satisfied depending on the precise
nature of the boundary terms.\cite{randommass} 

Upon averaging over disorder $\overline{ e^{S_V + \delta S_V}}^V =
e^{S + \delta S}$, the shift $\delta S$ obtained after transformation
on the disorder averaged MSR action. It is readily seen that for the
bare action, $\delta S$ is a sum of a single time integral cross
correlation boundary term (and one response field) and a term with no
time integral representing the second cumulant of the random
portential $V(u)$. More generally if one performs the transformation
(\ref{lambda}) on the coarse grained model, one must obtain a $\delta
S$ which is a sum of boundary terms, each containing less response
fields than time integrals.  The $n$-th cumulant of the renormalized
static disorder can then be retrieved from the corresponding boundary
term with no time integral and $1/T^n$ factor.

We are now prepared to discuss one how can construct the
effective action at finite temperature taking into account
the constraints from the FDT. It is useful to note that from the
fundamental fields $i \hat u$ and $u$ two linear combinations
transform simply under the symmetry (\ref{symfdt})
\begin{eqnarray}
\dot u & \to & - \dot u \\
 Y = 2 T i \hat u - \dot u & \to & 2 T i \hat u - \dot u
\end{eqnarray} 
Terms in the effective action which are exactly
invariant (i.e. whose variation do not produce any
boundary terms) must involve $\hat u$ only in the combination 
$2 T i \hat u - \dot u$. Examples will be constructed below.

It is clear from these considerations that non-zero $T$ 
terms can have time derivatives replaced by $T \hat u$. Therefore
it is natural to group terms which formerly (at $T=0$) were organized 
by the frequency power $m$ by the more general index:
\begin{eqnarray}
M = N_{\hat u} - n + m,
\end{eqnarray} 
where $N_{\hat u}$ is the number of $\hat u$ fields appearing in the
term and $n$ the number of independent times, as before. 
Terms with a given $M,n$ can mix under the
FDT transformation (\ref{symfdt}).

\begin{figure}
\centerline{\fig{6cm}{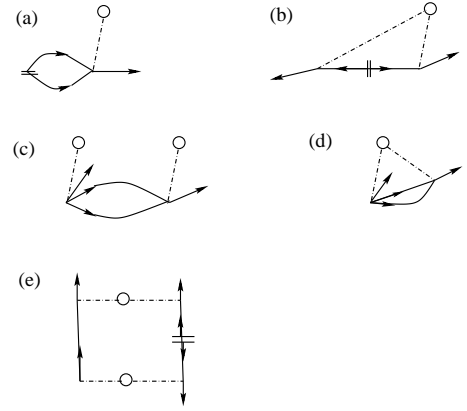}}
\caption{One loop diagrams which correct the effective action at $T>0$
  (in addition to the one for $T=0$): the internal lines contain the
  full response function and the graphs are 1P irreducible. (a -b) are
  the tadpoles using the full correlation function (the only possible
  ones as (d) should not be counted as it is a two loop diagram). (c)
  is the generic new one loop diagram at $T>0$ with two vertices (e)
  is one example of expanding the full correlation \label{graph26}}
\end{figure}

Let us start with $M=1$ and $n=2$. The only possible combination of the above invariants
is,
using symbolic notations as above:
\begin{eqnarray}
&& S^{(2)}_{T,1} =  \frac{( 2 T i \hat u_1 - \dot u_1)^2 - \dot u_1^2 )}{8 T} 
i \hat u_2 G(u_1-u_2) + (1 \leftrightarrow 2) 
\nonumber \\
&& = - \frac{1}{2} i \hat u_1 i \hat u_2 
[ (\dot u_1 - \dot u_2) - T (i \hat u_1 - i \hat u_2) ] G(u_1-u_2) \nonumber \\ &&
\end{eqnarray} 
recovering the zero temperature $G$ term together with a non-zero $T$
partner term. Note that at time $t_1$ we have used the invariant
combination $Y^2$, taking care to remove the piece proportional to
$\dot u_1^2$ since there must be at least one response field at every
time. This is not possible whenever there is an odd total number of
fields $\dot u$ and $\hat u$ at any particular time. Thus the $i \hat
u_2$ above cannot be embedded in a term exactly invariant.  Hence
$G(u)$ must be a total derivative and the above term gives a
non-vanishing boundary variation under the transformation
(\ref{symfdt}). This term can also be understood before disorder
averaging. It corresponds to replacing the naive zero T dynamical term
$\eta(u,r) i \hat u \dot u$ corresponding to the damping in (\ref{eq:reneom}) by the invariant combination:
\begin{eqnarray}
&& \int_{rt} \eta(u_{rt},r) \frac{(\dot u_{rt}^2 - ( 2 T i \hat u_{rt} -\dot u_{rt})^2) }{4 T} 
\end{eqnarray} 
Note that expanding out the factor in this term demonstrates that including $u$
dependence in the damping coefficient has given rise to a white thermal noise which
(for $D(u,r) = g(u,r) =0$ ) has $u$-dependent variance:
\begin{equation}
 \langle \zeta(r,t) \zeta(r',t') \rangle =
2 \eta(u,r) T \delta(t-t') \delta(r-r')
\end{equation} 
The fact that $G(u)$ is a total derivative then follows simply from
its interpretation as a cross cumulant of $\eta(u,r)$ and the
conservative random pinning force $f(u,r)$. Based on this reasoning it
is clear that the function $B(u)$ and $C(u)$ must also be total
derivatives.

Note that the finite $T$ partner of the $G$ term is
generated in parallel to it
from graphs of the form (e) in Fig.~\ref{graph26}. One easily
checks that it corrects the temperature term by
$\delta \eta T \hat{u} \hat{u}$ where $\delta \eta$ corresponds
to the correction (\ref{deta}) from $G'(0)$ so as to maintain FDT relation.

Let us now examine the terms $M=2$ and $n=2$. One can write the possible terms in a way
which makes apparent the invariants:
\begin{widetext}
\begin{eqnarray}
&& S^{(2)}_{T,2} = - \frac{1}{2} \frac{(\dot u_1^2 -( 2 T i \hat u_1 - \dot u_1)^2)}{4 T} 
\frac{(\dot u_2^2 -( 2 T i \hat u_2 - \dot u_2)^2)}{4 T} 
A(u_1-u_2)  \\
&&
+ \frac{( 2 T i \hat u_1 - \dot u_1) \dot u_1^2 
- ( 2 T i \hat u_1 - \dot u_1)^3}{8 T} 
i \hat u_2 B_1(u_1-u_2) +
(1 \leftrightarrow 2)  \\
&& - 
\frac{( 2 T i \hat u_1 - \dot u_1) \dot u_1^2}{4 T} i \hat u_2 \frac{C'(u_1-u_2)}{2} 
- \frac{( 2 T i \hat u_1 - \dot u_1) \ddot u_1}{4 T} i \hat u_2 C(u_1-u_2) +
(1 \leftrightarrow 2) 
\end{eqnarray} 
\end{widetext}
in a way such that unwanted terms (with no $\hat u$ field associated to
a time) cancel explicitly, apart from the last terms where they combine to
gives rise to a boundary term $\frac{1}{2} \partial_{t_1} (\dot u_1^2 C(u_1-u_2))$.
Accordingly, the $B$ has been splitted into $B(u)=B_1(u)+C'(u)/2$. The function 
$B_1(u)$ must be a total derivative (see above) and its variation
yields a boundary term. However, the invariance of the 
part cubic in the field in the $B_1$ term is exact, which can
be traced to an exactly invariant term in the unaveraged 
dynamical action:
\begin{eqnarray}
&& 
\frac{( 2 T i \hat u_{rt}- \dot u_{rt}) \dot u_{rt}^2 
- ( 2 T i \hat u_{rt} - \dot u_{rt})^3}{4 T} g(u_{rt},r) 
\end{eqnarray} 

Expanding all terms above one can write explicitly:
\begin{widetext}
\begin{eqnarray}
&& S^{(2)}_{T,2} = S^{(2)}_{T=0,2} 
+ 
\frac{T}{2} ((i \hat u_1)^2 i \hat u_2 \dot u_2 + (i \hat u_2)^2 i \hat u_1 \dot u_1) 
A(u_1-u_2) - \frac{T^2}{2} (i \hat u_1)^2 (i \hat u_2)^2 A(u_1-u_2) \\
&& + \frac{3}{2} T 
((i \hat u_1)^2 i \hat u_2 \dot u_1 + (i \hat u_2)^2 i \hat u_1 \dot u_2)
B_1(u_1-u_2) - T^2 ((i \hat u_1)^3 i \hat u_2 + i \hat u_1 (i \hat u_2)^3)
B_1(u_1-u_2)
\end{eqnarray} 
\end{widetext}
Note that $C$ does not give any bulk contribution at non-zero temperature. The
FDT constraint only requires some boundary noise term for $C$. This is 
because $C$ alone, with $A=B_1=0$ corresponds to a 
conservative dynamics.\cite{randommass} 

These considerations show that to the order studied the 
$T>0$ dynamical action is fully specified by the $T=0$
action and the FDT constraints. Thus we do not need to 
introduce at this stage any new operator associated to
finite $T$.

Having established that we are working with an appropriate action (and
hence have not neglected any pertinent coupling constants/functions),
we turn briefly to the effects of additional thermal contractions
upon the FRG equations.  Up to this point, the only such contractions we
have included are the ``diffusion'' terms ($\tilde{T}_l
\tilde\Delta''$ etc.) in the FRG equations for each coupling function.
It appears natural to neglect most effects of temperature since it is
an irrelevant variable under the FRG, while clearly these diffusion
terms are crucial, since they are necessary to stabilize the boundary
layer.  Within this treatment, the zero temperature rule of
conservation of powers of frequency still holds.  More generally,
however, one can a priori perform thermal contractions that feed
downward (i.e. reduce the number of time derivatives) in the frequency
hierarchy, in particular by thermally contracting fields containing
time derivatives.  For some simple such contractions, a preliminary
calculation shows that a cancellation in fact occurs amongst the
different ``partners'' required by the FDT, eliminating the unwanted
mixing.  We do not have, however, a general argument for such a
mechanism of cancellation.  Due to the complications of such a more
general analysis, we will however proceed assuming this is generally
true.  We comment briefly further on this in the conclusion.

\subsection{dynamical boundary layer analysis: terms associated with averaged
  relaxation time}

Having established that to this order no new terms arise due to
temperature we now attempt to study the structure of the thermal
boundary layer in the operators studied so far.

We consider the level $m=1$ in some detail. We add the effect
of temperature to lowest naive order which is to add 
to the right hand side of (\ref{dG}) the term
$\tilde T \tilde G''(u)$, originating from the
simple tadpole contraction of the $G$ vertex.

From examination of this equation we expect that 
$\tilde G(u)$ supports a thermal boundary layer 
form for $u \sim \tilde T_l/\epsilon$ 
\begin{eqnarray}
\tilde G(u) = \epsilon \tilde \chi g( \frac{\epsilon \tilde \chi u}{\tilde{T}_l} )
\end{eqnarray}
with $g(0) = 0$, $g(x)$ an analytic function at $x=0$, 
odd and positive for $x>0$. It should match the 
fixed point form
outside the boundary layer. For $u \sim O(1)$ and $\tilde T_l \ll \epsilon$ we expect
$\tilde \Delta(u), \tilde G(u) \ll \tilde \Delta''(0) \sim
\tilde G'(0) \sim \epsilon^2 \tilde \chi^2/\tilde T_l$. Thus in the
outer region only three out of the several terms involving 
$\tilde \Delta$ and $ \tilde G$ are non-negligible. For now we neglect
feeding from third cumulants functions $S$ and $H$, we return to
them below. The fixed point for $\overline{G}(u)$ outside the TBL is then trivially:
\begin{eqnarray}
\overline{G}^*(u) =  \left(  \frac{2 \tilde \Delta''(0)}{\tilde G'(0)} - 1 \right)
  \tilde{\Delta}'^*(u) 
\end{eqnarray}
Since at small argument $\tilde{\Delta}'^*(u) = - \epsilon \tilde \chi$ it
follows that $g(x) \to g_{+\infty}=1 - \frac{2 \tilde \Delta''(0)}{\tilde G'(0)}$
a constant, for large $x$. Using the TBL form to evaluate $\tilde G'(0)$
yields:
\begin{eqnarray}
g_{+\infty} = 1 + \frac{2}{g'(0)}
\end{eqnarray}

To analyze the boundary layer equation, we use the form 
(\ref{tblf}) for $\tilde{\Delta}$ and similar forms
for the third cumulant functions (\ref{bl}) for $\tilde S$
and:
\begin{eqnarray}
&& \tilde H(u_1,u_2,u_3) = (\epsilon \tilde \chi)^2
h(\tilde u_1,\tilde u_2,\tilde u_3) \\
&& \tilde u = \frac{\epsilon \tilde \chi u}{\tilde{T}_l} 
\end{eqnarray}
The TBL equation for $g(x)$ is then 
found to be:
\begin{widetext}
\begin{eqnarray}
&& 0 = 2 f'' g + 3 g' f' + g'(0) (f' - g) + g'' (1 + f)
+  2 f' (f''(0) +  f'') \\
&& +  s_{110}(0,0,\tilde u)  
+ \frac{1}{3} (h_{010}(\tilde u,0,0) - 2 h_{001}(0,\tilde u,0) - h_{100}(0,\tilde u,0) ) ,
\label{eq:BLdyn}
\end{eqnarray}
\end{widetext}
where $s(u_1,u_2,u_3)=s_d^{(3)}(u_1,u_2,u_3)$, 
all these terms being multiplied by $\epsilon^3 \chi/\tilde{T}_l$
while the terms originating from rescaling are proportional to
$\epsilon$. This form will thus hold at scales such that
$\tilde{T}_l \ll \epsilon^2$. 

For given functions $f,s,h$ this equation is an eigenvalue problem
for determining $g'(0)$. This can be seen since for large $x$ the
linear problem has one exponentially growing solution, in addition to
the one matching the outer solution which converges to a constant.
Thus $g'(0)$ must be tuned to select that solution. We illustrate this
behaviour in the approximation of neglecting all third cumulant
functions. Then we recall that $f$ satisfies:
\begin{eqnarray}
f'^2 + f'' (1 + f) = 1 . \label{eq:2cumbl}
\end{eqnarray}
whose solution is $f(x) = \sqrt{1 + x^2 } - 1$. Since,
as discussed earlier $G$ is a total derivative, it
is possible to integrate the boundary layer equation 
once, and defining $g=-f' + \gamma'$ one obtains:
\begin{eqnarray}
(1+f) \gamma'' + 2 f' \gamma' - g'(0) \gamma + (2 f - 1) (1 + g'(0)) = 0
\nonumber
\end{eqnarray}
with $\gamma(0)=\gamma'(0)=0$. One interesting solution but unrelated to the
physics of interest here is $\gamma=0$, $g'(0) = -1$, i.e. $g=-f'$. 
It corresponds to an integrable set of models with a single
exponential relaxation, which exactly obey
the full FRG equations to one loop, and is studied in Appendix.
A shooting procedure gives a solution $g(x)$ satisfying the
proper boundary conditions with $g'(0)=2.646 \pm 0.001$.

Thus we find that the growth of $\overline{\eta_l}$ 
is determined by:
\begin{eqnarray}
\partial_l \overline{\eta_l} = \frac{\epsilon^2 \tilde \chi^2}{\tilde T_l}
(f''(0) + g'(0)) \overline{\eta_l}
\end{eqnarray}
yielding:
\begin{eqnarray} 
\overline{\eta_l} \sim \exp( \alpha(1) \frac{\tilde \beta e^{\theta l}}{\theta} )
\end{eqnarray}
Clearly $\alpha(1)=3.646$ is a non-trivial number.

\subsection{terms associated with second moment of
relaxation time}
\label{sec:terms-assoc-with}

We now turn to the consideration of terms with $m=2$. As emphasized in 
Section II the principal quantities of interest are the
cumulants of the friction, the second one being embodied in $A(u)$.
The quantities $B_1$ and $C$ also appear at this order, complicating the
analysis. Since these embody somewhat different physics we will 
focus initially on $A(u)$ which fortunately satisfies equation (\ref{frgA}) which
is independent of $B_1$ and $C$.

We add the effect of temperature to lowest naive order which is to add 
to the right hand side of (\ref{frgA},\ref{frgB},\ref{frgC}) the terms
$\tilde T \tilde A''(u)$, $\tilde T \tilde B_1''(u)$, $\tilde T \tilde C''(u)$,
respectively. These originate from the simple tadpole contractions.

\subsubsection{second moment of relaxation time: $A(u)$}

In the previous Section III we pointed out the rapid divergence of
the moments of the friction (relaxation time) 
$\overline{\eta}, \eta^{(2)}=A(0), \cdots$ driven by the
low temperature divergence of $\tilde \Delta''(0)$.
In doing so we neglected all functional dependence 
(such as $A(u)$). In the previous subsection
we reconsidered the growth of the average friction $\overline{\eta}$,
which clearly does not itself has any functional dependence. 
Instead the deviations of its growth from the
prediction of Section III arise from a
secondary mechanism of the 
feedback of $\tilde G'(0)$ into $\overline{\eta}$.
Physically it corresponds to the cross correlation $G$
of the friction $\eta(u,r)$ with the random force $f(u,r)$ producing
an increased growth of $\overline{\eta}$. 

We would now like to reconsider the growth of the second
moment $A(0)=\eta^{(2)}$ including functional dependence. 
In this case already the leading effect of enhancement due to
the divergence of $\tilde \Delta''(0)$ is non-trivial. Thus we will
focus on it here primarily ignoring secondary effects of cross
correlations between the random friction coefficient and the
random force. In general these cross
correlation effects enter the flow of $A$ through 
$G$, $H$ and $W$ defined in (\ref{eq:cals123}). Terms 
involving $H$ and $W$ have already been dropped 
in  (\ref{frgA}) for $A$. We will initially
keep terms involving $G$ in (\ref{frgA})
but will drop them at a later stage of the analysis.
It is not clear at this stage whether keeping these 
terms without simultaneously including the
ones due to $H$ and $W$ would be consistent.

We then note that $\tilde A(u)$ satisfies a closed equation once
$\tilde G(u)$ is known. We first consider the nature of its solution
for $u \sim 1$ outside the TBL. Doing so one notes the presence of
several large terms proportional to $\tilde \Delta''(0)$,
$\tilde G'(0)$. Balancing these large terms, 
we obtain the solution outside the TBL:
\begin{equation} 
A_l(u)  \sim \overline{\eta}_l^2 
g_{\infty} \tilde G^{*\prime} = - \overline{\eta}_l^2 g^2_{\infty} \tilde  \Delta^{*\prime\prime}
\end{equation}
where $g_{\infty}$ was defined above. Note that, as was the case for $\tilde G$
the convergence is very rapid due to the homogeneous part $\partial A=-2 \tilde G'(0) A$.
The important feature of
this result is that $A(u)$ is of order $\overline{\eta}_l^2$ outside the
TBL.

We are going to search for a TBL solution for $A$ which grows faster:
\begin{eqnarray}
&& \tilde A(u) = \overline{\eta}_l^\lambda \frac{\epsilon^2 
\tilde \chi^2}{T} h(\frac{\epsilon \tilde \chi u}{\tilde T_l})
\end{eqnarray}
with $\lambda > 2$. In order to match the above solution 
outside the TBL one should have $h(\infty)=0$.
The TBL equation for $h$ then reads:
\begin{eqnarray}
&& ( - \lambda (f''(0) + g'(0)) + 2 f''(0) + 4 f'') h \nonumber \\
&& + (1 + f ) h'' + 4 f' h'  = 0 \label{tblh}
\end{eqnarray}
Due to the presumed faster growth of $\tilde A$ than $\overline{\eta}_l^2$
($\lambda > 2$) the feeding terms in (\ref{frgA}) are negligible and
have been dropped. As discussed above, since our principal interest is to compare the growth of
the second relaxation moment parameterized by $A(0)$ relative to the growth of the
mean $\overline{\eta}$, to be consistent we drop the analogous renormalization
of $\overline{\eta}$ by $\tilde G'(0)$, i.e. set $g'(0)=0$ in 
(\ref{tblh}). Numerical solution then yields:
\begin{eqnarray}
&& \lambda = 2.64..
\end{eqnarray}
The analysis is thus consistent since we find $\lambda >2$. To the order considered
we therefore have:
\begin{eqnarray}
&& \overline{\eta^2}_l \sim \overline{\eta}_l^{2.64..} \gg \overline{\eta}_l^{2} .
\end{eqnarray}
This gives, in the present approximation $\alpha(2)/\alpha(1) = 2.64$. As seen
in the previous subsection we expect both $\alpha(2)$ and $\alpha(1)$ to be both increased
by inclusion of the effect of cross-correlations of friction and
random force.

\subsubsection{growth of other $O(\omega^2)$ kinetic coefficients: $D, B, C$}

Due to the feedback of $A(0)$ into $D$ we expect $D$ to grow at least
as fast as $\overline{\eta}_l^{\lambda}$.  In the simplest scenario,
indeed, all $\omega^2$ quantities would scale the same in the same
manner.  However, we see no general reason why this need be the case.
Indeed, examination of $B_1$ and $C$ using the same truncation scheme
as for $A$, shows that they grow {\sl faster}.  We sketch this
analysis here.  Consider first
$\tilde{C}(u)$, which also feeds into the 
inertial mass $D$.  We assume $\tilde{C} \sim \overline{\eta}^\mu$, with
$\mu>\lambda>2$.  With such growth of $\tilde C$, the feedback of
$\tilde{C}'(0)$ into $\tilde{D}$ will overwhelm all other feeding
terms, and we expect $\tilde{D}=\overline{\eta}^\mu
\overline{D}$, with $\overline{D}$ scale independent.  It is then
natural to define $\tilde{C}(u)=- \tilde{D}
\overline{C}(u)$.  (\ref{dD}) becomes
\begin{eqnarray}
  \label{eq:overlineD}
  \partial_l \overline{D} =
  (-2+\mu(\tilde{G}'(0)-\tilde\Delta''(0))-\overline{C}'(0))\overline{D}. 
\end{eqnarray}
Thus to leading order in $1/\tilde{T}_l$, one needs
\begin{eqnarray}
  \label{eq:ccond}
  \overline{C}'(0)=\mu(\tilde{G}'(0)-\tilde\Delta''(0)) .
\end{eqnarray}
Using the above forms, $\overline{C}$ satisfies
\begin{widetext}
\begin{eqnarray}
   \partial \overline{C} & = & \big (2 - d  + \zeta \partial_u  -
   \tilde\Delta''(0) + \mu(\tilde\Delta''(0)-\tilde{G}'(0)) -   2
   \tilde \Delta'' 
   \big) \overline{C} +\tilde{T}_l \overline{C}''\nonumber
\\
&& - \tilde \Delta' \overline{C}'(0) -
   3 \tilde \Delta' \overline{C}'  +
 (\tilde  \Delta(0) - \tilde \Delta)\,\overline{C}''  
 - 2\,\tilde \Delta'\,\tilde \Delta''(0) 
 - 2 \tilde \Delta'\,\tilde \Delta''
\end{eqnarray}
\end{widetext}
to leading order, i.e. dropping terms $\sim
\overline{\eta}^{2-\mu},\overline{\eta}^{\lambda-\mu}$, and neglecting
feedback from higher cumulants as before.  As for $G$
and $A$, the outer solution for $u\sim O(1)$ is readily found equating
the large terms $\sim \tilde\Delta''(0) + \overline{C}'(0) \sim
1/\tilde{T}_l$:
\begin{eqnarray}
  \label{eq:couter}
  \overline{C} & \sim &  
    \frac{\tilde{C}'(0)+2\tilde\Delta''(0)}{(\mu-1)\tilde\Delta''(0)
-\mu\tilde{G}'(0)} \tilde\Delta'(u) \quad u\sim O(1).  \nonumber \\
&&
\end{eqnarray}
As before, for small $u$ we make a TBL ansatz,
\begin{eqnarray}
  \label{eq:cbl}
  && \overline{C}(u)  = \epsilon \tilde\chi c(\epsilon\tilde\chi
  u/\tilde{T}_l),
\end{eqnarray}
which yields an equation very similar to (\ref{eq:BLdyn}) for
$g(x)$:
\begin{widetext}
\begin{eqnarray}
  \label{eq:BLeqc}
 (1+f)c''+3f'c'  +(f''(0)-\mu (f''(0)+g'(0))+2f'' )c
 -f'(2f''(0)-c'(0)+2f'')=0.  
\end{eqnarray}  
\end{widetext}
We require, to match (\ref{eq:couter}), that $c$ goes at a
constant at large argument, and $c(0)=0$ since $c$ is an odd function.
Furthermore, from (\ref{eq:ccond}), we have
$c'(0)=-\mu(f''(0)+g'(0))$.  This formulates an eigenvalue problem for
$\mu$.  As above, to solve, we use the (approximate) form for $f(x)$
in (\ref{eq:2cumbl}) and, for consistency as before set $g'(0)=0$.
A shooting procedure gives $\mu=3.377$, indeed greater than $\lambda$
as required for consistency. In summary we find the growth of the kinetic coefficients:
\begin{eqnarray}
  \tilde{C}(u) \sim \tilde D \sim \overline{\eta}_l^{3.377}
\end{eqnarray}

Finally, we discuss the growth of $\tilde{B}_1$.  Since it is fed by
$\tilde{A}(0)$, it must grow at least as fast as
$\overline{\eta}^\lambda$, so all other feeding terms on the last line
of (\ref{frgB}) are certainly negligible.  Remarkably, even in the
presence of the thermal $\tilde{T}_l \tilde{B}''_1$ term an
asymptotically (for large $l$) exact solution can be found.  In
particular, one finds that the homogeneous (in $\tilde{B}_1$) part of
the $\tilde{B}_1$ equation has an exact eigenfunction which is just
$\tilde{B}_1(u,l)=\tilde{B}_1(l)$, a constant independent of $u$.
This turns out to be the most unstable eigenfunction, with eigenvalue
$-d -3 \tilde\Delta''(0)$.  The exponential growth of this unstable
eigenfunction is faster than that of $\tilde{A}(0)$, and hence
dominates the flow at large scales.  Hence, writing this relative in
terms of $\overline\eta_l$ (neglecting the $g'(0)$
renormalization of $\overline\eta$ as before), one finds
\begin{eqnarray}
  \label{eq:B1exact}
  \tilde{B}_1(u) = \tilde{B}_1(0) \overline{\eta}_l^3.
\end{eqnarray}

\section{conclusion}
\label{sec:conclusion}

We have through a series of successively better approximations arrived
at a description of the growth of the moments of relaxation times
(friction coefficient) encoded as eigenvalues of
functional FRG equations.  This final stage of analysis was carried
out only for the mean and variance -- the extension to higher
moments is a formidable technical challenge.  Nevertheless,
already at this level we have observed how these functional eigenvalue
problems provide a mechanism for describing a broad but non-trivial
(i.e. not log-normal) distribution of time scales. This is at
variance with numerous other existing examples of systems
exhibiting simpler log-normal tails which can be obtained
from simpler non-linear sigma model diagrammatic calculations,
such as in disordered conductors.\cite{russians}
A similar log-normal tail was indeed obtained in Section III
from an approximate truncation of the FRG equation.
A rather strong physical difference from the aforementioned
quantum diffusion problem is the rapid exponential scale dependence
of the relaxation times for $\theta >0$, very different from
the logarithmic dependence of two dimensional weak localization
corrections. It is an open question whether some less trivial
distribution might arise at the metal-insulator transition
in $d>2$ and whether similar functional renormalization 
ideas might be useful in this context.

Many outstanding issues and extensions remain.  Of these, the
most fundamental are germane to both the dynamics and the
statics.\cite{us_short,staticslong} In particular the very basic
problem of perturbative control of the theory (most interestingly
in the $\epsilon$-expansion) remains unsolved.  This question, and the
associated matching problem of relating e.g. random force quantities
like the $f_k$ in (\ref{bl}) defined deep within the boundary layer at
$u=0$ to the zero temperature ones occurring far outside for $u\sim
O(1)$ are better addressed in the simpler context of the statics.  
We will refrain from commenting further upon them here.

Of the problems specific to the dynamics, perhaps most important is a
systematic treatment of all thermal terms in the FRG.  We have begun
this program by classifying all operators associated with ``thermal
noises'' in the effective action (Sec.~\ref{sec:dynamical-action-at})
consistent with the FDT.  However, up to this point we have included
the effects of non-zero temperature only through the leading
``diffusion'' terms ($\tilde{T}_l \tilde\Delta''$ etc.) in the FRG
equations for each coupling function.  As discussed in
Sec.~\ref{sec:dynamical-action-at}, while this assumption is natural,
we do not at this stage have a general justification for it.  The
importance of additional thermal contractions thus remains an
important issue for further investigation.

Once these basic remaining issues in the FRG formulation are resolved,
the present methods offer the opportunity to explore numerous physical
problems.  Obviously, equilibrium response and correlation functions
are of considerable interest.  Perhaps the approximate techniques of
Section III (and Appendix~\ref{general}) may have an extension to the
full functional description.  It will also be valuable to
reinvestigate the response to a uniform applied force in the creep
regime,\cite{chauve_creep} in light of the full dynamical structure of
the thermal boundary layer exposed here.  Applications of these ideas
to non-equilibrium response and aging is also tantalizing.  Similar
approaches should be applicable to quantum problems in the Keldysh
formalism. These and other applications of the present formulation
certainly provide a broad scope for future progress in understanding
glassy dynamics.

\begin{acknowledgements}
  L.B. was supported by NSF grant DMR-9985255, and the Sloan and
  Packard foundations.  Both L.B. and P.L.D. were supported by the
  NSF-CNRS program through NSF grant INT-0089835, and CNRS Projet 10674.
\end{acknowledgements}

\appendix

\section{Single time scale calculations: Equilibrium}
\label{sec:onetimedetails}

\subsection{analytical results for the equilibrium response function}

It is interesting to observe how the two {\sl putative} scaling
regimes described in Sec.~\ref{sec:onetime}\ arise in a detailed
calculation of the mean response function.  To do so, we develop
an FRG scheme to calculate directly the response function at arbitrary
$\omega,k$ within the scaling regimes described above.  It is
necessary to follow the flow of the full wavevector and frequency
dependence of the ``kinetic'' part of the MSR action.  We therefore
generalize the form in (\ref{msr1}) to:
\begin{eqnarray}
S_0^{l}[u,\hat{u}] & = & \int_{r,r',tt'} ~ i \hat{u}_{rt}
(R^{-1}_l)_{rt,r't'}  u_{r't'}
 \nonumber \\
        && - \eta T \int_{r,t} (i \hat{u}_{rt}) (i \hat{u}_{rt}),
\label{freemsr}
\end{eqnarray}
where, in a slight abuse of notation, we have denoted the quadratic
MSR kernel by $R^{-1}$.  Using (\ref{freemsr}), we extend the FRG
analysis leading to (\ref{DeltaEq}) to derive an RG equation for
the response function.  As before, the strategy is to integrate out
spatial Fourier modes $\Lambda > k > \Lambda e^{-l}$, but now keeping
the explicit time dependence. At this stage, we will {\sl not} assume
time-translational invariance, though we will specialize to this at a
later stage.  The FRG equation for $R^{-1}_l$ is
\begin{eqnarray}
&& \partial_l R^{-1}_{q,l}(t,t') = - \Gamma_l \Lambda_l^4
( R_{\Lambda_l,l}(t,t') \! - \! \delta_{t,t'}\!\! \int_{t_i}^{t}
\!\! dt'' R_{\Lambda_l,l}(t,t'') ) \nonumber \\
&&  \label{Rflow1}
\end{eqnarray}
where $t_{i}$ is an initial time at which the system is prepared in some
as yet unspecified state (or distribution of states).
(\ref{Rflow1}) is obtained formally
by computing the correction to the (inverse) response function upon
integrating out the modes in the shell, and using definitions
(\ref{DeltaEq}) and (\ref{gammal}). We perform this integration
perturbatively in $\Delta$ (to first order), which gives the lowest
order term in $\epsilon$.

At the end we want the true response function
$R_q(t,t')$. It will be obtained by integrating the flow
from $l=0$ with the initial condition:

\begin{eqnarray}
R_{q,l=0}(t,t') = e^{- q^2 (t-t')} \theta(t-t') \label{init}
\end{eqnarray}
setting $\overline{\eta}_0=1$ for convenience,
up to the scale $l^*$ such that $\Lambda e^{-l^*} =q$.

\begin{eqnarray}
R_q(t,t') = R_{q,l= \ln(\Lambda/q)}(t,t')
\end{eqnarray}

This amounts to neglect contributions coming from the
modes $k<q$, as is usually done in the RG. These are
examined below.

Although the initial condition in (\ref{init}) is
time-translationally invariant (TTI), the solution of the RG equation
does not in general remain so, due to the presence of the initial time
$t_i$.  This leads to the aging properties to be discussed in the
Appendix~\ref{sec:onetimedetails2}. The TTI regime is recovered in the
limit $t_i \rightarrow \infty$ (for large but fixed finite size
system) where one can set $R_k(t,t') = R_k(t-t')$. Then
(\ref{Rflow1}) can be Fourier transformed in $t-t'$, $R_k(i
\omega)= \int_{t>0} R_k(t) e^{ - i \omega t}$ (i.e Laplace transformed
with $s=i \omega$) to obtain
\begin{eqnarray}
&& R^{-1}_{k}(i \omega) =  i \omega + k^2 + \Sigma_k(i \omega)
\label{sigdef}\\
&& \partial_k \Sigma_k(i \omega) = \tilde{\beta} k^{3-\theta} (
\frac{1}{i \omega + k^2 + \Sigma_k(i \omega)}
- \frac{1}{k^2 + \Sigma_k(0)} ) \nonumber \\
&& \label{sigtti}
\end{eqnarray}
where we have defined a ``self-energy'' $\Sigma_k(i\omega)$ with initial
condition $\Sigma_{k=1}(i \omega) =0$.  To obtain (\ref{sigtti}) one
writes
$\Sigma_k(i \omega) = \int_0^{\ln(\Lambda/k)} \partial_l R^{-1}_{k,l}(i
\omega)$,
uses the Fourier tranform of (\ref{Rflow1}) and differentiate w.r.t. $k$
(we set from now
on $\Lambda=1$). One can check that
consistently $\Sigma_k(0)=0$, as requested by the statistical tilt
symmetry, which we use from now on.
Apart special cases \cite{exactsolu} (\ref{sigtti}) does not admit
analytical
solution and we now analyze the various regimes of interest.

From (\ref{sigtti}) one first finds the small $\omega$ behaviour of $R^{-1}_{k}(i \omega)$ as:
\begin{eqnarray}
&& R^{-1}_{k}(i \omega) = k^2 + i \omega \eta_k + O(\omega^2)
\end{eqnarray}
where $\eta_k$ satisfies $\partial_k \eta_k = - \tilde{\beta} k^{-1
-\theta} \eta_k$
which yields $\eta_k = k^2 \tau_k$, i.e one recovers as expected the
single caracteristic time scale $\tau_k$ given by (\ref{tauk}).

To analyze further the higher order terms in $i \omega$ from
(\ref{sigtti}), we first consider the scaling regime $i \omega,k^2
\ll 1$
with $y=i \omega\tau_k $ fixed (which implies $i \omega \ll k^2$). 
Making
the scaling ansatz
\begin{equation}
\Sigma_k(i \omega) = k^2 g(y), \label{sf3}
\end{equation}
in (\ref{sigtti}) gives the closed differential equation $y g' =
g/(1+g)$, which has the implicit solution (taking into account the
behavior of $\Sigma$ for $i\omega \rightarrow 0$ from (\ref{sigtti}))
\begin{eqnarray}
g e^g = y. \label{sf3a}
\end{eqnarray}
(\ref{sf3}-\ref{sf3a}) correspond to the $Y$ scaling limit of
Sec.~\ref{sec:onetime}.

(\ref{sf3}) is valid for finite $y$. As
$y\rightarrow\infty$, we enter the logarithmic ($X$ scaling limit)
scaling regime, in which the scaling variable
$x=k(\ln(1/i\omega)/\tilde{\beta})^{1/\theta}$ is fixed and $i\omega,
k^2 \ll 1$.
Since $g(y) \rightarrow \infty$ in this limit, the first term on the
right hand side of (\ref{sigtti}) can be neglected, leading to the
ansatz
\begin{equation}
\Sigma_k(i \omega) = \tilde{\beta} k^{2-\theta} f(x). \label{sf4}
\end{equation}
with $(2-\theta) f + x f' = - 1$ from (\ref{sigtti})\,
which determines the form of the scaling function $f(x)$ as:
\begin{equation}
f(x) = {1 \over {2-\theta}}\left[ \left({x_c \over
x}\right)^{2-\theta} - 1 \right], \label{sig2}
\end{equation}
and the constant $x_c = (1/\theta)^{1/\theta}$ is determined by
matching to (\ref{sf3}).  This regimes exists only for:
\begin{equation}
k < x_c  (\ln(1/i\omega)/\tilde{\beta})^{- 1/\theta}
\end{equation}
and thus corresponds to the limit of small wavevectors at fixed
$\omega$,
or to relaxation times $\tau \ll \tau_k$ ($\tau \sim e^{\tilde{\beta}(
x/k )^{-\theta} }$ for
$x < x_c$). When $x \to x_c^{-}$ one crosses over to the ${\cal Y}$
regime.

We can check that these results merge smoothly with the result
directly obtained at the upper
critical dimension $d=4$. There the
equation for $\Sigma_k(i \omega)$ becomes:
\begin{eqnarray}
&& \partial_k \Sigma_k(i \omega) = \tilde{\beta} \frac{k}{(\ln(1/k))^2}
(\frac{1}{i \omega + k^2 + \Sigma_k(i \omega)}
- \frac{1}{k^2} ),\nonumber \\ &&
\label{sigtti2}
\end{eqnarray}
which yields the same two scaling regimes, the first one
with $\tau_k \sim  k^{-2} \exp(\tilde{\beta}/(2 k^2 (\ln(1/k))^2))$ and
the same
scaling function $g(y)$ (\ref{sf3a}) and the second one reading:
\begin{eqnarray}
\Sigma(k) = \frac{\tilde{\beta}}{\ln(1/k)} ( -1 + 2
\frac{\ln(1/k)}{\ln(\ln(1/i \omega)/\tilde{\beta})})
\end{eqnarray}

We now turn to the calculation of the response function in the time
domain.  Consider first the regime ${\cal Y}=t/\tau_k$ fixed and
$t\rightarrow\infty$, $k \ll 1$.  Inverse Laplace-Fourier transforming
(\ref{sigdef}) and using (\ref{sf1}) gives the scaling
form
\begin{equation}
R_k(t) = {1 \over {k^2\tau_k}} {\cal G}({\cal Y}), \label{sf5}
\end{equation}
with
\begin{equation}
{\cal G}({\cal Y}) = \frac{1}{2 \pi i} \int_{- i \infty + \gamma}^{i
\infty + \gamma}
\frac{e^{y {\cal Y}}}{1 + g(y)} .
\label{integral}
\end{equation}
While we have not performed a complete analysis of the integral in
(\ref{integral}), the large time behavior can be extracted
.\cite{footnoteint}  For
${\cal Y} \gg 1$, the integral is dominated by the vicinity of the
branch point on the real negative axis at $y=-1/e$, leading to
\begin{equation}
R_k(t) \sim {1 \over {k^2\tau_k}} \left({t \over \tau_k}\right)^{1/2}
\exp\left[ - {t \over {e\tau_k}} \right],\qquad t \gg \tau_k.
\label{asympt}
\end{equation}

In the logarithmic scaling regime, we cannot simply invert the Fourier
space result in (\ref{sf4}), as it does not extend over the
entire frequency interval.  Instead, we return to the defining RG
equation for $R^{-1}_k(t)$, (\ref{Rflow1}).  By inverting this
formal equation, and again integrating down to scale $k$, we obtain
an equation for $R_k(t)$ directly:
\begin{eqnarray}
&& \partial_k R_k(t) = - 2 k R_k *_t R_k \label{convolve} \\
&& - \tilde{\beta} k^{3-\theta} ( R_k *_t R_k *_t R_k - R_k *_t R_k
\int_{t'>0} R_k(t') )     \nonumber 
\end{eqnarray}
where $*_\tau$ denotes a convolution.  Note that, aside from the
momentum-dependence of the prefactors and the absence of derivative
terms, (\ref{convolve}) bears a formal similarity to the
mode-coupling equations of mean field models.  In order to match the
{\sl scaling} expected from the above logarithmic frequency regime, we
make the ansatz
\begin{eqnarray}
R_k(t) \sim \frac{1}{t (\ln t)^{2-2/\theta} \tilde{\beta}^{2/\theta}}
F[{\cal X} = \frac{1}{\tilde{\beta}} k^\theta \ln t]
\label{sf6}
\end{eqnarray}
with $F[0]$ a constant.  Inserting this in (\ref{convolve}), it is
permissible to drop the first two terms in the logarthmic scaling
regime, and moreover to approximate
$\int_0^t dt' R(t-t') R(t') \approx 2 R(t) \int_1^t dt' R(t')$.
This yields:
\begin{eqnarray}
\theta {\cal X} F'[{\cal X}]  = 2  F[{\cal X}] \int_0^{\cal X} dz
z^{-(2-2/\theta)} F[z].
\end{eqnarray}
The solution is
\begin{equation}
\int_{\ln ({\cal X}^{{{2-\theta} \over \theta}}F[{\cal X}])}^{+\infty}
\frac{d
h}{\sqrt{(2-\theta)^2 + 4 \theta e^h}}
= - \left({1 \over \theta}\right) \ln ({\cal X}/{\cal X}^*),
\label{ugly}
\end{equation}
where ${\cal X}^*=1/\theta$ is the boundary of the regime, at which
$F({\cal X})$ diverges,
signaling the onset of a regime of faster relaxation onto the regime
${\cal Y}$.

Having computed the response function $R_k(t)$ in the equilibrium TTI
regime, we also obtain the
time dependence of the connected correlation defined as
$C_k(t) = \overline{\langle u_k(t) u_{-k}(0) \rangle } -
\overline{\langle u_k(t) \rangle \langle u_{-k}(0) \rangle }$,
with $C_k(t \to +\infty)=0$ and $C_k(t=0)$ the equilibrium connected
correlation.
Indeed they are simply related through the fluctuation dissipation
relation
$\partial_t C_k(t) = - T R_k(t) $ or, in frequency space,
\begin{eqnarray}
C_k(i \omega) = \frac{-T}{i \omega} (R_k(i \omega) -
R_k(- i \omega))
\end{eqnarray}

\subsection{discussion}

We now pause and comment on the results of the FRG calculations
we have just obtained. Let us first mention the nice features before
stressing the unsatisfactory points below.

First we note that the Wilson scheme used directly on the response
function within the single time scale assumption
indeed yields, as we anticipated from general arguments in
Sec.~\ref{sec:onetime},
two distinct scaling regimes, the ${\cal Y}=t/\tau_k$ regime and the
${\cal X}=k^\theta \ln t$ regime, with scaling {\sl forms}
in (\ref{sf3},\ref{sf4},\ref{sf5},\ref{sf6}) . The scaling {\sl
functions} themselves
were found to be non-trivial, with interesting analytical structure.
While the existence of the ${\cal Y}=t/\tau_k$ regime seems to be a
straightforward
consequence of the assumption of a single time scale $\tau_k$, the
emergence of the
${\cal X} \sim k^\theta \ln t$ regime within this hypothesis
is less obvious. Within the Wilson scheme, it seems to
result from the system keeping a memory of a whole spectrum of
smaller relaxation times
$\tau \sim (\tau_k)^{x/x_c}$, $x<x_c$,
generated during the coarse graining procedure and naturally appears
here
(while one would naively expect the largest one only, $\tau_k$ to play a
role).
It does have the form of {\it activated dynamics} since the scaling
variable is truly ${\cal X}= (T/T^*) k^\theta \ln t $ and thus
corresponds to
crossing barriers of heigth $\sim k^{- \theta}$. That such an activated
regime
should exist is physically rather natural. Indeed we expect from simple
droplet arguments that the equilibrium dynamics of mode $k$ at large
time difference (in general
$t-t'$, denoted here $t$) is dominated
by the rare active configurations with (at least) two quasi-degenerate
low free energy states at scale $L \sim 1/k$ of the system. The
probability to find two nearly degenerate minima
(on the scale of the thermal energy $T$) at scale $1/k$ is $\sim T
k^\theta$.
One expects these two minima to be separated by a barrier $U_b$ also
scaling like
$U_b \sim k^{-\theta}$ and thus when $T \ln t > U_b$ {\it equilibrium}
thermally activated motion back and forth over this barrier
\cite{manyscales} becomes active and gives rise to time-dependent
correlations
on the scale $\ln t \sim k^{-\theta}$. Our analytical result thus
exhibits the correct scaling behaviour
and it is thus encouraging that such barrier crossing behaviour and
scaling comes out of the present RG calculation.

Upon a closer look to our results in the X regime, everything works as
if there is an effective distribution
of smaller barriers $U_b = x' k^{-\theta}$ with a distribution of
relaxation times $\tau = e^{\tilde{\beta} U_b}$
for $0<x'<x_c$. The total weight of this distribution being only $\sim T
k^\theta$ it can be written
as $k^\theta/\tilde{\beta} \phi(x')$. $\phi(x')$ diverges at $x'=x_c$.
This is easily seen, e.g. on the form
for the correlations. Indeed, using the above FDT relation one obtains:
\begin{eqnarray}
C_k(t_1) - C_k(t_2) = \frac{T}{k^2} \frac{k^\theta}{\tilde{\beta}}
\int_{k^\theta \ln t_1/\tilde{\beta}}^{k^\theta \ln t_2/\tilde{\beta}}
\frac{du}{u^{2 - \frac{2}{\theta}}} F(u)
\label{droplets}\
\end{eqnarray}
for the correlations in the logarithmic regime. In this expression, the
$T/k^2$ equilibrium
correlation is usually explained as $T/k^2 = (1/k^{d + 2 \zeta}) (T
k^\theta)$, i.e the
product of the size of a positional fluctuation between two degenerate
states at scale
$k$ and the probability of this active configuration to occur. Thus we
see that there is
here an additional reduction by an extra factor $T k^\theta$, the total
weight of
barriers much smaller than $\tau_k$ (note that the above correlation
variations within
regime X are subdominant compared to the ones in regime Y, which really
account for all
but a small fraction of the total variation). Similarly one sees that
the
response corresponding to a barrier $x' k^{-\theta}$ can be written as:
\begin{eqnarray}
\frac{1}{k^2} \frac{1}{1 + i \omega e^{\tilde{\beta} x' k^{-\theta}} }
= \frac{1}{k^2} \frac{1}{1 + e^{\tilde{\beta}  k^{-\theta} (x' - x)}}
\to \frac{1}{k^2} \theta(x-x')
\end{eqnarray}
as $k \to 0$ with $x=k^\theta \ln(1/i \omega)/\tilde{\beta}$.
Thus averaging with the weight $\frac{1}{\tilde{\beta}} k^\theta
\phi(x') dx'$ yields
exactly our result $R_k(i \omega)$ in regime X if one chooses:
\begin{eqnarray}
\int_0^{x} \phi(x') dx' = \frac{2-\theta}{( \frac{x_c}{x} )^{2-\theta} -
1 }
\end{eqnarray}

Another puzzling feature of the above results is the nonmonotonicity
of the above scaling functions. As discussed below this is directly
related to the assumption of a single time scale, and has prompted us
to reconsider the whole calculation (at a high price of technical
difficulty) in Section~\ref{sec:fterm}. We see from (\ref{sigtti}) that
$\Sigma_k(i\omega)$ is a decreasing function of $k$ always and that
$R_k(i \omega)$ at fixed $i \omega$ is an increasing function of $k$
for $k$ small enough. Correspondingly, the real-time solution $R_k(t)$
is an {\sl increasing} function of $k$ at fixed $t$ throughout the
logarithmic regime and also in the short-time portion of the $t \sim
\tau_k$ regime.  Similarly, (\ref{droplets}) implies that the
correlations are also increasing functions of $k$ at fixed $t$.  While
this behavior is unexpected, we are presently unsure whether it is in
fact unphysical. What is clear is that it is a consequence of the
single time scale assumption. Indeed, the rather simple and apparently
physical expression $R_k(t) = e^{-t/\tau_k}/(k^2 \tau_k)$ is also
increasing with $k$ for small $t/\tau_k$.  It appears that one can
argue fairly generally that, provided there exists a long-time regime
with a {\sl well-defined} $\tau_k$, the response must be increasing
with $k$ for $t/\tau_k \lesssim 1$.

As discussed at length in the text, one does expect that the single
time scale description is unsufficient and one should instead consider
a distribution of time scales. Let us examine the question of
monotonicity
when $R_k(t)$  is simply a superposition of elementary relaxation
processes.
Discarding the subdominant $k^{-2}$
prefactor and writing $\tau_k = e^{U_L}$ where $L=1/k$ is the scale,
one can consider the average:
\begin{eqnarray}
R_k(t) = \int dU P_L[U] e^{- t e^{-U} - U}
\end{eqnarray}
It is dominated by the saddle point $U^*(L,t)$ solution of:
\begin{eqnarray}
t e^{-U} - 1 + \partial_U \ln  P_L[U] = 0
\end{eqnarray}
and the condition for $R_k(t)$ to be an increasing function of $L$ is
\begin{eqnarray}
\partial_L P_L[U^*] >0
\end{eqnarray}
For a gaussian $\ln  P_L[U] = - (U-L^\theta)^2/(2 s L^\alpha)$ and $t=0$
(the worse point) one finds
$U^*=L^\theta - s L^\alpha$ and $\ln P_L[U^*]  = - L^\theta +
\frac{s}{2} L^\alpha$.
Thus one needs $\alpha \geq \theta$. Note however that this supposes the
gaussian to hold down to $U^* < 0$, which may not be the case in
general. On the other hand the only real condition concerns the
monotonicity of the scaling function itself. Thus, one way to reduce the
effect of nonmonotonicity is to increase the width of the distribution
of time scales.

To close this discussion, it is useful to contrast the present situation
of an elastic system with fast growing barriers with what happens in the
marginal case
$\theta=0$. This is realized for a periodic model in $d=2$, e.g. for
the line of fixed points of the Cardy Ostlund
model. There of course one expect a single scaling regime compatible
with
simple matching arguments.  Setting $\theta=0$ in (\ref{sigtti}) one
finds the exact solution:
\begin{eqnarray}
&& k^2 = (1 + \frac{\Sigma_k(i \omega)}{i \omega} )^{-2/\tilde{\beta}}
(1 + i \omega \frac{2}{2 + \tilde{\beta}}) \\
&&
- \frac{2}{2 + \tilde{\beta}} (i \omega + \Sigma_k(i \omega))
\end{eqnarray}
obtained writing $- d(k^2)/d\Sigma = (2/\tilde{\beta}) (1 + k^2 (i
\omega + \Sigma)^{-1})$. For
$\omega \ll 1$ this yields the scaling form
\begin{eqnarray}
&& \Sigma_k(i \omega) = k^2 g(y = i \omega k^{-z}) ,\quad z=2 +
\tilde{\beta} \nonumber \\
&& y = g (1 + \frac{2}{2 + \beta} g )^{\beta/2}
\end{eqnarray}
where $z$ is the equilibrium dynamical exponent (note that for $\beta
\to +\infty$ one recovers
(\ref{sf3a}). The self energy nicely interpolates between $\Sigma_k(i
\omega) \sim i \omega k^{-\tilde{\beta}}$
at small $i \omega \ll k^z$ (as also obtained from considering the flow
of the uniform $\overline{\eta}_l \sim e^{\tilde{\beta} l}$)
and $\Sigma_k(i \omega) \sim (\frac{z}{2})^{\tilde{\beta}/z} (i
\omega)^{2/z}$ at large $i \omega \gg k^z$.
>From there one obtains the response function $R_k(t) = k^{\tilde{\beta}} {\cal G}(t k^{-z})$, which is found to decay as in
(\ref{asympt}) with  $\tau_k \sim k^{-z}$ and a characteristic time
$(1+2/\tilde{\beta})^{\tilde{\beta}/2} \tau_k$
(instead of $e \tau_k$ obtained for $\tilde{\beta} \to +\infty$, and
which behaves as:
\begin{eqnarray}
&& R_k(t) \sim t^{-\tilde{\beta}/z}
\end{eqnarray}
in the limit $1 \ll t \ll k^{-z}$.  The function ${\cal G}$ obeys the
equation:
\begin{eqnarray}
&& \tilde{\beta} {\cal G} + (2 + \tilde{\beta}) {\cal Y} {\cal G}' =
(-2 + \tilde{\beta})  {\cal G} *_{\cal Y} {\cal G} -
\tilde{\beta}   {\cal G} *_{\cal Y} {\cal G} *_{\cal Y} {\cal
  G}\nonumber \\ &&
\end{eqnarray}
Note that such as scaling function ${\cal G}$ of
${\cal Y}=t k^z$ leads to a trivial scaling regime in ${\cal X}=\ln
t/\ln (1/k)$ reduced to
a delta function at ${\cal X}= -z$. Note finally that even in this case,
the scaling regime $R_k(t)$ is again
nonmonotonous: it vanishes at $k=0$, increases up to $k^*$, with $t
(k^*)^z = z \tilde{\beta}$ and
decreases beyond.

\section{Single time scale calculations: Nonequilibrium and aging}
\label{sec:onetimedetails2}

\subsection{response function and various regimes}

The RG recursion relation for the response function derived above was
not
restricted to equilibrium, and it is thus interesting to write down the
corresponding
equations (for response and correlations) in the full non-equilibrium
regime.
Within the intrinsic limitations of the single time scale approach, this
allows
in principle to acces the aging properties of the system.

To obtain closed equations for the two time response function
$R_k(t,t')$
one again iterates (\ref{Rflow1}) from the same (TTI) initial condition
$R_{k,l=0}(t,t') = \theta(t-t') e^{- k^2(t-t')}$ up to $l=\ln(1/q)$,
keeping
$t_i$ finite and making no TTI assumption. Thus the response satisfies
the differential equation:
\begin{eqnarray}
&& R_k^{-1}(t,t') = \delta_{tt'} ( \partial_{t'} + k^2 ) +
\Sigma_k(t,t') \\
&& \partial_k \Sigma_k(t,t')  = \tilde{\beta} k^{3-\theta}
( R_k(t,t') - \delta_{t t'} \int_{t_i}^t dt'' R_k(t,t'') ) \nonumber
\end{eqnarray}
where matrix multiplication and inversion is with respect to (t,t'). It
is more convenient
to avoid the two time self energy and write a closed equation for
$R_k(t,t')$ as:
\begin{eqnarray}
&& \partial_k R_k(t,t')  = - 2 k (R_k \cdot R_k)(t,t')  \\
&& - \tilde{\beta} k^{3-\theta}  ( (R_k \cdot R_k \cdot R_k )(t,t') 
\nonumber \\
&& - \int_{t'}^{t} dt_1 R_k(t,t_1) R_k(t_1,t')  \int_{t_i}^{t_1}
dt'' R_k(t_1,t'') ) \label{raging}
\end{eqnarray}
with initial condition $R_{k=1}(t,t') = \theta(t-t') e^{- k^2(t-t')}$. 
The full analysis of this equation is
quite complicated and we have not attempted it. We will give only a few
features, at a naive level,
which remain to be confirmed by a more detailed analysis left for the
future.

The function $R_k(t,t')$ depends on three variables but in the limit $k
\ll 1$, $t'-t_i \gg 1$, $t - t' \gg 1$
we expect that it takes scaling forms depending only on two variables.
What these variables really are
depends on the time regime, and one can identify several possible time
regimes and subregimes. They can be
classified as follows, where we indicate the form expected for the
response function $R_k(t,t')$,
by order of increasing time and time differences:

\begin{eqnarray}
&& (I) \quad \frac{\ln t'}{\ln \tau_k} < 1 ,  \nonumber  \\
&& (I a) \quad \frac{\ln(t-t')}{\ln t'} < 1 : \quad \frac{g(k^\theta
\ln(t-t'),\frac{\ln(t-t'))}{\ln t'}}{(t-t') \ln^\gamma(t-t')} 
\nonumber  \\
&& (I b) \quad t-t' \sim t' : \quad ~~~~~~  \frac{h(k^\theta
\ln(t-t'),\frac{t-t'}{t'} )}{(t-t') \ln^\delta(t-t')} \nonumber \\
&& (I c) \quad \frac{\ln t}{\ln t'} > 1, \frac{\ln t}{\ln \tau_k}< 1 :
\quad \frac{f(k^\theta \ln t,\frac{\ln t}{\ln t'})}{t' \ln^\alpha t'}
\nonumber \\
&& (I d) \quad  t \sim \tau_k : \quad ~~~~~~~~~~~~~~  \frac{m(k^\theta
\ln t',\frac{t}{\tau_k})}{t' \ln^\psi t'} \nonumber \\
&& \\
&&  (II) \quad t' \sim \tau_k, \nonumber \\
&& (II a) \quad \frac{\ln(t-t')}{\ln \tau_k} < 1  : ~~~ \frac{F(k^\theta
\ln(t-t'),\frac{t'}{\tau_k})}{(t-t') \ln^b(t-t')} \nonumber \\
&& (II b) \quad t-t' \sim \tau_k  \quad  ~~~~~~~~~~ \frac{{\cal
G}(\frac{t-t'}{\tau_k},\frac{t-t'}{t'})}{k^2 \tau_k}  \nonumber \\
&& \\
&&  (III) \quad \frac{\ln t'}{\ln \tau_k} > 1 \quad \text{equilibrium}
\quad R_k(t-t') \nonumber \\
&& ~~~~ (III a) \quad \frac{\ln(t-t')}{\ln \tau_k }< 1 \quad ~~ 
\frac{F(k^\theta \ln(t-t'))}{(t-t') \ln^{2-2/\theta}(t-t')} \nonumber \\
&& ~~~~ (III b) \quad t-t' \sim \tau_k  \quad ~~~~~~~~~  \frac{{\cal
G}(\frac{t-t'}{\tau_k})}{k^2 \tau_k} \nonumber
\end{eqnarray}

Regime (III) is the equilibrium TTI regime, where the only dependence is
in $t-t'$. There are
two scaling forms possible corresponding to the two subregimes X (IIIa)
and Y (IIIb)
studied in Appendix~\ref{sec:onetimedetails}. Fully equilibrated
regime (III) is expected here
for very large times $t> t' \gg \tau_k$, and is somehow at variance with
mean field models
(where one always expect aging, e.g. for $t \sim t'$, even for very
large t').  In regimes (I) and (II) the mode k
at $t'$ has not yet equilibrated, and the scaling functions ar now also
function of
$k^\theta \ln t'$ (regime I) or $t'/\tau_k$ (regime II), in addition of
being functions
of $t-t'$. In both regimes (I) and (II) if $t-t'$ is small one expects
some kind of
equilibrium regime. Indeed for $t-t' \sim O(1)$ we expect that there
will be a fully
TTI equilibrium regime, but it is also expected to be nonuniversal.
A universal, {\it quasi-equilibrium} regime is expected however for
$t-t'  \sim t'^u < t'$, $u<1$
(regimes (I a) and (II a)). As the time difference increases it should
crossover at $t - t' \sim t'$
to an intermediate aging regime (regimes (I b) and (II b)). Regime I is
most complex as
there one expects two later regimes as $t-t' \sim t \sim t'^v$, $v>1$
crossing over to yet
another scaling regime when $t$ reaches $\tau_k$. It is interesting to
note that either in
Sinai model \cite{sinai} or even more
clearly in the 1D random
field Ising model \cite{rfim1d} such regimes are also expected,
some have been studied
demonstrated and studied in details (there is also an equilibration time
scale analogous to
$\tau_k$).

In the determination of all the above regimes the quantity:
\begin{eqnarray}
\mu_k(t) = \int_{t_i}^{t}  dt'' R_k(t,t'')
\end{eqnarray}
which appears in  (\ref{raging}) plays an important role. It is a
function
of $t$ alone. It satisfies the equation:
\begin{eqnarray}
&& \partial_k \mu_k(t) =
\int_{t_i}^t dt_1 R_k(t,t_1) [ -2 k \mu_k(t_1) \label{muk}
\\
&& - \tilde{\beta} k^{3-\theta} ( \mu_k(t_1)^2
- \int_{t_i}^{t_1} dt' R_k(t_1,t') \mu_k(t'))]\nonumber
\end{eqnarray}
Its value can be understood by using the STS covariance
under $u_{r t} \to u_{r t} + v_r$, where $v_r$ is an arbitrary function.
In a general non-equilibrium situation, the STS gives constraints
relating
different initial conditions at $t=t_{i}$. It can be written as
$\ln Z[ h_{kt} , \hat{h}_{kt},u_{k,t=0}=0] = Z[ h_{kt} + k^2 v_k,
\hat{h}_{kt}, u_{k,t=0}=v_k]
- \int_{kt} \hat{h}_{kt} v_k$, where the initial condition is explicitly
indicated. It thus immediately yields:
\begin{eqnarray}
\mu_k(t)  = \frac{1}{q^2} (1 -  \frac{\delta <u_q(t)>}{ \delta u_q(t_i)
} )
\end{eqnarray}
When $t_1 \gg \tau_k$ we expect that the influence of the initial
condition on mode $k$ has
been washed out, and we find the equilibrium constraint $\lim_{t \to
+\infty} \mu_k(t)
= \int_0^{+\infty} R_k(\tau) d \tau = 1/k^2$ (which, combined with FDT
gives $C_k(0) - C_k(\infty) =T/k^2$).
Thus, we expect $\mu_k(t)$ to take the form:
\begin{eqnarray}
&& \mu_k(t)   = k^{-2 + \sigma} m(k^\theta \ln t) \quad  \frac{\ln
t}{\ln \tau_k} < 1\\
&& \mu_k(t)   = k^{-2} {\cal M}(t/\tau_k) \quad t \sim \tau_k
\end{eqnarray}
with ${\cal M}(\infty) = 1$ and a reduction $k^{\sigma}$ in the short
time regime compared to asymptotic one,
with an interpretation in terms of the susceptibility to initial
condition being almost $1$, presumably
with some rare (droplets ?) configurations exhibiting decorrelation.

To discuss the specific choice of the scaling functions and prefactors
we proceed as follows. Let us consider regime (I). We have found that
with the forms of the prefactors in subregimes (I c) and (I a)
indicated above we could obtain from (\ref{muk}) non-trivial equations for the
scaling functions.  The regime (I b) is then necessary to match (I c)
and (I a). Next, with the forms conjectured for (Ia,b,c) the terms in
(\ref{muk}) scale respectively as $1$, $k^{2 - \theta(1-\alpha_i)}$,
$k^{4-\theta - 2 \theta(1-\alpha_i)}$, $k^{2-\theta -
  \theta(1-\alpha_i) + \sigma}$ (where the first term is the
derivative) with $\alpha_i=\gamma,\delta,\alpha$ respectively in each
subregime.  Thus either $\sigma \leq \theta/2$ and
$\theta(1-\alpha)=2-\theta + \sigma$ and only the last term counts or
$\sigma \geq \theta/2$ and $\theta(1-\alpha) = 2 - \theta/2$ (in
equilibrium regime III one had $\sigma=0$ leading to $\alpha=2 -
\theta/2$). Here, we see that if $\mu_k(t)$ is determined by an
integration over regimes (Ia,b,c) as is natural, it implies $\sigma =
2 - \max_{\alpha_i=\gamma,\delta,\alpha} \theta (1-\alpha_i)$, and one
sees that $\sigma=\theta/2$ and:
\begin{eqnarray}
&& \gamma = \delta = \alpha = \frac{3}{2} - \frac{2}{\theta}
\end{eqnarray}
is the only solution. Note that this contradicts the naive expectation
that the form
in the quasi equilibrium regime (I a) would scale as the equilibrium
form IIIa (they differ
by a power of a $\ln(t-t')$): to get $\gamma=2-2/\theta$ would require
$\sigma=0$, and
some argument that the value of $\mu_k(t)$ is controlled by $t-t''$ in
the short time non-universal
regime.

Accepting the above scenario as reasonable we find the equation for the
scaling function $f(x,u)$
as:
\begin{eqnarray}
&& \theta x \partial_x f(x,u) = \tilde{\beta} x^{2(1-\alpha)} \times (
\\
&&
\int_u^1 \frac{du_1}{u_1^\alpha} f(x,u_1) f(x u_1, \frac{u}{u_1})
\int_0^{u_1}  \frac{du_2}{u_2^\alpha} f(x u_1, \frac{u_2}{u_1})
\nonumber \\
&& -\int_u^1 \frac{du_1}{u_1^\alpha} \int_u^{u_1} 
\frac{du_2}{u_2^\alpha}
f(x,u_1) f(x u_1, \frac{u_2}{u_1} ) f(x u_2, \frac{u}{u_2} ) )\nonumber
\end{eqnarray}
The $0$ bound in the integral really comes from the ration $\ln t_i/\ln
t$ assumed to be very
small. The first term in the right hand side of (\ref{raging}) gives a
subdominant contribution.
Similar equations hold for the other regimes. We have not attempted to
analyze further
these equations at this stage. This would be necessary to fully confirm
the self consistency of
the scenario proposed here.

\subsection{correlation function}

Let us now indicate the RG equation obeyed by the correlation function.
It is obtained
from considering the full local quadratic term in the running effective
action:
\begin{eqnarray}
- \frac{1}{2} \int_{r,t>t_i,t'>t_i} (i \hat{u}_{rt}) (i \hat{u}_{r t'})
U_l(t,t')
\end{eqnarray}
One can also decompose it as $U_l(t,t') = V_l(t,t') + \Delta_l(0)$ by
extracting the persistent part (disorder), requiring that
$\lim_{t,t',t-t' \to +\infty} V_l(t,t')=0$. To lowest order
$O(\Delta)$ $U_l$ is
corrected and flows as follows:
\begin{eqnarray}
&& \partial_l U_l(t,t') = - \Gamma_l \Lambda_l^4
(\frac{1}{2} C_{\Lambda e^{-l},l}(t,t)
+ \frac{1}{2} C_{\Lambda e^{-l},l}(t',t') \nonumber
\\
&& - C_{\Lambda e^{-l},l}(t,t') ) \label{u}
\end{eqnarray}
where we assumed a flat initial condition $u_{r t=0}=0$ (otherwise it
should be added) and to this order $U_l$ remains local. The
persistent part of (\ref{u}) yields $\partial_l \Delta_l(0) = - T
\Gamma_l \Lambda_l^2$
in agreement with (\ref{DeltaEq}), (\ref{frgeq}) (using that the
persistent part of the parenthesis in
(\ref{u}) is the equilibrium
connected correlation $C^{eq,c}_{\Lambda e^{-l},l}  = T_l \Lambda^{-2}
e^{2 l}$).
Substracting it yields the flow of $V_l$. One closes the equations
determining $C_l$,
$U_l$ using $C_{k,l}(t,t') = [R_{k,l} \cdot U_l \cdot R_{k,l}](t,t')$.
Equivalently
one can separate the effect of the random force part of the
disorder in the correlation and write $C_{k,l}(t,t') =
\tilde{C}_{k,l}(t,t') + \Delta_l(0) \mu_{k,l}(t) \mu_{k,l}(t')$
(with $\mu_{k,l}(t) = \int_0^t dt' R_{k,l}(t,t')$) write two closed
equations for
$\tilde{C}_l$ and $V_l$ using $\tilde{C}_{k,l}(t,t') = [R_{k,l} \cdot
V_l \cdot R_{k,l}](t,t')$.

Proceeding as above, to determine the correlation one defines
$U_k=U_{l=\ln(\Lambda/k)}$ (similarly
for $V_k$) and obtains:
\begin{eqnarray}
&& \partial_k U_k(t,t') = \tilde{\beta} k^{3-\theta}
(\frac{1}{2} C_{k}(t,t)  + \frac{1}{2} C_{k}(t',t')  - C_{k}(t,t') ) 
\nonumber \\
&&
\end{eqnarray}
with $C_{k,l=\ln(\Lambda/k)}(t,t') =C_k(t,t')$, which should be solved
along with:
\begin{eqnarray}
&& C_{k}(t,t')= \int_{t_i}^t dt_1 \int_{t_i}^{t'} dt_2 R_{k}(t,t_1)
U_k(t_1,t_2) R_{k}(t',t_2) \nonumber\\
&&
\label{corr}
\end{eqnarray}
with initial conditions at $k=1$:
\begin{eqnarray}
&& U_{k}(t,t') = 2 \eta T \delta_{t t'} + \Delta(0)  \quad (k=1) \\
&& C_{k}(t,t') = \frac{T}{k^2}
( e^{- k^2 |t-t'|} - e^{- k^2 (t+t')} ) \nonumber
\\
&& +
\frac{\Delta(0)}{k^4} (1 - e^{- k^2 t})(1 - e^{- k^2 t'}) \quad (k=1)
\label{inic}
\end{eqnarray}
Using the equation for $\partial_k R_k$ one can also write the closed
equation for $C_k(t,t')$ as:
\begin{widetext}
\begin{eqnarray}
&& \partial_k C_{k}(t,t') = \int_{t_1,t_2>t_i}\hspace{-0.2in}
[R_{k}(t,t_1) C_{k}(t_2,t') + R_{k}(t',t_1) C_{k}(t_2,t)] 
\times [-2 k \delta_{t_1,t_2}  - \tilde{\beta} k^{3-\theta}
(R_{k}(t_1,t_2) - \delta_{t_1,t_2} \int_0^{t_1} dt''
R_{k}(t_1,t'') ) ] \nonumber \\
&&
+ \tilde{\beta} k^{3-\theta} R_{k}(t,t_1)
R_{k}(t',t_2)
(\frac{1}{2} C_{k}(t_1,t_1)  + \frac{1}{2} C_{k}(t_2,t_2) - C_{k}(t_1,t_2) )  
\end{eqnarray}
\end{widetext}
and initial condition (\ref{inic}). Alternatively, one can work with
$\tilde{C}_k$ and
$V_k$, which have a more complicated equation but simpler initial
conditions:
\begin{eqnarray}
&& V_{k}(t,t') = 2 \eta T \delta_{t t'}  \quad (k=1) \\
&& \tilde{C}_{k}(t,t') = \frac{T}{k^2}
( e^{- k^2 |t-t'|} - e^{- k^2 (t+t')} ) \quad (k=1) \nonumber
\end{eqnarray}

One easily checks that upon the assumption of time translational
invariance as should hold in the equilibrium regime, the equation for
$C_k(t,t')= C_k(t-t')$ becomes, as expected, equivalent to the one for
$R_k(t-t')$ via the FDT relation. Further study of the nonequilibrium
equations, including the determination of the FD violation ratio
$X(t,t')$ in the various regimes is left for forthcoming publications.

\section{Corrections to $F$-term by pinning disorder}
\label{app:fterms}

Let us give some details about the calculation of the graphs in
Figs.~\ref{graph2},\ref{graph3}. The correction to the effective action
to lowest order in $T$ coming from the cross term $F \Delta$ reads:
\begin{widetext}
\begin{eqnarray}
&& \delta \Gamma = \langle \frac{1}{2} \int_{r,r_1,t,t'}
(i \hat{u}_{rt} + i \delta \hat{u}_{rt})
(i \hat{u}_{rt'} + i \delta \hat{u}_{rt'})
\Delta(u_{rt} - u_{rt'} + \delta(u_{rt} - u_{rt'}))  \nonumber \\ &&
\times F[z_{r_1} 
+ \int_{t_1} 
 i \delta \hat{u}_{r_1 t_1} \partial_{t_1} u_{r_1 t_1}
 + \int_{t_1}
i \hat{u}_{r_1 t_1} \partial_{t_1} \delta \ u_{r_1 t_1}
+ \int_{t_1}
i \delta \hat{u}_{r_1 t_1} \partial_{t_1} \delta \ u_{r_1 t_1}]
\rangle^{1PI}_{\delta u, \delta \hat{u}}
\end{eqnarray}
with $z_r =  \int_{t}  i \hat{u}_{r t} \partial_{t} u_{r t}$ and
the averages over $\delta u$, $\delta \hat{u}$ are restricted to
1-particle irreducible graphs.
This splits into contributions corresponding to graphs (a,b,c) in Fig.
(\ref{graph2})
which evaluate respectively as (dropping all terms which do not correct
$F$):
\begin{eqnarray}
&& \delta \Gamma^{(a)} =  \frac{1}{2} \int_{r,r_1,t,t'}
i \hat{u}_{rt} i \hat{u}_{rt'} \frac{1}{2} \Delta''(0)
< (\delta u_{rt} - \delta u_{rt'})^2 F[z_{r_1} + \int_{t_1}
 i \delta \hat{u}_{r_1 t_1} \partial_{t_1} u_{r_1 t_1} ] >
 \nonumber \\ && = -
\frac{1}{2} \Delta''(0) R_{q,\omega=0}^2 \int_{r} ( z_r^2 F''[z_r]
- \int_{t t'} i \hat{u}_{rt'} i \hat{u}_{rt} (\partial_{t} u_{r t})^2)
 \\
&& \delta \Gamma^{(b)} =  \frac{1}{2} \int_{r,r_1,t,t'}
\Delta(u_{rt}-u_{rt'})
<i \delta \hat{u}_{rt} i \delta \hat{u}_{rt'} 
F[z_{r_1} + \int_{t_1}
 i \hat{u}_{r_1 t_1} \partial_{t_1} \delta u_{r_1 t_1} ] > \nonumber  \\
&& =  \frac{1}{2} R_{q,\omega=0}^2 \int_{r,t,t'}
\partial_t \partial_{t'} (\Delta(u_{rt}-u_{rt'})) i \hat{u}_{r t}
i \hat{u}_{r t'}  F''[z_{r}]  \\
&& \delta \Gamma^{(c)} =  \frac{1}{2} 2 \int_{r,r_1,t,t'}
i \hat{u}_{rt} <i \delta \hat{u}_{rt'} \Delta(u_{rt}-u_{rt'} + \delta
u_{rt} - \delta u_{rt'}) \nonumber  \\
&&
F[z_{r_1} + \int_{t_1} \delta i \hat{u}_{r_1 t_1} \partial_{t_1} u_{r_1
t_1}
+ \int_{t_1}  \partial_{t_1} \delta u_{r_1 t_1}  i \hat{u}_{r_1 t_1} ] >
 = R_{q,\omega=0}^2 \int_{r,t,t'} i \hat{u}_{rt} i \hat{u}_{r t'}
\partial_{t} u_{r t}
\partial_{t'} \Delta'(u_{rt} - u_{rt'}) F''[z_{r}]
\end{eqnarray}
as well as the graph in Fig. \ref{graph3}:
\begin{eqnarray}
&& \delta \Gamma^{(3)} = \frac{1}{2} 2 \int_{r,r_1,t,t'}
i \hat{u}_{rt} <i \delta \hat{u}_{rt'} \Delta(u_{rt}-u_{rt'} + \delta
u_{rt} - \delta u_{rt'}) 
F[z_{r_1} + \int_{t_1} \delta i \hat{u}_{r_1 t_1} \partial_{t_1} \delta
u_{r_1 t_1} ] > \nonumber  \\
&& = \int_{r,t,t'}
i \hat{u}_{rt} \partial_{t'} \Delta'(u_{rt}-u_{rt'}) F'[z_{r}]
\int_{t_1} R_{q,t-t_1} R_{q,t_1-t'}
\end{eqnarray}
This yields the result (\ref{eq:Fflow}) in the text.
\end{widetext}

\section{mapping of random friction model
onto polymer and related problems}

The random friction model (in its non-trivial $T=0$ limit) can be mapped
formally onto various other problems, such as, after disorder averaging,
the statistical mechanics of a pure self interacting chain
(e.g. a self avoiding walk problem) or, prior to averaging, to
some random diffusion models, e.g. depolarisation of a spin diffusing in
a random magnetic field.
Concerning its behaviour one should distinguish between the genuine
model
(with a fixed distribution $P(\eta)$) and the effective one which appear
as a coarse grained version of the pinning problem, in which $P(\eta)$
flows and
becomes very broad.

First setting $P(r,t) \sim \eta(r) u_{rt}$ one sees that (\ref{eom2})
(with $f(r,u)=0$ and $T=0$) is the Fokker-Planck equation $\partial_t
P=\nabla D(r) (\nabla + \nabla V(r)) P$ for the diffusion of a
particle with a random diffusion coefficient $D(r)=1/\eta(r)$ in a
random potential $V(r)=-ln \eta(r)$, of equilibrium measure
$P_{eq}(r)=e^{-V(r)}=\eta(r)$. When $\eta(r)$ is uncorrelated from
site to site one does not expect any anomalous behaviour in any
dimension, except if the distribution of $\eta$ has broad tails (e.g.
algebraic would yield anomalous power law diffusion). In the effective
model $V(r)$ becomes gaussian and grows with scale which corresponds
to a particle localized in some regions of space.

A complementary picture can be developed based on a mapping onto
a self-interacting chain. The response function $R_{rt,r't'}=d\delta
u_{rt}/dh|_{h=0}$ of this model is obtained by solving:
\begin{eqnarray}
(\eta(r) \partial_t - \nabla^2 ) \delta u_{rt} = h \delta(r-r')
\delta(t-t')
\end{eqnarray}
with initial condition $\delta u_{r t=0}=0$. This implies $\delta u_{r
t}=0$ for all $t<t'$. Thus
the response is a function of $t-t'$ alone and its Laplace-Fourier
transform $s=i \omega$
in any given random environment, can be written as:
\begin{eqnarray}
&& R_{r r'}(s)= <r | \frac{1}{-\nabla^2 + s \eta(r) } |r'>
\nonumber \\ && = \int_{0}^{+\infty} du <r | e^{- u H}  |r'> 
\end{eqnarray}
where $H=-\nabla^2 + s \eta(r)$, which has a positive spectrum for
$s>-s^*$. The
value $s^*$ at which $H$ develops an eigenstate of zero eigenvalue (e.g.
$s^*=1/\eta_{max}$
in the ``classical'' limit $c\to 0$) gives the large time decay of $R_{r
r'}(t) \sim e^{-s^* t}$.

We can also write, in the Fourier domain, using the Feynman Kac formula:
\begin{eqnarray}
&& R_{r r' }(i \omega) = \int_0^{+\infty} du e^{- (\mu + i \omega
\overline{\eta}) u}
\int_{x(0)=r'}^{x(u)=r} Dx(v) \nonumber \\
&&
\exp( - \int_0^u dv ( \frac{1}{4} (\frac{d x}{d v})^2 + i \omega \delta
\eta(x(v)) ))
\end{eqnarray}
with the ``time'' variables $u$ and $v$. We have splitted
$\eta(x)=\overline{\eta}+
\delta \eta(x)$ for convenience and added a small mass term $\mu$ for
convenience.
In this form the problem has the form of a spin decoherence problem, the
integral
being dominated by paths which average well over the random relaxation
times, rather than paths with multiple returns to the same region which
average poorly and then cancel incoherently (details about the mapping
and special distribution of noise can be found in\cite{mitra_pld}).

Averaging over disorder leads, for small disorder, to:
\begin{eqnarray}
&& R_{r-r',\omega} = \int_0^{+\infty} du e^{- (\mu + i \omega
\overline{\eta}) u}
Z(r-r' , u, g= \frac{1}{2} \omega^2 \eta_2) \nonumber \\
&& Z(r-r' , u, g) = \int_{x(0)=r'}^{x(u)=r} Dx(v) \\
&&
\exp( - \int_0^u dv \frac{1}{4} (\frac{d x}{d v})^2 - \int_0^u dv dv'
g \delta(x(v) - x(v')))\nonumber
\end{eqnarray}
which is the partition function of a self avoiding walk in the Edwards
representation.
We have retained only the second moment $\eta_2$ of $\delta \eta(x)$,
but of course the
full interaction could be written using $F[z]$. The theory is described
by a
non-trivial fixed point in $d<4$ (related to the $n=0$ $O(n)$ model
with mass $(\mu + i \omega \overline{\eta})$ and coupling $\tilde{g}
\sim g \Lambda^{4-d}$).
This is compatible with the previous conclusions for the $F$ theories
using perturbation theory
if we take $t$ as a simple index, with no power counting dimension (it
plays a role somewhat
similar to the replica index $a$). The correction to $g$ by $g^2$
comes from the contraction of two $\eta_2$ vertices, of the form
$\delta(\omega^2 \eta_2)= \omega^4 \eta_2^2$
and is indeed logarithmically divergent in $d=4$. Note that the mass
term is thus relevant at the fixed point $\tilde{g}=g^*$.

This analysis yields information at finite time. The Ginsburg criterion
gives
the critical regime as $r > g^{-1/(4-d)}$ or $u >  g^{-2/(4-d)}$, in
which
$Z(r-r',u,g)$ takes the scaling form
\begin{eqnarray}
&& Z(r,u,g) = u^{-\nu d} F[r u^{-\nu}] Z(q=0,u,g)   \\
&& Z(q=0,u,g) \sim u^{\gamma-1} e^{- s_c(g) u}
\end{eqnarray}
with $s_c(g)=c g$ for $d>2$, $s_c(g)=c g \ln(1/g)$ for
$d=2$ and  $s_c(g)=c g^{2/3}$ for $d=1$.
Thus for $d>2$ we expect that:
\begin{eqnarray}
R_{q,\omega} \approx \frac{1}{i \overline{\eta} \omega +
c \eta_2 \omega^2 + q^2}
\end{eqnarray}
in the non-critical regime, while we expect:
\begin{eqnarray}
R_{q=0,\omega} \approx \left(\frac{1}{i \overline{\eta} \omega +
c \eta_2 \omega^2}\right)^{\gamma}
\end{eqnarray}
i.e, $R_{q=0,t} \sim t^{\gamma-1} e^{- c \eta_2 t/\overline{\eta}}$
in the critical regime, and for $q>0$,
the appropriate scaling function of $r t^{-\nu}$.

The critical regime correspond to:
\begin{eqnarray}
\overline{\eta} \omega > (\frac{\overline{\eta}^2}{\eta_2})^{2/d}
\end{eqnarray}
which for the genuine model gives a singularity only at finite (true)
time
(see however \cite{mitra_pld} for possibly more radical effect of
non-gaussian disorder).  However in the limit of very broad disorder, as
in the effective model,
one has  $\eta_2 >> \overline{\eta}^2$  and thus the critical
singularity
moves to small $\omega$.

Note that for $d<2$ the behaviour is more radical as one expects,
e.g. in $d=1$:
\begin{eqnarray}
R_{q,\omega} \approx \frac{1}{i \overline{\eta} \omega +
c (\eta_2)^{2/3} \omega^{4/3} + q^2}
\end{eqnarray}

Thus to conclude, in the absence of pinning disorder at $T=0$
the $F$ term is preserved but generate higher order time derivative
terms.
The theory can be rescaled so as to possess a non-trivial finite
$\omega$, finite
disorder term, which presumably in $d>2$
produces only preexponential algebraic corrections
to the leading behaviours given by the most relevant
$\overline{\eta}$ term. For the effective model this critical behaviour
should
be observable even at small $\omega$.

\section{Full flow of the truncated effective action}

\label{general}

In this appendix we discuss an approach to the calculation of the mean
response function extending the approximate FRG scheme of Section III,
but still neglecting the functional dependence of operators considered
in Section IV.  In Section III the broad distribution of time scales
was embodied in the so-called $F$ term. We uncover here an interesting
structure of additional operators in the spirit of the more complete
set of moments $\overline{\langle t^{p_1} \rangle_L \cdots \langle
  t^{p_N}\rangle_L}$ discussed in the introduction.  Recall that in
Sec.~\ref{sec:rf}, we showed that, although the $F$ term did not
renormalize itself, it did generate higher derivative terms.  Such
terms {\sl can} contribute to the frequency dependence of the response
function.  We studied in the previous Appendix the response function
of the pure random friction model, which does generate higher
derivative terms, but neglects the scale dependence generated by the
pinning disorder.  Here we consider these two effects in tandem, hence
modifying the results for $R_k(t), R_k(\omega)$ within the purely
random friction model.  While we have not been able to obtain simple
expressions in this more complete (albeit still non-functional)
approximation, one can go quite far in reducing the problem to one of
applied mathematics.

\subsection{Generalized random friction model}

To proceed further one needs to construct a more systematic approach
where all possible important terms in the dynamical action functional
are included. We will generalize the F term (at zero temperature for
simplicity)
in the form:
\begin{eqnarray}
  S_{\rm kin} & = & \sum_{n=1}^\infty \sum_{p_1 \cdots p_n \geq 1}
  F^{(n)}_{p_1\cdots p_n} \int_{r t_1 \cdots t_n} \!
 i \hat{u}_{r t_1} \cdots i \hat{u}_{r t_n} \nonumber \\
  && \times \partial_{t_1}^{p_1} u_{r t_1} \cdots \partial_{t_n}^{p_n}
  u_{r t_n},
\end{eqnarray}
and we will often denote by $m=p_1+\cdots +p_n$ the total number of
time derivative in a given term of the sum.  This form of the action
neglects terms with products of time derivatives of $u_{rt}$ at the
same time, as well as statistically translationally invariant
functional dependence, e.g. on $u_{rt_1}-u_{rt_2}$.  However, it does
include considerably more physics, and as we will see, enough
generality to approach the problem of computing the averaged response
functions.  This kinetic part of the action corresponds to the
following generalization of the random friction model:
\begin{eqnarray}
\sum_{m=1}^{+\infty} \eta_m(r) \partial_t^m u_{rt} = \nabla^2 u_{rt} +
F(u_{rt},r)
\end{eqnarray}
with $[\overline{\eta_{p_1}(r_1) ... \eta_{p_n}(r_n)}]_C =
n! (-1)^{n+1} F^n_{p_1,..p_n} \delta_{r_1,\cdots,r_n}$.

The response function is related to the
lowest ($n=1$) member of this hierarchy via
\begin{eqnarray}
&& R^{-1}_k(i\omega) = k^2 + \tilde{\Sigma}_k(i\omega) \\
&& = \tilde{\Sigma}_k(i\omega) = \sum_{m=1}^\infty F_m^{(1)} ~
(i\omega)^m. \label{ssum}
\end{eqnarray}
and within the Wilson scheme the true physical response function
$R^{-1}_k(i\omega)$
is obtained via the same formula using the running $F_m^{(1)}|_{l =
\ln(\Lambda/k)}$.
This is represented graphically in Fig. \ref{graph13}.

\begin{figure}
\centerline{\fig{4cm}{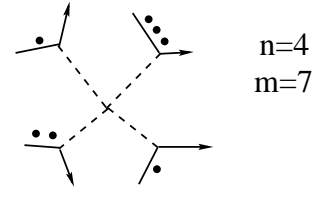}}
\caption{ Graphical representation of an $F$
vertex $F^{n}_{p_1,..p_n}$ with $n=4$, $p_1=1$, $p_2=2$, $p_3=1$,
$p_4=3$, $m = \sum_i p_i=7$.
Dots represent the number of time derivatives (i.e power of
frequency factor), each leg has a different frequency.
\label{graph5}}
\end{figure}

\begin{figure}
\centerline{\fig{4cm}{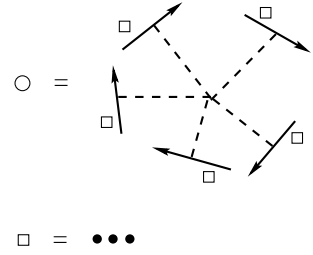}}
\caption{ Compact notation for a generic $F$ vertex.  The open circle
represents an $F$ vertex with an arbitrary number of legs $n$ not shown.
On incoming ($u$) lines, an arbitrary number of time derivatives
(powers of $\omega$) are indicated by an open square. \label{graph6}}
\end{figure}

\begin{figure}
\centerline{\fig{6cm}{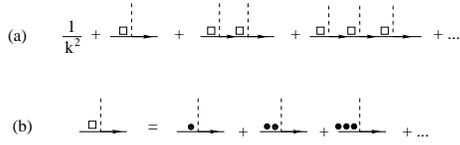}}
\caption{Graphs with only $n=1$ vertices which enter in the response
function (a) and
its self energy (b). \label{graph13}}
\end{figure}

\begin{figure}
\centerline{\fig{6cm}{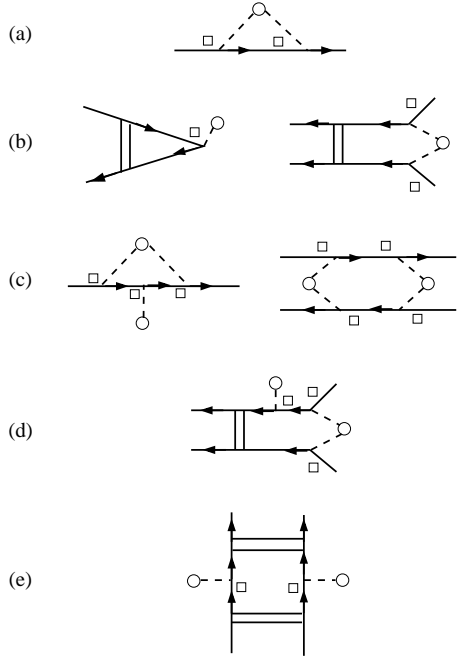}}
\caption{One-loop diagrammatic contributions to the $F$ terms.  The
diagrams in (a)-(e) represent contributions to $F$ of order $F$,
$\Delta F$, $F^2$, $F^2\Delta$, and $F^2 \Delta^2$, respectively.
\label{graph7}}
\end{figure}

One can carry perturbation theory using the generalized $F$.  The
vertices are shown in Figs.~\ref{graph5},\ref{graph6}.  In this
notation, there are a variety of important one loop diagrams to be
considered.  These are shown in Fig.~\ref{graph7}.
Schematically, these contributions give rise to an RG equation for the
$F_m^{(n)}$ of the form
\begin{eqnarray}
\partial_l F_m^{(n)} & = & \Delta F_m^{(n)} + F_m^{(n+1)} +
F_{m'}^{(n')} F_{m-m'}^{(n+2-n')} \nonumber \\
        & & + \Delta F_{m'}^{(n')}
F_{m-m'}^{(n+1-n')} + \Delta^2 F_{m'}^{(n')}
F_{m-m'}^{(n-n')} + \cdots \label{master}
\end{eqnarray}
In (\ref{master}), we have neglected coefficients, powers of $l$,
and fine distinctions such as the precise form of $\Delta$ which
appears in a given term.  Repeated (primed) indices other than $m$ and
$n$ are summed.  The pure random friction model contains only those
diagrammatic contributions with no pinning disorder $\Delta=0$.  Thus
only the second and third terms in (\ref{master}) are taken into
account for instance by the mapping to a polymer problem in the
previous appendix.   Note that there are no terms of $O(F\Delta^2)$
contributing to the renormalization of $F$, which follows
diagrammatically because one needs at least two $F$'s to put boxes
(time derivatives) on the incoming legs originating from the two
pinning disorder vertices.

A beautiful simplicity arises due to the extremely broad distribution
of timescales in the RG, and can be seen from the structure of
(\ref{master}).  In particular, we note that the total number of
powers of $\omega$ is conserved by all terms.  Moreover, from the
prior analysis, we expect $\eta^{(m)} \sim \eta_l^{2m^2-m}$.  We thus
conjecture that {\sl all} $F_m^{(n)}$ scale in this way {\sl
independently of $n$}, i.e. $F_m^{(n)} \sim
\eta_l^{2m^2-m}$ (see below to see that this scaling is indeed
self-consistent).  Under this assumption, the terms involving more
than two $F's$ in (\ref{master}) can be seen to be strongly
subdominant, which follows from the convexity of the $2m^2-m$ factor
in the exponential.  Thus the super-exponential (Gaussian) growth of
the moments of the timescales, which is directly connected to the
broad distribution of relaxation times, plays a key role in
simplifying the structure of the RG.

One can thus restrict the analysis to the linear part in $F$ of the
full one-loop RG equation.  Note that this indeed combines the effects
of pinning disorder and the ``upward'' feedback of the random friction
model -- the first two terms in (\ref{master}).  The linearized RG
equation reads:
\begin{eqnarray}
  \partial_l F^{(n)}_{p_1\cdots p_n} & = & \Gamma_l (2n^2 - n)
  F^{(n)}_{p_1\cdots p_n}  \label{diffflow} \\
    & & \hspace{-0.8in} + (n+1)\alpha e^{-(d-2)l} \sum_{r=1}^n
\sum_{q=1}^{p_r-1}
    F^{(n+1)}_{p_1\cdots p_{r-1} (p_r-q) p_{r+1}\cdots p_n q} \; ,
\nonumber
\end{eqnarray}
where $\alpha = \Lambda^{d-2} A_d$. The general study of this equation
is
again highly difficult but we see that it does have special solutions
where the
$F^{(n)}_{\{ p_i \}}$ depend only upon $m=\sum_{i} p_i$. From the above
consideration
about the asymptotic behaviour it is rather natural to look for such
solutions.

Thus we let $F^{(n)}_{\{ p_i \} } = F^{(n)}_{l,m}$.  Then the linear RG
equation
(\ref{diffflow}) becomes:
\begin{equation}
\partial_l F^{(n)}_{l,m} = \Gamma_l (2n^2-n) F^{(n)}_{l,m} + (n+1)
\alpha e^{-(d-2)l} (m-n) F^{(n+1)}_{l,m}. \label{simplerRG}
\end{equation}
This is much easier to solve and the attentive reader will
easily find that this infinite hierarchy of differential flow equations
is solved
asymptotically by the ansatz
\begin{equation}
F^{(n)}_{p_1\cdots p_n} = \left[{\tilde{\beta} \over \alpha}
e^{(d-2+\theta)l}
\right]^{n-m}\left({\eta_l \over \eta_0}\right)^{2m^2-m}
\!\!\! \tilde\eta_0^{(m)} (-1)^{m+1} a_m^n, \label{ansatz}
\end{equation}
where $m=\sum_{i=1}^n p_i$.  The $a_m^n$ coefficients are given by
\begin{equation}
a_m^n = {m! \over n!} \prod_{r=n}^{m-1} {1 \over {2m+2r-1}} = {{m!
(2m+2n-3)!!} \over {n!(4m-3)!!}}. \label{coeffs}
\end{equation}
The $\tilde\eta_0^{(m)}$ coefficients are not determined by the
asymptotic analysis.  In principle, they should be matched at some
scale $l^*>0$ to the form of the $F^{(n)}_m$ coefficients determined
by the early stages of renormalization, in which the linearized RG
equation used to obtain them is not valid.  One might imagine
beginning with a model in which the bare relaxation time was
distributed with cumulants $\eta_0^{(m)}$, and naively
$\tilde\eta_0^{(m)} \approx \overline{\eta}_0^{m}$.  We will use this
prescription below purely in order to simplify notation.  However, a
more detailed analysis of the early stages of renormalization is in
fact required to discern whether this is indeed correct.

\subsection{Equation for the equilibrium response function}

We are now in a position to extract the response function.  Applying
(\ref{ssum},\ref{ansatz}-\ref{coeffs}) for $n=1$ gives
the response as an infinite series in $i \omega$:
\begin{eqnarray}
R^{-1}_k(i\omega) & = & \sum_m (-1)^{m+1} \eta_0^{(m)} \exp\bigg[
(2m^2-m) U_k \nonumber \\
        && + [\gamma - (d-2+\theta)l - \ln(1/i\omega)]m\bigg],
\end{eqnarray}
where $U_k \equiv U_{l=\ln (\Lambda/k)}$.

Since it is not easy to resum this series, and its convergence
properties
are unclear, we now reformulate the above calculation in a functional
way, in hopes of surmounting the limitations of the expansion in terms
of moments of relaxation times.
Let us introduce the generating
function
\begin{equation}
G(i\omega,z) = \sum_{m \geq n \geq 1} (i\omega)^{m-n} z^n F_{m l}^n,
\end{equation}
which conveniently captures both the $F$ term (and hence distribution
of relaxation times) and the response function:
\begin{eqnarray}
F(z) & = & G(0,z), \\
i \eta_0 \omega + \Sigma_k (i\omega) & = & {\partial \over {\partial
z}}  \left. G(i\omega,z) \right|_{z=0}.
\end{eqnarray}
Multiplying (\ref{simplerRG}) by the appropriate powers of $z$ and
$i\omega$ and summing gives the flow equation
\begin{eqnarray}
\partial_l G & = & \Gamma_l (z \partial_z G + 2 z^2 \partial^2_z G)
 \\
        & & + \alpha e^{-(d-2)l} \left\{ i\omega
\partial_{i\omega}\left[ i\omega (\partial_z G - \left.\partial_z
G\right|_{z=0})
\right] \right\} .\nonumber
\end{eqnarray}
This is somewhat simplified by defining the derivative $H(i\omega,z) =
\partial_z G(i\omega,z)$,
\begin{equation}
\partial_l H = \Gamma_l (H+ 5z \partial_z H + 2z^2 \partial_z^2 H) +
\alpha e^{-(d-2)l} i\omega \partial_{i\omega}\left(i\omega \partial_z
H\right).
\end{equation}
A hopefully illuminating change of variables is to define
\begin{eqnarray}
u & = & \ln(1/i\omega), \\
v & = & \ln (z/i\omega) - (\theta+d-2)l \pm \ln (\alpha/\chi).
\end{eqnarray}
Then $K(u,v) = H(i\omega,z)$, and obeys
\begin{equation}
\partial_{U_l} K = K + 3\partial_v K + 2 \partial_v^2 K +
(\partial_u-\partial_v) \left( e^{-v} \partial_v K \right),
\label{unsolved}
\end{equation}
where $\partial_l U_l = \chi e^{\theta l}$ as before.  Note that the
$F$ term is recovered in the limit $v\rightarrow \infty$ and $u-v =
\ln z + {\rm const.}$ is fixed, in which the second term is
negligible.  In that limit, we recover the diffusion with drift
equation (\ref{diffdrift}), and the appropriate solution is $K(u,v) =
\Phi(v-u) = F'(e^{v-u})$.  More general solutions of (\ref{unsolved})
remain to be found.

\begin{widetext}

\section{One loop hierarchy: method of calculation}
\label{sec:one-loop-hierarchy}
In this Appendix we show how the systematic calculation of
the one loop correction to the dynamical effective action
can be organized, and sketch explicit calculation on
the simplest examples. We focus on $T=0$.

The schematic form of the dynamical action $S$ is given in the text in 
(\ref{eq:sigmas}) as a sum of terms containing an increasing number of
independent times (cumulants):
\begin{eqnarray}
S = i\hat{u}_1 {\cal S}_{1} - \frac{1}{2} i\hat{u}_1 i\hat{u}_2 {\cal S}_{12} 
    - \frac{1}{6} i\hat{u}_1 i\hat{u}_2 i\hat{u}_3 {\cal S}_{123} - .. 
\end{eqnarray}
We use the same schematic notation where the indices $1,2,3..$ are short hand
notations for $t_1,t_2,t_3..$, space coordinate and 
all time and space integrations are implicit. From (\ref{eq:sigmas})
${\cal S}_{1}$ is parameterized by an infinite set of kinetic coefficients
$\overline{\eta}, D,..$, ${\cal S}_{12}$ by a set of 
second cumulant functions $\Delta,G,A,B,C,.. $, ${\cal S}_{123}$
by a set of third cumulants $H,W,..$ etc.. (from (\ref{eq:cals12})).

In a first stage we write the total one loop corrections to the action as
the sum of tadpoles, two vertex loop, three vertex triangles etc.. with
either ${\cal S}_{12}$ or ${\cal S}_{123}$ (and so on..) type vertices
using the full response function $R_{12}$, inverse of $i\hat{u}_1 {\cal S}_{1}$,
to contract the vertices (internal lines). Enumerating possible contractions
and performing some combinatorics, yields upon grouping resulting terms
by number of independent times:
\begin{eqnarray}
&& \delta {\cal S}_1 = - < i\hat{u}_2 {\cal S}_{12} > \label{s1} \\
&& \delta {\cal S}_{12} = < i\hat{u}_3 {\cal S}_{123} > + [ \frac{1}{2} {\cal S}_{34} <
{\cal S}_{12} i\hat{u}_3 i\hat{u}_4 > +
 < {\cal S}_{24} i\hat{u}_3 > < {\cal S}_{13} i\hat{u}_4 > ] \label{s2} \\
&& \delta {\cal S}_{123} = 
[ \frac{1}{2} {\cal S}_{45} <S_{123} i\hat{u}_4 i\hat{u}_5 > +
3 < {\cal S}_{14} i\hat{u}_5 > <S_{235} i\hat{u}_4  >
+  \frac{3}{2} <{\cal S}_{12} i\hat{u}_4 i\hat{u}_5 > {\cal S}_{345} ] \\
&& +
[ 2 <{\cal S}_{34} i\hat{u}_5 >  <{\cal S}_{15} i\hat{u}_6 > <{\cal S}_{26}
i\hat{u}_4 >
+ 3 <{\cal S}_{12} i\hat{u}_4 i\hat{u}_6 > <{\cal S}_{34} i\hat{u}_5 >
{\cal S}_{56} ]\nonumber
\end{eqnarray}
where additional (time) indices are integrated over 
Note that ${\cal S}_1$ is corrected only by tadpoles,
${\cal S}_{12}$ by tadpoles and two vertex loops, and so on.. 

In this formula notations such as, e.g. $<{\cal S}_{12} \hat{u}_3 \hat{u}_4 >$
denote the sum of all possible contractions of the $\hat{u}$ fields with the
$u$ fields inside the bracket (at $T=0$ these are the only
possible contractions). Since ${\cal S}_{12} = {\cal S}_{12}[u_{12},\dot u_1,\dot u_2,
\ddot u_1, \ddot u_2,..]$ is an explicit function of $u_{12}=u_1-u_2$, and
time derivatives of $u_1$ and $u_2$ (and similarly for all other vertices)
one may write the sum of all possible contractions as:
\begin{eqnarray}
&& < i\hat{u}_2 {\cal S}_{12} > =  \left(R_{12} \partial_{u_{12}} + 
(\partial_1 R_{12}) \partial_{\dot{u}_{1}} +
(\partial_1^2 R_{12}) \partial_{\ddot{u}_{1}} \right) {\cal S}_{12} + .. \label{contract1}
\end{eqnarray}
We recall that here causality $R_{22}=0$ restricts contractions 
only with $u_1$ (first term), $\dot u_1$ (second term) etc.. 

The next stage is to make apparent $\overline{\eta}$, $D$ etc..
and thus to define the expansion:
\begin{eqnarray}
&& R_{12} = \frac{1}{k^2} (\delta_{12} + \sum_{p=1}^{+\infty} A_p
\partial_1^p \delta_{12})  \label{exp} \\
&& k^2 R(s) = \frac{k^2}{k^2 + \Sigma(s)} = 1 + \sum_{n=1}^{+\infty}
(-1)^n k^{-2 n} \Sigma(s)^n = 1 + \sum_{p=1}^{+ \infty} A_p s^p \\
&& A_1 = - k^{-2} \overline{\eta} \quad, \quad A_2 = - k^{-2} D + k^{-4}
\overline{\eta}^2
\end{eqnarray}
The momentum structure of the one loop diagrams being trivial,
within Wilson one can replace $k=\Lambda_l$ everywhere. As explained in
the text this expansion in power of frequency can be done consistently
and corresponds diagrammatically to expansion in number of dots. 

One then evaluate the contractions shifting time integrations.
Let us illustrate this on simple examples. 

The corrections to $\eta$ and $D$ can be obtained from (\ref{s1})
using (\ref{contract1}). One has:

\begin{eqnarray}
&& \delta {\cal S}_1 = - ( R_{12} \partial_{u_{12}} +
(\partial_1 R_{12}) \partial_{\dot{u}_{1}} +
(\partial_1^2 R_{12}) \partial_{\ddot{u}_{1}} ) {\cal S}_{12} \\
&& =  - [ (\delta_{12} + A_1 \partial_1 \delta_{12} +
A_2 \partial_1^2 \delta_{12} ) \partial_{u_{12}} {\cal S}_{12} +
(\partial_1 \delta_{12} +
A_1 \partial_1^2 \delta_{12} ) \partial_{\dot{u}_{1}}  {\cal S}_{12} +
\partial_1^2 \delta_{12} \partial_{\ddot{u}_{1}} {\cal S}_{12} ] \\
&& = - \delta_{12} [ (1 + A_1 \partial_2 + A_2 \partial_2^2 )
\partial_{u_{12}} {\cal S}_{12} +
(\partial_2 + A_1 \partial_2^2 ) \partial_{\dot{u}_{1}}  {\cal S}_{12}
+ \partial_2^2 \partial_{\ddot{u}_{1}} {\cal S}_{12} ]
\end{eqnarray}
In the second line we have used the expansion (\ref{exp}) and in the 
last line we have used that $\partial_1 R_{12} = - \partial_2 R_{12}$
and integrated by parts over $t_2$. Then time derivatives on the
vertex ${\cal S}_{12}$ can be evaluated and tranformed into derivatives 
w.r.t. fields as:
\begin{eqnarray} \label{idi}
&& \partial_2 \equiv  \partial_2 {\cal S}_{12} = ( - \dot{u}_2
\partial_{u_{12}}
+ \ddot{u}_2 \partial_{\dot{u}_{2}} ) {\cal S}_{12} \\
&& \partial_2^2 \equiv  \partial_2^2 {\cal S}_{12} =
( - \ddot{u}_2 \partial_{u_{12}}
+ \dot{u}_2^2 \partial_{u_{12}}^2 ) {\cal S}_{12}
\end{eqnarray}
At all stages of the calculation we can drop all terms containing more than
a fixed number (here 2) of time derivatives since they will contribute
only to higher order terms in the effective action. Putting everything
together we obtain:
\begin{eqnarray}
&& \delta \eta = G'(0) - \overline \eta \Delta''(0) \\
&& \delta D = -A(0) + C'(0) - 2 \overline \eta k^{-2} G'(0)
-\Delta''(0) (k^{-2} D - k^{-4} \overline \eta^2)
\end{eqnarray}
which, in terms of rescaled quantities, yield the (\ref{deta},\ref{dD})
in the text.

Next we want to evaluate the corrections to second cumulant
functions $\Delta,G,A,B,C$ encoded in $\delta {\cal S}_{12}$.

We start by the simplest, the tadpole, which yields the
feedback of third cumulants into second ones. 
\begin{eqnarray}
&& \delta|_{tadpole} {\cal S}_{12} = < {\cal S}_{123} \hat{u}_3  > 
= (R_{13} \partial_{u_1} S_{123} + \partial_1 R_{13}
\partial_{\dot{u}_1} S_{123} + R_{23} \partial_{u_2} S_{123} +
\partial_2 R_{23}
\partial_{\dot{u}_2} S_{123} \\
&& = 2 \delta_{13} [
(1 + A_1 \partial_3) \partial_{u_1} + \partial_3 \partial_{\dot{u}_1} ]
S_{123} \\
&& = 2  \delta_{13} [
\partial_{u_1} + A_1 \dot{u}_3 \partial_{u_3} \partial_{u_1} +
 \dot{u}_3 \partial_{u_3} \partial_{\dot{u}_1} ] S_{123}
\end{eqnarray}
with $\partial_3 =  \dot{u}_3 \partial_{u_3}$. This gives the 
$2 S'_1(0,0,u)$ term in the equation (\ref{DelEqn}) for $\Delta$
and the $H$ feeding term in the equation for $G$ (we have not
explicitly computed the feeding of $W$ into $A,B,C$ but it is
easily obtained from the above).

Next we need the ${\cal S}_{12}^2$ corrections to ${\cal S}_{12}$
which yield all non linear terms in
(\ref{dG},\ref{frgA},\ref{frgB},\ref{frgC}) for $G$,$A$,$B$,$C$.  The
corresponding correction to $\delta {\cal S}_{12}$ consists in the two
terms in the square bracket in (\ref{s2}).  The full calculation being
tedious we only indicate here how one shuffles time integrals in the
first term (denoted $\delta_1 {\cal S}_{12}$). Starting from:
\begin{eqnarray}
&&\delta_1 {\cal S}_{12} = \frac{1}{2} {\cal S}_{34} < {\cal S}_{12} \hat{u}_3 \hat{u}_4 > =
\frac{1}{2}{\cal S}_{34} [ (R_{14} - R_{24}) \partial_{u_{12}} +
\partial_1 R_{14}  \partial_{\dot{u}_{1}} + \partial_2 R_{24}
\partial_{\dot{u}_{2}} + \partial^2_1 R_{14} \partial_{\ddot{u}_{1}}
+ \partial^2_2 R_{24} \partial_{\ddot{u}_{2}} ] \\
&&
\times [ (R_{13} - R_{23}) \partial_{u_{12}} +
\partial_1 R_{13}  \partial_{\dot{u}_{1}} + \partial_2 R_{23}
\partial_{\dot{u}_{2}} + \partial^2_1 R_{13} \partial_{\ddot{u}_{1}}
+ \partial^2_2 R_{23} \partial_{\ddot{u}_{2}} ] {\cal S}_{12} \nonumber
\end{eqnarray}
using again identities such that
$\partial_1 R_{14} = - \partial_4 R_{14}$, expanding the $R$
and integrating by parts over $t_4$ and $t_3$ one can
write
\begin{eqnarray}
&&\delta_1 {\cal S}_{12} = \frac{1}{2} {\cal S}_{34}
[ (1 + A_1 \overleftarrow{\partial_4} + A_2
\overleftarrow{\partial_4}^2) (\delta_{14} - \delta_{24})
\partial_{u_{12}} +
(\overleftarrow{\partial_4} + A_1 \overleftarrow{\partial_4}^2)
(\delta_{14} \partial_{\dot{u}_{1}} + \delta_{24}
\partial_{\dot{u}_{2}})
+ \overleftarrow{\partial_4}^2 (\delta_{14} \partial_{\ddot{u}_{1}} +
\delta_{24} \partial_{\ddot{u}_{2}}) ]  \nonumber\\
&&
\times
[ (1 + A_1 \overleftarrow{\partial_3} + A_2
\overleftarrow{\partial_3}^2) (\delta_{13} - \delta_{23})
\partial_{u_{12}} +
(\overleftarrow{\partial_3} + A_1 \overleftarrow{\partial_3}^2)
(\delta_{13} \partial_{\dot{u}_{1}} + \delta_{23}
\partial_{\dot{u}_{2}})
+ \overleftarrow{\partial_3}^2 (\delta_{13} \partial_{\ddot{u}_{1}} +
\delta_{23} \partial_{\ddot{u}_{2}}) ]  {\cal S}_{12}
\end{eqnarray}
This is then in the form where, as above, all time derivatives can
be replaced by derivatives over fields acting either on
${\cal S}_{12}$ or ${\cal S}_{34}$ using identities such as
(\ref{idi}) together with $\partial_4 \partial_3 {\cal S}_{34} = - \dot u_3 \dot u_4
\partial^2_{u_{34}} {\cal S}_{34}$. The evaluation of the
second term in (\ref{s2}) proceeds similarly and the sum
of the two yields the (\ref{dG},\ref{frgA},\ref{frgB},\ref{frgC}) in the text.
Note that causality must be enforced at each step of the
calculation.

\end{widetext}

\newpage

\section{Integrable ``unirelaxational model''}
\label{marvelous} 

In this section we introduce a set of integrable models
in various dimensions which can be used as a check of
the FRG equations derived in this paper. This models
have a remarkable property that despite being random the
dynamics is extremely simple and the relaxation 
time scales are simply that of a single mode generalized
oscillator. 

Let us consider first the toy model in zero dimension:
\begin{eqnarray}
&& \eta(u_t) \partial_t u_t = f(u_t) - m^2 u_t \label{marvelous2}
\end{eqnarray}
with $f(u)=- V'(u)$, when:
\begin{eqnarray}
&& \eta(u) = \overline{\eta} ( 1 - m^{-2} f'(u) ) 
\end{eqnarray}
it can be rewritten:
\begin{eqnarray}
&& \partial_t (f(u_t) - m^2 u_t) = - \frac{m^2}{\overline{\eta}} (f(u_t) - m^2 u_t)
\end{eqnarray}
which can be integrated exactly, yielding
a pure exponential relaxation with a single time scale:
\begin{eqnarray} \label{g4}
&& f(u_t) - m^2 u_t = e^{- t \frac{m^2}{\overline{\eta}}}  (f(u_0) - m^2 u_0)
\end{eqnarray} 
Drawing $F(u)=f(u)- m^2 u$ as a function of $u$ we see that 
all initial conditions starting in an interval between two 
adjacent minima and maxima and which contains a zero of 
$F(u)$, will converge exponentially to this zero of $F(u)$.\cite{footnote13} 
Thus, if $F(u)$ has several zeroes the convergence will be
to minima and maxima of the potential
energy $H(u)=V(u) + \frac{1}{2} m^2 u^2$ 
(depending on initial condition). This should not be a 
surprise since 
in that case $\eta(u)$ changes sign so the dynamics is no more dissipative.
One can then extend the system to non-zero temperature $T>0$, imposing FDT with a
stationary measure $e^{- H(u)/T}$ 
by adding
$\zeta_t$ to the right hand side of (\ref{marvelous2}) with correlations:
\begin{eqnarray}
&& \langle \zeta_t \zeta_{t'} \rangle = 2 T \eta(u)
\end{eqnarray}
Note that this means imaginary noise at points where $\eta(u)$ is negative. Thus
the convergence to the maxima of $V(u)$ is killed by interference effects.

The exact response function associated to the $u$ field can be
obtained for this model at $T=0$. Upon adding an infinitesimal
perturbation $h_t$ on the r.h.s. of (\ref{marvelous2}) 
the change $\delta u$ is such that:
\begin{eqnarray}
&& (f'(u_t) - m^2 ) \delta u_t = - m^2 \int_{t'} R^{(0)}_{t t'} h_{t'} 
\end{eqnarray}
where $u_t$ is given by (\ref{g4}) and 
$R^{(0)}_{t t'} = \frac{1}{\overline{\eta}} e^{- t \frac{m^2}{\overline{\eta}}}$.
The disorder averaged response function is thus:
\begin{eqnarray}
&& R_{t t'} = \overline{(1 - \frac{f'(u_t)}{m^2})^{-1} }
R^{(0)}_{t t'}
\end{eqnarray}
In the large time limit time translational invariance is restored and
one finds:
\begin{eqnarray}
R_{t t'} = R^{(0)}_{t t'}
\end{eqnarray}
This a consequence of the following property:
\begin{eqnarray}
&& \overline{(1 - \frac{f'(u_t)}{m^2})^{-1}} = 1 
\end{eqnarray}
where for each realization of the random function $f(u)$, $u_t$
is the solution of:
\begin{eqnarray}
f(u_t) = m^2 u_t
\end{eqnarray}
and the average is taken w.r.t. any translationally invariant
distribution for $f(u)$. 

A similar model may be introduced in arbitrary dimension. It is defined as:
\begin{eqnarray}
&& \overline{\eta} \dot F_{rt} = - F_{rt} + \zeta_{rt} \label{marv3} \\
&& F_{rt} = \nabla^2 u_{rt} + f(u_{rt},r) 
\end{eqnarray} 
This yields the equation of motion:
\begin{eqnarray}
&& - \overline{\eta} \nabla^2 \dot u_{rt} - 
\overline{\eta} f'(u_{rt},r) \dot u_{rt} =
\nabla^2 u_{rt} + f(u_{rt},r)  + \zeta_{rt} \nonumber \\
&& 
\end{eqnarray}
This is identical in form to the model discussed in the
text with the exception that the damping coefficient 
is wave vector $q$ dependent and vanishes as $q^2$. 
Upon averaging over disorder one obtains an MSR action
identical to the model studied in the text apart from the
$q^2$ mean damping with:
\begin{eqnarray} \label{eq:crosscorrels2}
&& G(u) = \overline{\eta} \Delta'(u) \\
&& A(u) = - \overline{\eta}^2 \Delta''(u) 
\end{eqnarray}
and no other higher order vertex 
for a gaussian distributed $f$ (more general expressions 
can be easily obtained for non-gaussian distributions).
The bare response function of this model factors as:
\begin{eqnarray} \label{response2}
&& R_{q \omega} = \frac{1}{q^2 (1 + \overline{\eta} i \omega)} 
\end{eqnarray}
Similar arguments as above yield that this is also the
exact response. 

The one loop Wilson FRG of this model is very similar to the
one performed for the model in the text. Since the relaxation time
is dimensionless in this model vertices such as $G$ and 
$A$ scale identically to $\Delta$. Hence the appropriate 
rescaled functions for these vertices are $\tilde G \sim \Lambda_l^{- \epsilon} 
G$, $\tilde A \sim \Lambda_l^{- \epsilon} A$. The one loop FRG equations are
identical to the one given in the text for $\tilde G$ and $\tilde A$ apart from the
(linear) rescaling part (not involving $\zeta$)
being identical to that for $\tilde \Delta$. One can then check that the
relation:
\begin{eqnarray}
&& \tilde G(u) = \tilde \Delta'(u) \\
&& \tilde A(u) = - \overline{\eta}^2 \tilde \Delta''(u) 
\end{eqnarray}
specific to this model, are indeed exactly 
preserved by the FRG, as announced in the text. Computing 
the correction to the self energy yields:
\begin{eqnarray}
&& \partial_l \Sigma(\omega) = A_d \Lambda_l^{d-2} [
 \Delta''(0) (R(\omega) - R(0)) 
\\
&& + G'(0) (2 i \omega R(\omega) - i \omega R(0)) 
+ A(0) \omega^2 R(\omega) ] \nonumber 
\end{eqnarray}
One then checks that this exactly vanishes using
$A(0) = - \overline{\eta}^2 \Delta''(0)$, $G'(0) =  \overline{\eta} \Delta''(0)$
and the above exact form for $R(\omega)$.

This model can be further generalized to include second
time derivative terms $D \neq 0$. Adding the term
$D \ddot F_{rt}$ to the l.h.s. of (\ref{marv3})
one obtains the model in the text with
$\overline{\eta} \to q^2 \overline{\eta}$ and
$D \to q^2 D$ in the bare inverse response function.
Similar arguments yield invariance of the FRG function 
within the manifold (\ref{manifold} ) given in the text (third cumulants 
have also been included).


\begin{thebibliography}{10}

\bibitem{fisher_huse}
D.S. Fisher, D.A. Huse Phys. Rev. B {\bf 38} 373 (1988).

\bibitem{Cuku}
L. F. Cugliandolo and J. Kurchan; Phys. Rev. Lett. {\bf 71}, 173 (1993)
and J. Phys. {\bf A27}, 5749 (1994). H. Kinzelbach and H. Horner; J.
Phys. I (France) {\bf 3}, 1329 (1993), {\it
  ibid} {\bf 3}, 1901 (1993).

\bibitem{Cuku2}
L. Cugliandolo, J. Kurchan, P. Le Doussal, Phys. Rev. Lett. {\bf 76}
2390
  (1996).

\bibitem{Cukureview}
J.P.Bouchaud, L.F. Cugliandolo, J. Kurchan and M. Mezard in {\em Spin
glasses and random fields}
Ed. A. P. Young, World Scientific, Singapore, 1998.

\bibitem{aging_traps}
M. Feigelman and V. M. Vinokur, J. Phys. France {\bf 49} 1731 (1988).
J.-P. Bouchaud; J. Phys. I (France) {\bf 2}, 1705 (1992).

\bibitem{sibani} P. Sibani Phys. Rev. B {\bf 35} 8572 (1987) K. H.
Hoffmann and P. Sibani Z. Phys. B 80 429 (1990)

\bibitem{rfim1d}
D.S. Fisher, P. Le Doussal and C. Monthus, Phys. Rev. Lett.
{\bf 80} 3539 (1998) and Phys. Rev. E 64, 066107 (2001).

\bibitem{Vihaoc}
E. Vincent, J. Hammann e M. Ocio; in {\it `Recent progress in Random
Magnets'},
ed. D. H. Ryan, World Scientific, Singapore (1992).

\bibitem{anderson}
see e.g. S.R. Anderson Phys. Rev B {\bf 36} 8435 (1987).

\bibitem{blatter_vortex_review}
G. Blatter, M. V. Feigel'man, V. B. Geshkenbein, A. I Larkin and V. M.
Vinokur, Review of Modern Physics {\bf 66} 1125 (1994).

\bibitem{Legi}
T. Giamarchi and P. Le Doussal Phys. Rev. {\bf B52} 1242 (1995).

\bibitem{nattermann_book_young}
T. Nattermann,  {\em Statics and dynamics of disordered elastic
systems} in {\em Spin glasses and random fields}
  edited by A.~P. Young (World Scientific, Singapore, 1998), p.\ 277.

\bibitem{creepexp}
S.~Lemerle et al. Phys. Rev. Lett. {\bf 80} (1998) 849.

\bibitem{gruner_revue_cdw}
G. Gr{\"u}ner, Rev. Mod. Phys. {\bf 60},  1129  (1988).

\bibitem{andrei_wigner_2d}
E.~Y. Andrei and {al.}, Phys. Rev. Lett. {\bf 60},  2765  (1988).

\bibitem{fisher_functional_rg}
D.~S. Fisher, Phys. Rev. Lett. {\bf 56},  1964  (1986).

\bibitem{balents_frg_largen}
L. Balents and D.~S. Fisher, Phys. Rev. B {\bf 48},  5959  (1993).

\bibitem{balents_loc}
L. Balents, Europhys. Lett. {\bf 24},  489  (1993).

\bibitem{narayan_fisher}
O. Narayan and D.~S. Fisher, Phys. Rev. B {\bf 46},  11520  (1992),
Phys. Rev. B {\bf 48},  7030  (1993).

\bibitem{nattermann_depinning}
T. Nattermann, S. Stepanow, L.~H. Tang, and H. Leschhorn, J. Phys.
(Paris) {\bf 2},  1483  (1992). H. Leschhorn, 
T. Nattermann, S. Stepanow, and L.~H. Tang, Annalen der
Physik {\bf 6},  1 (1997).

\bibitem{frg2loop}
P.~Chauve, P.~Le Doussal and K.J. Wiese,
Phys. Rev. Lett. {\bf 86} (2001) 1785
P.~Le Doussal and K.J. Wiese and P.~Chauve
cond-mat/0205108, Phys. Rev. B 66, 174201 (2002)
and cond-mat/0304614.

\bibitem{chauve_creep}
P. Chauve, T. Giamarchi, and P. {Le Doussal}, Europhys. Lett. {\bf 44},
110 (1998) and Phys. Rev. B {\bf 62} 6241 (2000).

\bibitem{staticslong}
L. Balents and P. Le Doussal, in preparation.

\bibitem{us_short}
L. Balents and P. Le Doussal, cond-mat/0205358.

\bibitem{denis}
Denis A. Gorokhov, Daniel S. Fisher, Gianni Blatter,
cond-mat/0205416.

\bibitem{creep_pheno}
T. Nattermann, Europhys. Lett. {\bf 4},  1241  (1987),
L.~B. Ioffe and V.~M. Vinokur, J. Phys. C {\bf 20},  6149  (1987),
T. Nattermann, Y. Shapir, and I. Vilfan, Phys. Rev. B {\bf 42},  8577
(1990), M. Feigelman, V.~B. Geshkenbein, A.~I. Larkin, and V. Vinokur, Phys.
Rev. Lett. {\bf 63},  2303  (1989).

\bibitem{dropevid} See, e.g. A. A. Middleton, Phys. Rev. {\bf B63}
  60202 (2001); T. Hwa and D. S. Fisher, Phys. Rev. {\bf B49}, 3136
  (1994).

\bibitem{drossel_barrier}
B. Drossel and M. Kardar, Phys. Rev. E {\bf 52},  4841  (1995).

\bibitem{mikheev_barrier_2d}
L.~V. Mikheev, B. Drossel, and M. Kardar, Phys. Rev. Lett. {\bf 75},
1170 (1995).

\bibitem{drossel_barrier_3d}
B. Drossel, J. Stat. Phys. {\bf 82},  431  (1996).

\bibitem{mezpar}
M. M\'ezard, G. Parisi J. Phys. I (France) {\bf 1} 809 (1991).
L. Cugliandolo et al. Phys. Rev. Lett. {\bf 76} 2390 (1996).

\bibitem{frgN}
P.~Le Doussal and K.J. Wiese, Phys. Rev. Lett. 89, 125702 (2002)
and cond-mat/0305634, Phys. Rev. E. in press.

\bibitem{infinited}
D.S. Fisher Phys. Rev. {\bf B 31} 1396 (1985); 
J.~Vannimenus, B.~Derrida
J. Stat. Phys. {\bf 105} 1 (2001).

\bibitem{msr}
P.~C. Martin, E.~D. Siggia, and H.~A. Rose, Phys. Rev. A {\bf 8},  423
(1973). H.~K. Janssen, Z. Phys. B {\bf 23},  377  (1976).

\bibitem{footnoteq0}
except for the massless $q=0$ mode, which is not important.

\bibitem{footnotek} The idea is that by inversion symmetry of the
averaged
action spatial gradient corrections must start as $k^2$ multiplied by
some
power of $\omega$ and can thus always be neglected compared to the
elastic term $k^2$

\bibitem{footnote10} 
in that case 
$D > \overline{\eta}^2/4$ corresponds to underdamped
dynamics and no determined sign for the response function
(consistent with the connected moment (\ref{conn1}) becoming
negative for $D > \overline{\eta}^2/2$). 


\bibitem{footnotepert}
Perturbation theory (e.g. within the Wilson mode integration) can be
equivalently
performed, splitting as usual $S=S_{quad}+S_{int}$, either (i) including
no kinetic term in the quadratic part
of the action $S_{quad}$ (using then $R^{quad}_{ktt'}=\delta_{tt'}/q^2$
but considering
in all internal response lines the chain of graphs represented in Fig.
\ref{graph13},
with multiple $n=1$, $m \ge 1$ F vertices on internal lines) (ii)
including just $\overline{\eta}$
in the quadratic part (using then $R^{quad}_{k}(i\omega)=1/(q^2+ i
\overline{\eta} \omega)$
and considering all graphs with multiple $n=1$, $m \ge 2$ F vertices on
internal lines)
or (iii) considering the total quadratic part (e.g. using the full
$R_{k,l}(i \omega)$ and
only $n \ge 2$ vertices).

\bibitem{ERG}
G. Schehr, P. Le Doussal cond-mat/0304486, Phys. Rev. E. in press. 
P.~Chauve et al. Phys. Rev. E 64, 051102 (2001). 


\bibitem{depinning} 
Note that for depinning
the $G$ term in (\ref{dG}) yields $G \sim \eta \Delta^2$ and gives a part of the
two loop $O(\epsilon^2)$ correction
to $z$ of the same order than the one coming from the direct correction
$\eta \Delta^2$ to $\eta$ to two loops.

\bibitem{randommass}
Consider the equation of motion
$\frac{1}{2} \partial_u D(u_{rt},r) \dot u_{rt}^2 + D(u_{rt},r) \ddot u_{rt} =\nabla^2 u_{rt} + 
f(u_{rt},r)$. Multipying by $\dot u_{rt}$ 
we note that it has conserved ''total energy''
$\partial_t E=0$ with $E= \int_r [ H[u_{rt},r] + \frac{1}{2} D(u_{rt},r) \dot u_{rt}^2 ]$.
It thus corresponds to some Hamiltonian classical dynamics (with a non
linear mass term). The associated MSR dynamical action obeys the
continuous global $\lambda$ symmetry. This however is 
not FDT, even if we choose $\lambda=1/T$, meaning that if we add
the standard terms (\ref{msr1}) it would not satisfy FDT. 
The reason for that can be traced to the fact that the
boundary term obtained upon applying the symmetry (\ref{symfdt})
is $(E(t_f)-E(t_i))/T$ and thus contain extra time derivatives
which invalidate the arguments made in Section. It is possible 
however to add additional noise at the boundary, i.e. a bulk
term $ - T \int_{rt} \partial_t ( D(u_{rt},r) (i \hat u_{rt})^2)$,
so that FDT is satisfied, i.e. the extra time derivative term is
cancelled upon integration over $\hat u_{rt}$ at the boundary.
This is relevant also for the disorder case, since 
averaging over disorder
one gets, an $m=2$ term with $C$ a first derivative, 
$B=C'/2$ and $A=0$. This indeed is the condition for 
invariance of the $m=2$ term under the
continuous global $\lambda$ symmetry.



\bibitem{kpz}
another symmetry often considered associated to FDT is 
$i \hat u_{rt} \to - i \hat u_{r,-t} + (T \eta)^{-1} 
\frac{\delta H}{\delta u}|_{u_{r,-t}}$,
$u_{rt} \to  u_{r,-t}$. This is a symmetry of the unaveraged
dynamical effective action. It is useful e.g. to 
study \cite{frey} consequences of the accidental FDT property of the
one dimensional KPZ equation. It seems however a priori less
useful in order to study the effective action once 
one integrates over modes, i.e. in the
context of Wilson FRG (also further complications 
arise after averaging over disorder). It may be
worth studying within the exact RG context, which is
beyond the
scope of the present paper. 

\bibitem{exactsolu} as a curiosity note the exact (implicit) solution
for $\theta=2$
obtained by writing (\ref{sigtti}) as $\tilde{\beta} d(k^{-2})/d\Sigma=2
k^{-2} + 2 (i \omega + \Sigma)^{-1}$,
which yields $\Sigma_k(i \omega) = - \frac{\tilde{\beta}}{2}
\ln( k^2 (1 + 2 \int_0^\Sigma d \lambda e^{-2 \lambda/\tilde{\beta}} (i
\omega + \lambda)^{-1})$.


\bibitem{footnoteint} Note that since $dy/dg = (1+g) e^g$ formally
  this integral reads simply $\int dg e^{g + {\cal Y} g e^g}$ but the
contour remains to be worked
out.

\bibitem{manyscales}
Ideally one should also take into account the influence of modes with
wavevectors $q<k$ on the relaxation of mode $k$ since these are
obviously left out of the RG scheme. However simple arguments
suggest that this influence decays fast: the influence of the jumps in
the mode $q$ on the mode $k$ may decay as fast as $e^{- k/q}$.
At least this is necessary condition for such droplet arguments
to make sense.

\bibitem{sinai}
Pierre Le Doussal, Cecile Monthus,Daniel S. Fisher
cond-mat/9811300, Phys. Rev. E, {\bf 59} 4795 (1999).
L. Laloux, P. Le Doussal cond-mat/9705249
Phys. Rev. E. {\bf 57} 6296 (1998).

\bibitem{mitra_pld}
P.P. Mitra, P. Le Doussal, Phys. Rev. B  {\bf 44} 12035 (1991).


\bibitem{footnote13}
Note however that if the initial condition starts in
an interval two adjacent minima and maxima which does not
contains a zero of $F(u)$, it instead converges in finite time 
to one of the endpoint of the interval at which $\eta(u)$ vanishes.
The solution in (\ref{g4}) analytically continues the
relaxation of the force to zero. 

\bibitem{toy}
P. Le Doussal and C. Monthus, cond-mat/0204168
Physica A 317 (1-2): 140-198 JAN 1 2003. 


\bibitem{frey}
E. Frey, U. Tauber and T. Hwa cond-mat/9601049. 

\bibitem{russians} B. Altshuler et al. JETP Letters {\bf 45}
199 (1987), Zh Eksp. Teor. Fiz. {\bf 94}
258 (1988). 

\end{thebibliography}

\end{document}